\documentclass[twocolumn, tighten, twocolappendix]{aastex7}

\shorttitle{\texttt{pop-cosmos}}
\shortauthors{Thorp et al.}

\usepackage{amsmath}	
\usepackage{bm}
\usepackage{booktabs}
\usepackage{threeparttable}
\usepackage[normalem]{ulem}

\defcitealias{leistedt23}{L23}
\defcitealias{alsing24}{A24}
\defcitealias{thorp24}{T24}
\defcitealias{gallazzi05}{G05}
\defcitealias{gallazzi14}{G14}

\allowdisplaybreaks

\begin{document}

\title{\texttt{pop-cosmos}: Insights from generative modeling of a deep, infrared-selected galaxy population}

\correspondingauthor{Stephen Thorp}
\email{stephen.thorp@ast.cam.ac.uk}

\author[0009-0005-6323-0457]{Stephen Thorp}
\affiliation{The Oskar Klein Centre, Department of Physics, Stockholm University, AlbaNova University Centre, SE 106 91 Stockholm, Sweden}
\affiliation{Institute of Astronomy and Kavli Institute for Cosmology, University of Cambridge, Madingley Road, Cambridge CB3 0HA, UK}
\email{stephen.thorp@ast.cam.ac.uk}

\author[0000-0002-2519-584X]{Hiranya V.\ Peiris}
\affiliation{Institute of Astronomy and Kavli Institute for Cosmology, University of Cambridge, Madingley Road, Cambridge CB3 0HA, UK}
\affiliation{The Oskar Klein Centre, Department of Physics, Stockholm University, AlbaNova University Centre, SE 106 91 Stockholm, Sweden}
\email{hiranya.peiris@ast.cam.ac.uk}

\author[0009-0004-7935-2785]{Gurjeet Jagwani}
\affiliation{Institute of Astronomy and Kavli Institute for Cosmology, University of Cambridge, Madingley Road, Cambridge CB3 0HA, UK}
\affiliation{Research Computing Services, University of Cambridge, Roger Needham Building, 7 JJ Thomson Ave, Cambridge CB3 0RB, UK}
\email{gj329@cam.ac.uk}

\author[0000-0003-1943-723X]{Sinan Deger}
\affiliation{Institute of Astronomy and Kavli Institute for Cosmology, University of Cambridge, Madingley Road, Cambridge CB3 0HA, UK}
\email{sd2062@cam.ac.uk}

\author[0000-0003-4618-3546]{Justin Alsing}
\affiliation{The Oskar Klein Centre, Department of Physics, Stockholm University, AlbaNova University Centre, SE 106 91 Stockholm, Sweden}
\email{justin.alsing@fysik.su.se}

\author[0000-0002-3962-9274]{Boris Leistedt}
\affiliation{Astrophysics Group, Imperial College London, Blackett Laboratory, Prince Consort Road, London, SW7 2AZ, UK}
\email{b.leistedt@imperial.ac.uk}

\author[0000-0002-0041-3783]{Daniel J.\ Mortlock}
\affiliation{Astrophysics Group, Imperial College London, Blackett Laboratory, Prince Consort Road, London, SW7 2AZ, UK}
\affiliation{Department of Mathematics, Imperial College London, London SW7 2AZ, UK}
\email{d.mortlock@imperial.ac.uk}

\author[0000-0002-0352-9351]{Anik Halder}
\affiliation{Institute of Astronomy and Kavli Institute for Cosmology, University of Cambridge, Madingley Road, Cambridge CB3 0HA, UK}
\email{ah2425@cam.ac.uk}

\author[0000-0001-6755-1315]{Joel Leja}
\affiliation{Department of Astronomy \& Astrophysics, The Pennsylvania State University, University Park, PA 16802, USA}
\affiliation{Institute for Computational \& Data Sciences, The Pennsylvania State University, University Park, PA 16802, USA}
\affiliation{Institute for Gravitation \& the Cosmos, The Pennsylvania State University, University Park, PA 16802, USA}
\email{joel.leja@psu.edu}



\begin{abstract}
\noindent
We present an extension of the \texttt{pop-cosmos} model for the evolving galaxy population up to redshift $z\sim6$. The model is trained on distributions of observed colors and magnitudes, from 26-band photometry of $\sim420,000$ galaxies in the COSMOS2020 catalog with \textit{Spitzer} IRAC $\textit{Ch.\,1}<26$. The generative model includes a flexible distribution over 16 stellar population synthesis (SPS) parameters, and a depth-dependent photometric uncertainty model, both represented using score-based diffusion models. We use the trained model to predict scaling relationships for the galaxy population, such as the stellar mass function, star-forming main sequence, and gas-phase and stellar metallicity vs.\ mass relations, demonstrating reasonable-to-excellent agreement with previously published results. We explore the connection between mid-infrared emission from active galactic nuclei (AGN) and star-formation rate, finding high AGN activity for galaxies above the star-forming main sequence at $1\lesssim z\lesssim 2$. Using the trained population model as a prior distribution, we perform inference of the redshifts and SPS parameters for 429,669 COSMOS2020 galaxies, including 39,588 with publicly available spectroscopic redshifts. The resulting redshift estimates exhibit minimal bias ($\text{median}[\Delta_z]=-8\times10^{-4}$), scatter ($\sigma_\text{MAD}=0.0132$), and outlier fraction ($6.19\%$) for the full $0<z<6$ spectroscopic compilation. These results establish that \texttt{pop-cosmos} can achieve the accuracy and realism needed to forward-model modern wide--deep surveys for Stage IV cosmology. We publicly release \texttt{pop-cosmos} software, mock galaxy catalogs, and COSMOS2020 redshift and SPS parameter posteriors.
\end{abstract}



\section{Introduction}
The next generation of imaging surveys, such as those which will be undertaken by the Vera C.\ Rubin Observatory \citep{ivezic19}, \textit{Euclid} \citep{mellier24}, and the \textit{Nancy Grace Roman Space Telescope} \citep{spergel15}, will produce photometric catalogs containing many billions of galaxies.  The scale of these data-sets means that traditional data analysis techniques, typically based on parametric models and individual object likelihoods, will not be viable.  To properly account for the complexity of these data-sets, including selection, will require the use of non-parametric models, forward simulation of the data, and fast emulation of observable galaxy properties.

The need for accurate photometric redshifts for Stage IV cosmology surveys \citep[see][]{mandelbaum18, newman22} has motivated the development of sophisticated forward-modeling and simulation based inference techniques for photometric surveys. To infer a population-level redshift distribution, or obtain accurate redshifts for individual objects, it is essential to understand how galaxy spectral energy distributions (SEDs) evolve over cosmic time. Moreover, modeling the full galaxy population provides a way of controlling systematics such as intrinsic alignments \citep[see, e.g.,][]{mandelbaum06, mandelbaum11, troxel15, joachimi15}, supernova--host correlations \citep[see, e.g.,][]{childress14, wiseman22, wiseman23, jones24, karchev25}, and the effect of baryonic feedback \citep[see, e.g.,][]{mccarthy17, schaye23}. In this way, redshift inference and cosmology are inextricably linked to the evolution of the galaxy population (for an overview, see \citealp{alsing23}). 

Numerous parallel developments over the past decade have been moving toward addressing these challenges for Stage IV cosmology. These have included: fully differentiable implementations of physical models \citep[e.g.,][]{hearin21, hearin23, alarcon23}; emulators for stellar population synthesis (SPS) models \citep[e.g.,][]{han12, han14, alsing20, kwon23, matthews23, tortorelli25}; machine-learning accelerators for hydrodynamical or radiative transfer models \citep[e.g.][]{lovell19, gilda21, qiu22}; parametric population models combined with rapid image simulation based on morphology indicators \citep{herbel17, moser24, fischbacher24_py, fishbacher24_ufig, fishbacher24, tortorelli25}; flow-based models for the color--redshift relation \citep[e.g][]{crenshaw24}; SPS-based forward-modeling for spectro-photometric data \citep{hahn23, hahn24}; flow-based population models \citep{li24}; galaxy- and population-level Bayesian model comparison \citep{han19, han23}; explicit physically-motivated priors on galaxy properties in template fitting \citep[e.g.][]{tanaka15}; hierarchical models for clustering/cross-correlation redshift distributions \citep[e.g.,][]{sanchez19, rau20, rau23}; stratified learning techniques \citep{autenrieth24}; and neural posterior estimators \citep{hahn22, wang23_sbi}.

Our work on galaxy population modeling was also driven initially by the problem of photometric redshift distributions, starting with a Bayesian hierarchical model for inferring galaxy redshift distributions \citep{leistedt16}.  One significant limitation was the use of a finite set of predefined template galaxy SEDs, motivating more flexible data-driven approaches to SED modeling \citep{leistedt17, leistedt19}.  The limited constraining power of photometric data alone also motivated the inclusion of physics constraints via SPS (for a review of SPS see, e.g., \citealp{tinsley80, conroy13, iyer25}), for which we developed a fast neural network emulator, \texttt{Speculator} \citep{alsing20}. \texttt{Speculator} was combined with a parametric model of the galaxy population (including SPS parameters) to infer galaxy redshift distributions with sufficient fidelity to meet Stage III survey requirements \citep{alsing23} and to self-calibrate the data model, e.g., zero points \citep{leistedt23}.  However, for Stage IV surveys the use of parametric distributions for the galaxy population and SPS parameters is too restrictive.  To overcome this limitation, in \citet[hereafter \citetalias{alsing24}]{alsing24} we used a flexible diffusion model which was calibrated on the COSMOS2020 catalog \citep{weaver22} from the Cosmic Evolution Survey (COSMOS; \citealp{scoville07}) to produce the \texttt{pop-cosmos} galaxy population model. We also developed a complementary framework for rapid SPS parameter inference for individual objects using the \texttt{pop-cosmos} population model as a prior \citep[hereafter \citetalias{thorp24}]{thorp24}.

In this paper, we present an extended version of the \texttt{pop-cosmos} generative model, calibrated on a \textit{Spitzer} IRAC-selected galaxy sample from the COSMOS2020 catalog \citep{weaver22}, which is a magnitude deeper than the sample used in \citetalias{alsing24}. We also present a first public release of software for sampling from and evaluating the \texttt{pop-cosmos} population model, along with COSMOS-like mock galaxy catalogs drawn from it, and an update to the posterior samples from \citetalias{thorp24} for individual COSMOS galaxies.  

In Section~\ref{sec:data}, we describe the selection of our training data. In Section~\ref{sec:methods}, we give an overview of the \texttt{pop-cosmos} modeling framework and the improvements made since \citetalias{alsing24}, with the performance of the new model evaluated in Section~\ref{sec:validation}. Astrophysical insights extracted from the model are presented in Section~\ref{sec:results_astro}. We discuss the wider context of our results, and directions for future developments in Section~\ref{sec:discussion}. This paper will be followed by both a more detailed astrophysical application of the new model (S.\ Deger et al., in preparation) and two different applications to the Kilo Degree Survey (KiDS; \citealp{kuijken19}) catalogs (B.\ Leistedt et al.\ and A.\ Halder et al., both in preparation).

Where relevant, we assume a flat $\Lambda$CDM cosmology with $H_0=67.66$~km\,s$^{-1}$\,Mpc$^{-1}$ and $\Omega_{\textrm{m}}=0.3097$ \citep{planck18}. Magnitudes and relative fluxes are all based on the AB system with a reference flux density of $f_\text{AB}=3631$~Jy \citep{oke83, fukugita96}.

\section{Data}
\label{sec:data}

We now describe the multi-wavelength photometric data we use for calibrating the generative model presented in this work, and the spectroscopic data we use for validation.

\subsection{Photometric Data}
\label{sec:photometry}
As in \citetalias{alsing24}, we use 26-band photometry from the COSMOS2020 catalog \citep{weaver22}; specifically v2.1 of the \texttt{Farmer} catalog that uses a profile-fitting method for the photometric extraction \citep{weaver23}. From COSMOS2020, we use $u$-band data from the Canada--France--Hawai'i--Telescope \citep{laigle16, sawicki19}, $grizy$ data from Subaru Hyper Suprime-Cam \citep[HSC;][]{aihara19}, $YJHK_S$ data from UltraVISTA \citep{mccracken12}, Channel 1 \& 2 of \textit{Spitzer} IRAC \citep{steinhardt14, ashby13, ashby15, ashby18, moneti22}, and optical narrow- and medium-band data from Suprime-Cam \citep{taniguchi07, taniguchi15}. We select our sample within the 1.27~deg$^2$ ``combined'' mask for COSMOS2020, which includes HSC, UltraVISTA, and IRAC coverage, and avoids bright stars \citep{weaver22}.

We follow \citet{weaver23_smf} in limiting our sample to $<26$~mag in \textit{Spitzer} IRAC Channel 1 (\textit{Ch.\,1}). Based on deeper data available from the Cosmic Assembly Near-infrared Deep Extragalactic Legacy Survey (CANDELS; \citealp{grogin11, koekemoer11}) in a subset of the COSMOS field, specifically the catalog from \citet{nayyeri17}, \citet{weaver22, weaver23_smf} estimate that the COSMOS2020 detection efficiency\footnote{\citet{weaver22,weaver23_smf} find that their photometric extraction routines recover 75\% of $\textit{Ch.\,1}\approx26$ sources in the CANDELS survey area within the COSMOS field.} at $\textit{Ch.\,1}=26$ is 75\%. The loss of efficiency towards faint \textit{Ch.\,1} magnitudes stems from the fact that detection is based on a co-added (as in \citealp{szalay99, drlica18}) $izYJHK_S$ image, which does not lead to a significant detection for all sources with $25<\textit{Ch.\,1}<26$ (see also \citealp{davidzon17})\footnote{Detection of sources in the mid-infrared (MIR) alone is inefficient for $\textit{Ch.\,1}>22$ \citep[see][]{moneti22} due to the difficulty of deblending the low-resolution IRAC images without a prior informed by the near-infrared (NIR) data.}.

To enable the best use of low signal-to-noise (S/N) data, and to avoid applying a lower flux limit of zero (as was done in \citetalias{alsing24}), we convert the COSMOS2020 catalog fluxes into asinh magnitudes \citep{lupton99}\footnote{See also \citet{matthews23} for a recent motivating application in an SED fitting context.}. These retain the key feature of logarithmic magnitudes \citep{pogson56} --- a reduced dynamic range between the brightest and faintest sources --- while remaining well-behaved mathematically close to a survey's detection limit. All training data take the form of asinh magnitudes and colors. The asinh magnitude corresponding to flux\footnote{This is formally a flux density; flux is used here for brevity.} $f$ is defined as 
\begin{equation}
    m = -\frac{5}{2} \log_{10}(e) \left[\mathrm{asinh}\!\left(\frac{f}{2f_b}\right) + \ln\left(\frac{f_b}{f_\mathrm{AB}}\right)\right],
    \label{eq:asinh}
\end{equation}
where $f_\mathrm{AB} = 3,631 ~\textrm{Jy}$ is the standard reference flux in the AB system \citep{oke83, fukugita96}, and $f_b$ is the flux below which asinh magnitudes transition from being logarithmic to linear, referred to as the softening parameter \citep{lupton99}. For reference, the corresponding definition of a logarithmic AB magnitude would be
\begin{equation}
    m = -\frac{5}{2}\log_{10}\left(\frac{f}{f_\text{AB}}\right).
    \label{eq:log}
\end{equation}
The value of $f_b$ in \ref{eq:asinh} can be different for each band and survey; it is typically chosen to be a factor of a few times the background flux uncertainty.

We set the softening parameter $f_b$ by estimating an effective depth from the reported COSMOS flux uncertainties and logarithmic AB magnitudes; the resultant values are tabulated in Table~\ref{tab:band_calibration} in Appendix \ref{sec:calibration}. These are generally fainter than the $3\sigma$ depths reported by \citet{weaver22} based on a $3''$ aperture. Our choices of $f_b$ correspond to a transition between linear and logarithmic behavior at a signal-to-noise ratio (S/N) of $\sim\!5$. We do not propagate the COSMOS flux uncertainties into asinh magnitude uncertainties at any stage of our pipeline. When training our uncertainty model (Section \ref{sec:unc}), and performing galaxy-level inference (Section \ref{sec:mcmc}), we work with the COSMOS flux uncertainties directly.

A total of 423,262 galaxies from COSMOS2020 pass the $\textit{Ch.\,1}<26$ cut, three times as many as the $r<25$ training set in  \citetalias{alsing24}. This is the sample on which we calibrate our generative model in this work. As in \citetalias{alsing24}, this cut also includes the elimination of objects considered likely stars by \citet{weaver22}\footnote{From the COSMOS2020 public catalog, we remove objects with \texttt{lp\_type=1} and \texttt{lp\_type=-99}, but not objects with \texttt{lp\_type=2} (X-ray AGN identified in \textit{Chandra}; \citealp{civano16}).}. Unlike \citetalias{alsing24}, we apply no further cuts on low flux or S/N objects. We do not omit galaxies that are likely AGN hosts, flagged as having a coincident \textit{Chandra} X-ray detection in the \citet{civano16} catalog; within the $\textit{Ch.\,1}<26$ sample there are 1,951 such galaxies ($\sim0.5\%$ of the catalog). We find 6,407 COSMOS2020 sources with $\textit{Ch.\,1}\geq26$ but $r<25$, implying that there is a small MIR-faint but optically bright population that will be missed by our sample selection criterion\footnote{These do not all pass the selection criteria used by \citetalias{alsing24} and \citetalias{thorp24}. Due to the additional flux and S/N cuts applied there, the number of sources with $\textit{Ch.\,1}\geq26$ is 824 out of 140,745.}.

\subsection{Spectroscopic Data}
\label{sec:spectroscopy}
For our update to the \citetalias{thorp24} SPS parameter and redshift inference for individual galaxies, we validate against the new COSMOS spectroscopic archive from \citet{khostovan25}. Of the 97,929 unique objects in \citet{khostovan25}, 60,203 have an identified counterpart in the COSMOS2020 \texttt{Farmer} catalog from \citet{weaver22}, with 48,291 of these being in our $\textit{Ch.\,1}<26$ sample. The cumulative numbers of $\textit{Ch.\,1}<26$ sources assigned a given confidence level (CL) by \citet{khostovan25} are: $\text{CL}\geq97\%$, 16,902 sources;  $\text{CL}\geq95\%$, 22,848 sources; $\text{CL}\geq80\%$, 33,368 sources; $\text{CL}\geq50\%$, 39,588 sources. The $\text{CL}\geq50\%$ sources correspond to 39,579 unique COSMOS2020 \texttt{Farmer} galaxies; further validation of the cross-match is described in Appendix \ref{sec:ambiguous_spec}.

\section{Methods}
\label{sec:methods}
The full \texttt{pop-cosmos} pipeline is outlined in Section~\ref{sec:overview}, with Sections~\ref{sec:sps}--\ref{sec:training} containing more specific details about some of the components. Parameters and notation are summarised in Tables \ref{tab:notation} \& \ref{tab:sps_notation}.

\subsection{Overview of the Generative Model}
\label{sec:overview}
The \texttt{pop-cosmos} generative model contains the following elements:
\begin{enumerate}
    \item A population distribution $P(\bm{\vartheta}|\bm{\psi})$ over 16 SPS parameters including redshift $\bm{\vartheta}=(\bm{\varphi},z)$. This is a score-based diffusion model \citep{song21} defined by a set of hyperparameters $\bm{\psi}$. The architecture and initialization of this model are given in Section~\ref{sec:init}.
    \item A primary emulator (\texttt{Speculator}; \citealp{alsing20}) for the Flexible Stellar Population Synthesis library (FSPS; \citealp{conroy09, conroy10a, conroy10b}). This deterministically predicts model fluxes $\bm{f}_\text{SPS}(\bm{\vartheta})$ in the 26 COSMOS passbands given a set of input parameters drawn from $P(\bm{\vartheta}|\bm{\psi})$. This is described in Section~\ref{sec:sps}.
    \item A secondary emulator \citep{alsing20, leistedt23} for the flux contributions of 44 key nebular emission lines $\bm{F}_\text{EM}(\bm{\vartheta})$. 
    \item A set of population-wide calibration parameters $(\bm{\alpha}_\text{ZP}, \bm{\beta}_\text{EM}, \bm{\gamma}_\text{EM})$ that model respectively: a zero-point offset in each passband; corrections to the strength of the 44 key nebular emission lines; and variances in the strength of the emission lines.
    \item A non-parametric uncertainty model $P(\bm{\sigma}_P|\bm{f},\bm{\chi})$ that predicts flux uncertainty $\bm{\sigma}_P$ conditional on the calibrated model fluxes, $\bm{f}=\bm{\alpha}_\text{ZP}\circ[\bm{f}_\text{SPS}(\bm{\vartheta}) + \bm{\beta}_\text{EM}\cdot\bm{F}_\text{EM}(\bm{\vartheta})]$. The uncertainty model is a score-based diffusion model \citep{song21}, with hyperparameters $\bm{\chi}$. This model is defined in Section~\ref{sec:unc}.
    \item A Student's $t$ error model $P(\bm{d}|\bm{f}, \bm{\sigma})$ that predicts observable fluxes given model fluxes $\bm{f}$ and a total uncertainty $\bm{\sigma} = \sqrt{\bm{\sigma}_P^2 + \bm{\sigma}_\text{EM}^2}$, where $\bm{\sigma}_\text{EM} = [\bm{\gamma}_\text{EM}\circ(1+\bm{\beta}_\text{EM})]\cdot\bm{F}_\text{EM}(\bm{\vartheta})$, and $\circ$ denotes an element-wise product.
\end{enumerate}
In our main training loop (Section~\ref{sec:training}), we simultaneously optimize the population level parameters $(\bm{\psi}, \bm{\alpha}_\text{ZP}, \bm{\beta}_\text{EM}, \bm{\gamma}_\text{EM})$ by comparing the mock photometry to the COSMOS2020 photometry. 

\subsection{Emulated Stellar Population Synthesis}
\label{sec:sps}

\begin{table}
    \centering
    \caption{Summary of main notation.}
    \label{tab:notation}
    \begin{tabular}{l l}
        \toprule\toprule
        Symbol & Description \\
        \midrule
        & \emph{Population-level quantities} \\
        \cmidrule(l){2-2}
        $\bm{\psi}$ & Population model hyperparameters \\
        $\bm{\chi}$ & Uncertainty model hyperparameters \\
        $\bm{\alpha}_\text{ZP}$ & Fractional zero-point corrections \\
        $\bm{\beta}_\text{EM}$ & Fractional emission line corrections \\
        $\bm{\gamma}_\text{EM}$ & Fractional emission line std.\ devs.\ \\
        \midrule
        & \emph{Galaxy-level quantities} \\
        \cmidrule(l){2-2}
        $\bm{d}$ & Noisy model fluxes \\
        $\bm{f}$ & Calibrated model fluxes \\
        $\bm{m}$ & Calibrated model asinh magnitudes \\
        $\bm{f}_\text{SPS}(\bm{\vartheta})$ & SPS model fluxes \\
        $\bm{F}_\text{EM}(\bm{\vartheta})$ & Emission line model fluxes \\
        $\bm{\sigma}$ & Model flux uncertainties \\
        $\bm{\sigma}_P$ & Model photometric uncertainty \\
        $\bm{\sigma}_\text{EM}$ & Emission line uncertainty \\
        $\bm{\vartheta}$ & Galaxy level parameters $(\bm{\varphi},z)$\\
        $\bm{\varphi}$ & Non-redshift SPS parameters \\
        \midrule
        & \emph{Photometry-related} \\
        \cmidrule(l){2-2}
        $f$ & Flux in the AB system \\
        $f_\text{AB}$ & Flux of the AB standard source \\
        $f_b$ & Flux softening parameter \\
        $m$ & asinh magnitude \\
        \midrule
        & \emph{Diffusion-related} \\
        \cmidrule(l){2-2}
        $\bm{w}(t)$ & Brownian motion process \\
        $\bm{x}(t)$ & State vector at time $t$ \\
        $t$ & Time index in the diffusion process \\
        $T$ & Maximum time, $\bm{x}(T)\sim\text{base density}$\\
        $p_t(\bm{x})$ & Marginal density of $\bm{x}(t)$ at time $t$\\
        $\nabla_{\bm{x}}p_t(\bm{x})$ & Score at time $t$ \\
        $\bm{s}_{\bm{\Theta}}(\bm{x},t)$ & Score network with hyperparameters $\bm{\Theta}$ \\
        $\beta(t)$ & Scaling factor in VP SDE \\
        $\beta_\text{min}$, $\beta_\text{max}$ & Minimum/maximum scaling factor \\
        \bottomrule
    \end{tabular}
\end{table}

\begin{table*}
    \centering
    \caption{Summary of SPS parameters.}
    \label{tab:sps_notation}
    \begin{tabular}{l c c l}
        \toprule\toprule
        Symbol / Unit & Lower Bound & Upper Bound & Description\\
        \midrule
        & & & \emph{Base parameters} \\
        \cmidrule(l){4-4}
        $\log_{10}(M_\text{form}/M_\odot)$ & - & - & Logarithm of stellar mass formed\\
        $\log_{10}(Z/Z_\odot)$ & $-1.98$ & 0.19 & Logarithm of stellar metallicity \\
        $\Delta\log_{10}(\text{SFR})_{\{1:6\}}$ & $-5.0$ & 5.0 & Ratios of SFR between adjacent bins of SFH \\
        $\tau_2/\text{mag}$ & 0.0 & 4.0 & Diffuse dust optical depth \\
        $n$ & $-1.0$ & 0.4 & Power-law index for diffuse dust attenuation law \\
        $\tau_1/\tau_2$ & 0.0 & 2.0 & Birth cloud dust optical depth relative to $\tau_2$ \\
        $\ln(f_\text{AGN})$ & $-5\ln(10)$ & $\ln(3)$ & Logarithm of AGN bolometric luminosity fraction \\
        $\ln(\tau_\text{AGN})$ & $\ln(5)$ & $\ln(150)$ & Logarithm of AGN torus optical depth \\
        $\log_{10}(Z_\text{gas}/Z_\odot)$ & $-2.0$ & $0.5$ & Logarithm of gas-phase metallicity \\
        $\log_{10}(U_\text{gas})$ & $-4.0$ & $-1.0$ & Logarithm of gas ionization, $U_\text{gas}=\frac{\text{photon density}}{\text{H density}}$ \\
        $z$ & $0.0$ & $6.0$ & Redshift \\
        \midrule
        & & & \emph{Derived quantities} \\
        \cmidrule(l){4-4}
        $t_\text{age}/\text{Gyr}$ & - & - & Mass-weighted age \\
        $\log_{10}(M/M_\odot)$ & - & - & Logarithm of stellar mass remaining\\
        $\log_{10}(\text{SFR}/M_\odot\,\text{yr}^{-1})$ & - & - & Logarithm of SFR\\
        $\log_{10}(\text{sSFR}/\text{yr}^{-1})$ & - & - & Logarithm of specific SFR per unit mass remaining\\
        \bottomrule
    \end{tabular}
\end{table*}

Our 16-parameter SPS model largely follows the configuration used in \citetalias{alsing24}, which is in turn based on Prospector-$\alpha$ model \citep{leja17, johnson21} and \texttt{v3.2} of FSPS \citep{conroy09, conroy10a, conroy10b}. The free parameters that describe a galaxy are given in Table \ref{tab:sps_notation}. The core ingredients of the FSPS model are: the \citet{chabrier03} initial mass function (IMF); MILES stellar templates \citep{sanchez06, falcon11}; isochrones from the MESA Isochrones and Stellar Tracks library (MIST; \citealp{dotter16, choi16}); the \citet{draine07} polycyclic aromatic hydrocarbon (PAH) model for dust emission\footnote{We use this model with fixed parameters, $U_\text{min}=1$, $\gamma_e=1\%$, and $Q_\text{PAH}=2\%$. The choice of $U_\text{min}=1$ and $\gamma_e=1\%$ corresponds to a minimum starlight intensity of $1$, which $1-\gamma_e=99\%$ of dust mass is exposed to; the remaining $\gamma_e=1\%$ of dust mass will be exposed to a power law distribution of intensities $>U_\text{min}$ (as in \citealp{dale02}). The choice of $Q_\text{PAH}=2\%$ corresponds to a dust model with $2\%$ of total dust mass in PAHs. As our reddest data (IRAC \textit{Ch.\,2}) covers $\sim4$--5~\textmu m in the observer frame, we expect to be relatively insensitive to these choices, as these features would be strongest further into the IR \citep[see][]{leja17}.}; the \citet{nenkova08i, nenkova08ii} models for the dusty torus around active galactic nuclei \citep[see][]{leja18}; and the \citet{madau95} model for extinction by the intergalactic medium. We use a 7-bin piecewise constant star-formation history \citep[SFH; see][]{leja19_sfh}. MIR emission due to dust shells around asymptotic giant branch (AGB) stars is included \citep{villaume15}.

Our base nebular emission model uses the standard FSPS treatment \citep{byler17}, computed using the \texttt{Cloudy} photoionization code \citep{ferland13} for the MIST stellar evolutionary tracks. We use the (recommended) dust-free configuration from \citet{byler17}, based on a stellar ionizing spectrum within a spherical shell of pure gas. Dust attenuation is included separately using a modified \citet{calzetti00} dust law \citep{noll09} with the \citet{kriek13} prescription for the UV bump strength. An additional birth-cloud attenuation component is included for stars younger than 10~Myr \citep{charlot00}, with attenuation proportional to $\lambda^{-1}$.

We use \texttt{Prospector} \citep{johnson21}, \texttt{python-fsps} \citep{pythonfsps}, and \texttt{sedpy} \citep{sedpy} to generate FSPS photometry in 26 passbands for 6.4 million galaxies using the configuration described above, with the 16 SPS parameters ($\bm{\vartheta}$) sampled randomly for each galaxy within the ranges given in Table \ref{tab:sps_notation} (where the ranges for the non-redshift parameters, $\bm{\varphi}$, are the same as in \citetalias{alsing24}). To ensure coverage of the deeper IRAC-selected sample, we allow $z\in[0,6]$, compared to $z\in[0,4.5]$ for the $r$-band-selected sample used in \citetalias{alsing24}. Using the generated FSPS photometry, we train an emulator using \texttt{Speculator} \citep{alsing20} to predict magnitude as a function of the input parameters $\bm{\vartheta}=(\bm{\varphi},z)$. The emulator is a four-layer dense neural network with 128 units per layer, and the bespoke activation function introduced by \citet{alsing20}. Training uses the \texttt{Adam} optimizer \citep{kingma14} to minimize the mean squared error between predicted and true magnitude. Unlike in \citet{alsing20, alsing24}, where this loss was computed for absolute magnitudes per unit mass, here we use apparent asinh magnitudes \citep{lupton99} to avoid giving excessive weight to parts of parameter space that would be too faint ($\gtrsim30$~mag) to observe.

In addition to the photometry emulator, we also use a secondary \texttt{Speculator}-based emulator to compute the strengths of the 44 nebular emission lines whose strengths we modify relative to the \citealp{byler17} \texttt{Cloudy} model grids. We set up this emulator following the procedure detailed in \citet[\S3.4.2; hereafter \citetalias{leistedt23}]{leistedt23}, modeling the emission lines as delta functions, and the COSMOS photometric bandpasses as generalized Gaussian mixture models.

\subsection{Population Model Initialization}
\label{sec:init}

In \citetalias{alsing24}, the population model was a score-based diffusion model \citep{song21}. The free parameters in this model are the weights and biases, $\bm{\psi}$, of a score network, $\bm{s}_{\bm{\psi}}(\bm{x},t)$, which approximates a score function $\nabla_{\bm{x}}p_t(\bm{x})$. The role of the score function is to define a continuous-time transform of a vector $\bm{x}(t)$ between a base density $p_T(\bm{x})$ at time $t=T$, and a target density $p_0(\bm{x})$ at time $t=0$. In \texttt{pop-cosmos}, the target density at $t=0$ is the population distribution over 16 SPS parameters, $P(\bm{\vartheta}|\bm{\psi})$.

We use a diffusion model based on a variance-exploding stochastic differential equation (VE SDE; \citealp{song21}), with a unit Gaussian base density as in \citetalias{alsing24}. The VE SDE can be thought of as a continuous-time version of score matching with Langevin dynamics \citep{song19}. Our score model $\bm{s}_{\bm{\psi}}(\bm{x},t)$ is a four-layer dense neural network with 128 units per layer, and sigmoid linear unit (SiLU) as an activation function. We have found that the initialization and predictive performance are comparable when using a variance-preserving SDE (VP SDE) as the basis of the diffusion model, but that the VE SDE is faster to train and sample from. 

We initialize the model based on a set of maximum a posteriori (MAP) estimates of $\bm{\vartheta}$, obtained under the Prospector-$\alpha$ prior \citep{leja17} for the COSMOS2020 \texttt{Farmer} catalog \citep{weaver22}. From this distribution of MAP estimates, $\bm{\psi}$ is optimized based on the denoising score-matching loss function \citep{hyvarinen05, vincent11, song19, song21}. 

\subsection{Non-parametric Uncertainty Model}
\label{sec:unc}

In \citetalias{alsing24}, the uncertainty model was a single-component mixture density network \citep[MDN;][]{bishop06} of the form $P(\bm{\sigma}_P|\bm{f},\tilde{\bm{\chi}})=\mathcal{N}(\bm{\sigma}_P|\bm{\mu}(\bm{f}), \bm{\Sigma}(\bm{f}))$, where $\bm{\mu}$ and $\bm{\Sigma}$ were small dense neural networks with weights and biases $\tilde{\bm{\chi}}$. The use of a single-component model was motivated by the need for something that can be represented by a bijective transform of a simple base density. Such a representation is necessary for gradient-based optimization of \texttt{pop-cosmos}. However, such a model is unable to capture multimodal distributions of $\bm{\sigma}_P$ given $\bm{f}$. Such multimodality is most pronounced for photometric bands that do not have uniform depth across the COSMOS field: $u$-band (including data from \citealp{sawicki19} with a new $u$ filter, and data from \citealp{laigle16} with an older filter); $YJHK_S$ bands (due to the four ultra-deep stripes in UltraVISTA, detailed in \citealp{mccracken12}); and the \textit{Spitzer} IRAC channels (which include deeper regions tracking the UltraVISTA ultra-deep stripes from \citealp{ashby18}).

In this work, we use a score-based diffusion model \citep{song21} to replace the MDN used in \citetalias{alsing24}. Specifically, we set up an uncertainty model of the form $P(\ln\bm{\sigma}_P|\bm{m}, \bm{\chi})$ that predicts the logarithm of flux uncertainty conditional on calibrated model magnitudes $\bm{m}$. The free parameters $\bm{\chi}$ define a conditional score network $s_{\bm{\chi}}(\bm{x},t;\bm{m})$ that approximates the time-dependent score of the diffusion process for a given set of input magnitudes $\bm{m}$. Our score network in this case is a five-layer densely connected neural network with 256 units per layer and tanh activation functions. The diffusion process follows a VP SDE,
\begin{equation}
    \mathrm{d}\bm{x} = -\frac{1}{2}\beta(t)\bm{x}\,\mathrm{d}t +\sqrt{\beta(t)}\,\mathrm{d}\bm{w},
\end{equation}
where $\bm{w}$ is a Brownian motion process and $\beta(t)=\beta_\text{min} + \frac{t}{T}(\beta_\text{max} - \beta_\text{min})$. We use $\beta_\text{min=0.1}$ and $\beta_\text{max}=20.0$ \citep{ho20, song21}, and sample $\bm{x}(t=T)$ from a Gaussian base density with unit variance. The VP SDE can be viewed as a continuous generalization of a denoising diffusion probabilistic model \citep{sohl15, ho20, jsong21, song21}.

We fit the uncertainty model to the reported magnitudes and flux errors from the COSMOS2020 \texttt{Farmer} catalog \citep{weaver22}, by minimizing a denoising score-matching loss function \citep{hyvarinen05, vincent11, song19, song21} using the \texttt{Adam} optimizer \citep{kingma14}. In the main population model training loop (Section~\ref{sec:training}), we fix $\bm{\chi}$.

\subsection{Population Model Training}
\label{sec:training}

We use a training procedure similar to \citetalias{alsing24}, minimizing a loss that measures similarity between the observed COSMOS2020 photometry and the model photometry produced by the generative model. Our loss function is based on a set of summary statistics of the color and magnitude distributions. From the forward simulated model magnitudes, and observed COSMOS magnitudes, we compute 11 broadband and 13 narrow-band colors from all adjacent pairs of magnitudes. This gives $26+11+13=50$ colors and magnitudes for each model or observed galaxy. We use these colors and magnitudes to construct a vector of summary statistics containing: the mean of each color and magnitude; the standard deviation of each color and magnitude; the full set of color--color, magnitude--magnitude, and color--magnitude correlation coefficients; and a set of 19 percentiles (5th, 10th, \dots, 95th) of each color and magnitude distribution. Our loss function is then the mean squared error across the collected set summary statistics. We arrived at this set of summaries through numerical experiments, finding that these gave the best predictive performance. In future work, it would be worthwhile to formulate a compressed set of summaries, e.g., using a principal component analysis \citep{thorp25}, or by making use of non-linear correlations identified using mutual information \citep[e.g.][]{chartab23}.

We minimize our loss function using the \texttt{Adam} optimizer \citep{kingma14}. We optimize the parameters $(\bm{\psi}, \bm{\alpha}_\text{ZP}, \bm{\beta}_\text{EM}, \bm{\gamma}_\text{EM})$ simultaneously for 120 iterations. We initialize the calibration parameters $(\bm{\alpha}_\text{ZP}, \bm{\beta}_\text{EM}, \bm{\gamma}_\text{EM})$ to the \citetalias{alsing24} values.

\subsection{Individual-Galaxy Inference}
\label{sec:mcmc}
As demonstrated in \citetalias{thorp24}, the trained population model can be used as a prior over the 16 SPS parameters in downstream inference tasks; for example SED fitting for individual galaxies. Using an empirically calibrated population distribution for this task breaks degeneracies and leads to more accurate parameter constraints (see \citetalias{thorp24}; or \citealp{wang23} for a different approach to the construction of a more informative prior in SPS parameter space; see also \citealp{jespersen25} for further motivation). Since the population distribution over SPS parameters in \texttt{pop-cosmos} is a score-based diffusion model (Section~\ref{sec:init}), it is possible to evaluate the prior probability of a set of SPS parameters as described in \citetalias{thorp24} (for further background, see \citealp{grathwohl18, chen18, song21}).

Using the trained model as a prior in this way, we perform posterior inference of the SPS parameters for the 423,262 COSMOS2020 galaxies with $\textit{Ch.\,1}<26$, as well as 6,407 galaxies with $\textit{Ch.\,1}\geq26$ but $r<25$ that were included in the \citetalias{alsing24} training sample. As a baseline, we also perform this inference under a broader, \texttt{Prospector}-$\alpha$-like prior \citep{leja17}, as was done in \citetalias{thorp24}. Under both priors, we compute model fluxes using the \texttt{Speculator} emulator \citep{alsing20} described in Section~\ref{sec:sps} (including zero-point and emission line strength corrections set using the $\bm{\alpha}_\text{ZP}$ and $\beta_\text{EM}$ values fitted in Section~\ref{sec:training}). Our likelihood is a Student's $t$-distribution \citepalias{leistedt23} in flux, with a scale parameter that includes a quadrature sum of the reported COSMOS2020 flux errors from \citet{weaver22}, the emission line variances ($\bm{\gamma}_\text{EM}$) fitted in Section~\ref{sec:training}, and the fractional error floors estimated in \citetalias{leistedt23}. As we work directly with the reported fluxes at this stage, we do not need to propagate the flux uncertainty to an asinh magnitude uncertainty.

Under both priors, we generate $\sim5,000$ independent posterior samples for each galaxy using a bespoke batched and GPU-accelerated affine-invariant MCMC sampler, \texttt{affine}\footnote{\url{https://github.com/justinalsing/affine/tree/torch}}, implemented in PyTorch (based on the algorithm from \citealp{goodman10, foreman13}). We run the sampler with two ensembles of 256 walkers, using the parallel stretch move from \citet{foreman13}, for 1000 iterations. This yields 512,000 posterior samples per galaxy, which we thin by a factor of 100. The MCMC is initialized using maximum a posteriori (MAP) estimates obtained using the \texttt{Adam} optimizer. Under the \texttt{Prospector}-$\alpha$-like prior, we batch computations over 2,000 galaxies simultaneously, with a single batch executing in $\sim23$~min on an NVIDIA A100 GPU card. Under the \texttt{pop-cosmos} prior we batch over 500 galaxies, with a single batch taking $\sim82$~mins to complete on an A100 GPU. This gives an average throughput of $0.7$~GPU-sec per galaxy under the \texttt{Prospector} prior, or $<10$~GPU-sec per galaxy under the \texttt{pop-cosmos} prior. The total processing time for the 429,669 galaxies was around 1,000 GPU-hrs.

\subsection{Mock Catalog Generation}
\label{sec:mocks}
For all subsequent analyses in the paper, we use two mock galaxy catalogs (one $r$-band limited at 25~mag; one \textit{Ch.\,1}-limited at 26~mag) that we have made publicly available on Zenodo. For each of these catalogs, we generated galaxies from the forward model until 2 million galaxies were produced are above the relevant magnitude limit (based on their noisy model magnitudes). For the $\textit{Ch.\,1}<26$ catalog, this process takes around 2 GPU-hrs on a single NVIDIA A100 card. 

Each galaxy in the mock catalogs has 16 SPS parameters and 26-band COSMOS-like photometry associated with it. When we show photometric predictions from these catalogs, we show logarithmic magnitudes\footnote{Although we carry out our fits using asinh magnitudes, we show predictions in logarithmic magnitudes for consistency with \citetalias{alsing24} and other literature. Above the effective depths listed in Table \ref{tab:band_calibration}, the two systems are almost equivalent \citep[see also][]{lupton99}.}. In the public release we include fluxes, logarithmic magnitudes, and asinh magnitudes. We include both noiseless and noisy model fluxes for each galaxy, as well as the model photometric uncertainties that were applied to obtain the latter from the former. The magnitudes in our plots and in the public catalogs are noisy magnitudes, obtained by applying either Equation \ref{eq:asinh} or \ref{eq:log} to the noisy model fluxes. We do not propagate the model flux uncertainties to magnitude uncertainties, or report noiseless model magnitudes (although these can be easily obtained from the noiseless model fluxes).

From the 16 base SPS parameters (listed in the upper part of Table \ref{tab:sps_notation}), we compute several derived quantities that are used in our downstream analyses and included in the public catalogs; these are listed in the lower part of Table \ref{tab:sps_notation}. We compute the stellar mass remaining using a neural network emulator that approximates the fraction of surviving stellar mass (calculated by FSPS) as a function of the 16 base parameters\footnote{For our parametrization, the fraction of mass remaining is set by the star-forming history and stellar metallicity; i.e.\ the salient parameters are $z$, $\Delta\log_{10}(\text{SFR})_{\{1:6\}}$, and $\log_{10}(Z/Z_\odot)$.}. The (s)SFR and mass weighted age are computed directly using the star-formation rate ratios, $\Delta\log_{10}(\text{SFR})_{1:6}$, with the (s)SFR being averaged over the last 100~Myr of a galaxy's evolution (corresponding to the two most recent bins of the SFH; \citealp{leja19_sfh}). A full description is given in Appendix \ref{sec:derived}.

\section{Validation}
\label{sec:validation}
Given the algorithmic and numerical complexity of the \texttt{pop-cosmos} model described in Section~\ref{sec:methods}, it is critical to carefully validate all the different aspects of the model's performance. In Section~\ref{sec:results_photo}, we show the colors and magnitudes predicted by the \texttt{pop-cosmos} model, compared to those in the \citet{weaver22} COSMOS2020 \texttt{Farmer} catalog. Section \ref{sec:unc_validation} demonstrates the performance of the uncertainty model relative to the COSMOS2020 flux errors reported by \citet{weaver22}. Sections \ref{sec:results_mcmc_redshift}--\ref{sec:results_mcmc_fagn} validate the inferences made for individual galaxies under the new \texttt{pop-cosmos} prior, compared to the COSMOS spectroscopic archive \citep{khostovan25}, the results under a \texttt{Prospector}-like prior, and the \texttt{LePhare} results from \citet{weaver22}. Finally, in Section \ref{sec:mass_completeness}, we examine the stellar mass vs.\ redshift relation predicted by the model to estimate the redshift-evolving mass completeness limit of our predictions. Supplementary results are included in Appendices \ref{sec:photo_validation}--\ref{sec:results_photo_extra}.

\begin{figure*}
    \centering
    \includegraphics[width=\linewidth]{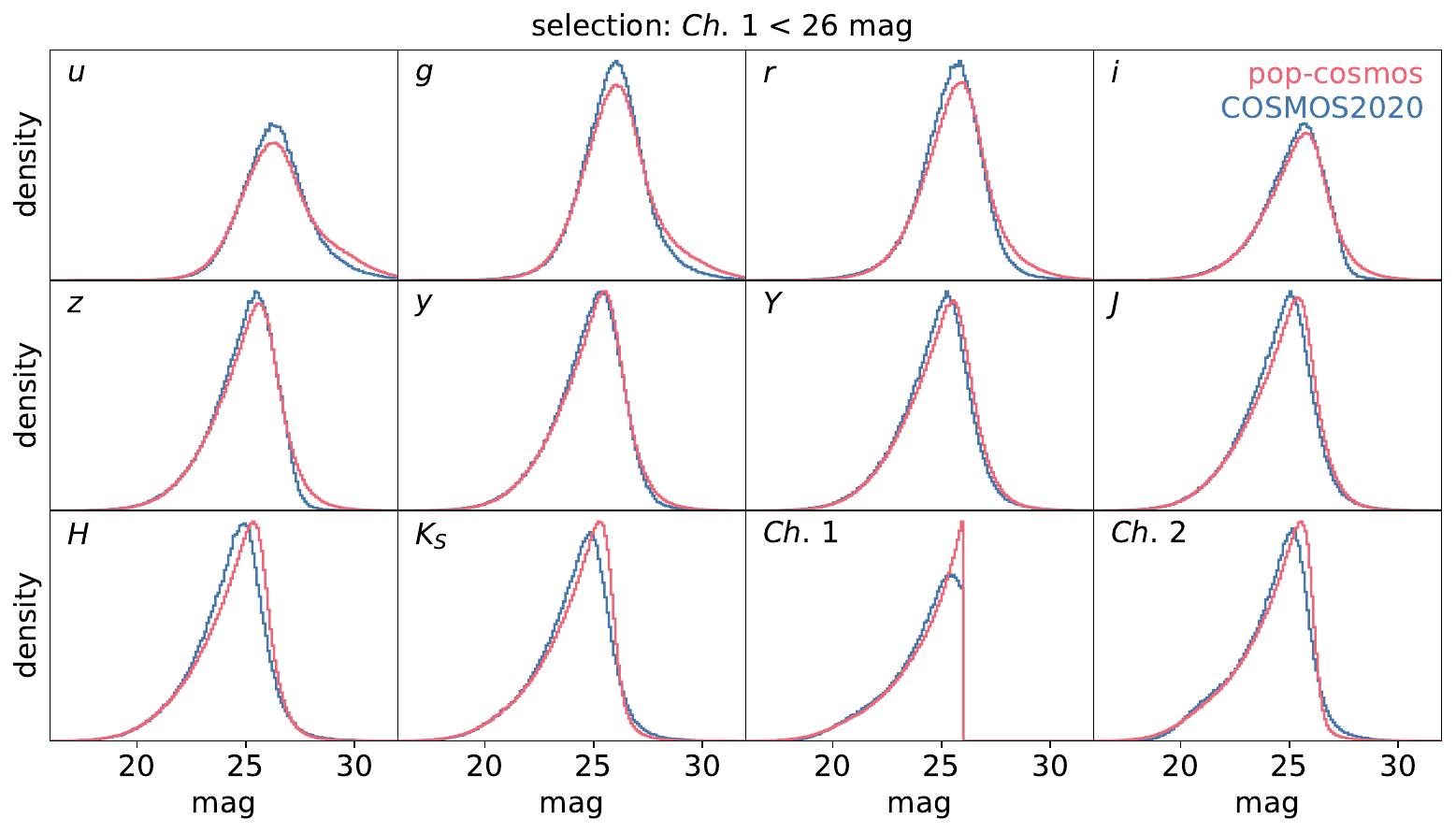}
    \caption{Broadband magnitudes (logarithmic) predicted by the new \texttt{pop-cosmos} generative model, compared to COSMOS2020 \citep{weaver22} for $\textit{Ch.\,1}<26$.}
    \label{fig:mags_ch126}

    \centering
    \includegraphics[width=\linewidth]{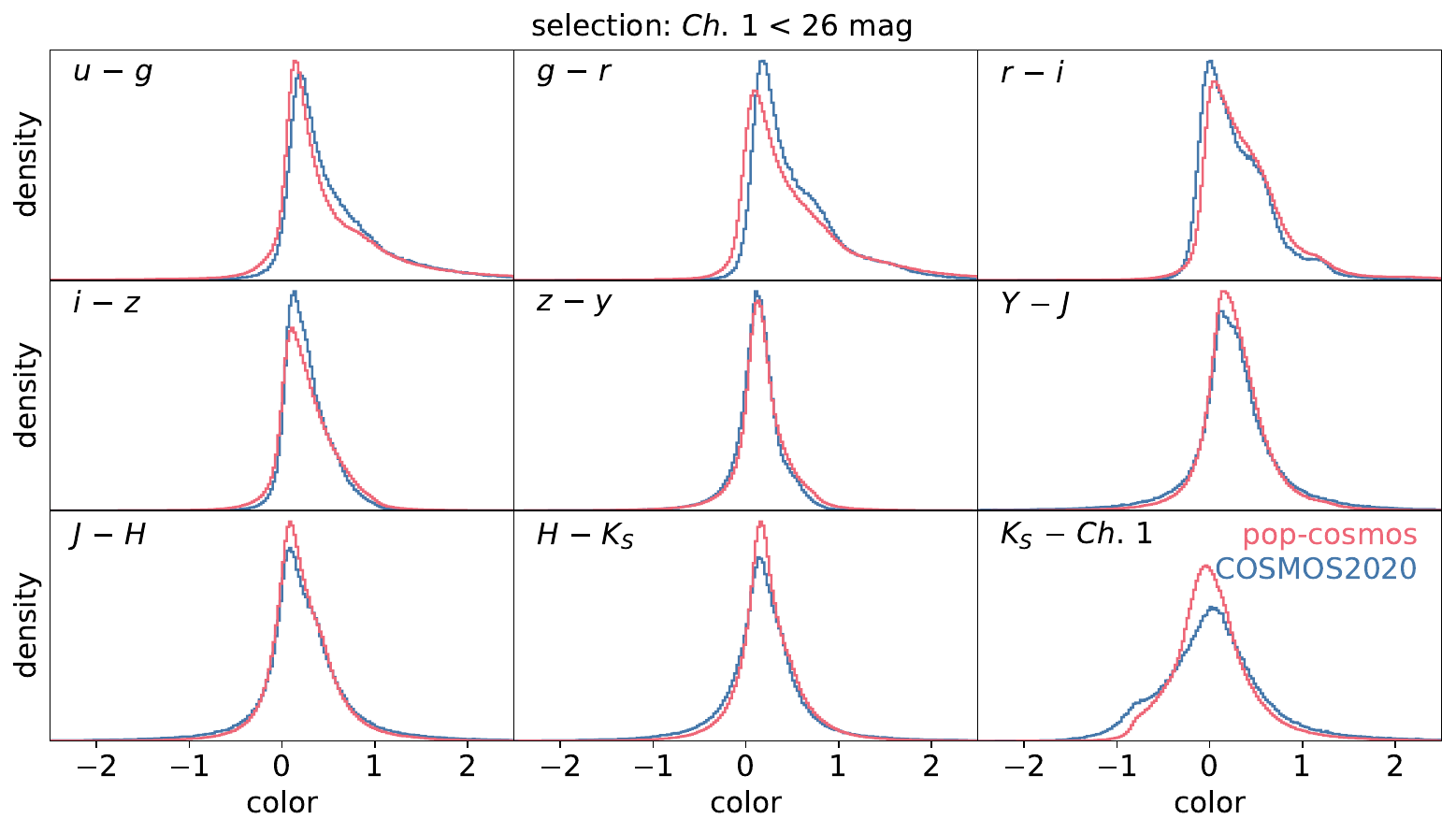}
    \caption{Broadband colors (logarithmic, adjacent bands) predicted by the new \texttt{pop-cosmos} generative model, compared to COSMOS2020 \citep{weaver22} for $\textit{Ch.\,1}<26$.}
    \label{fig:colors_ch126}
\end{figure*}

\begin{figure*}
    \centering
    \includegraphics[width=\linewidth]{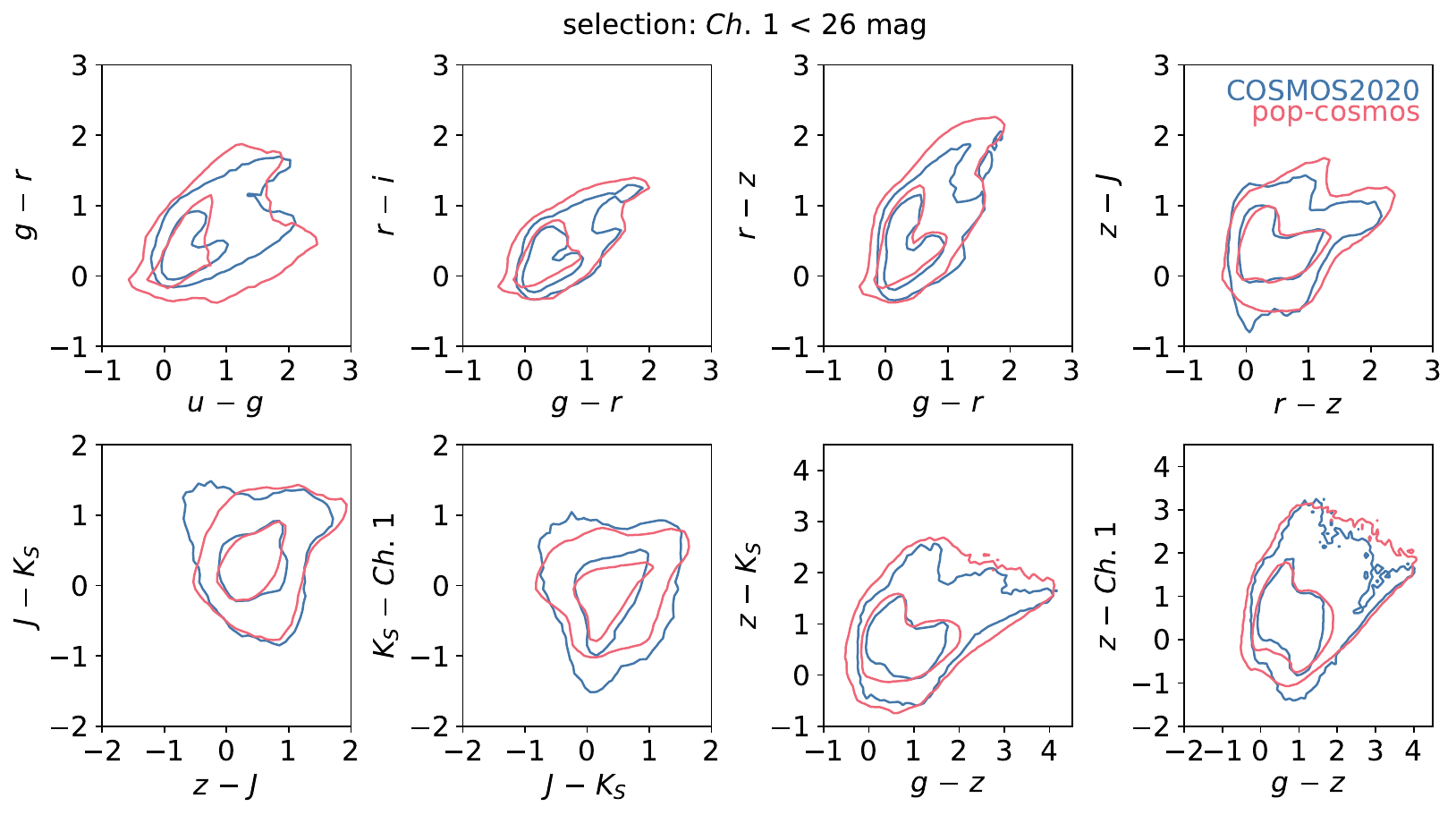}
    \caption{Broadband colors (logarithmic, \citetalias{alsing24} band combinations) predicted by the new \texttt{pop-cosmos} generative model, compared to COSMOS2020 \citep{weaver22} for $\textit{Ch.\,1}<26$. Contours enclose the 68 and 95\% highest density regions.}
    \label{fig:color_color_ch126}
\end{figure*}

\begin{figure*}
    \centering
    \includegraphics[width=\linewidth]{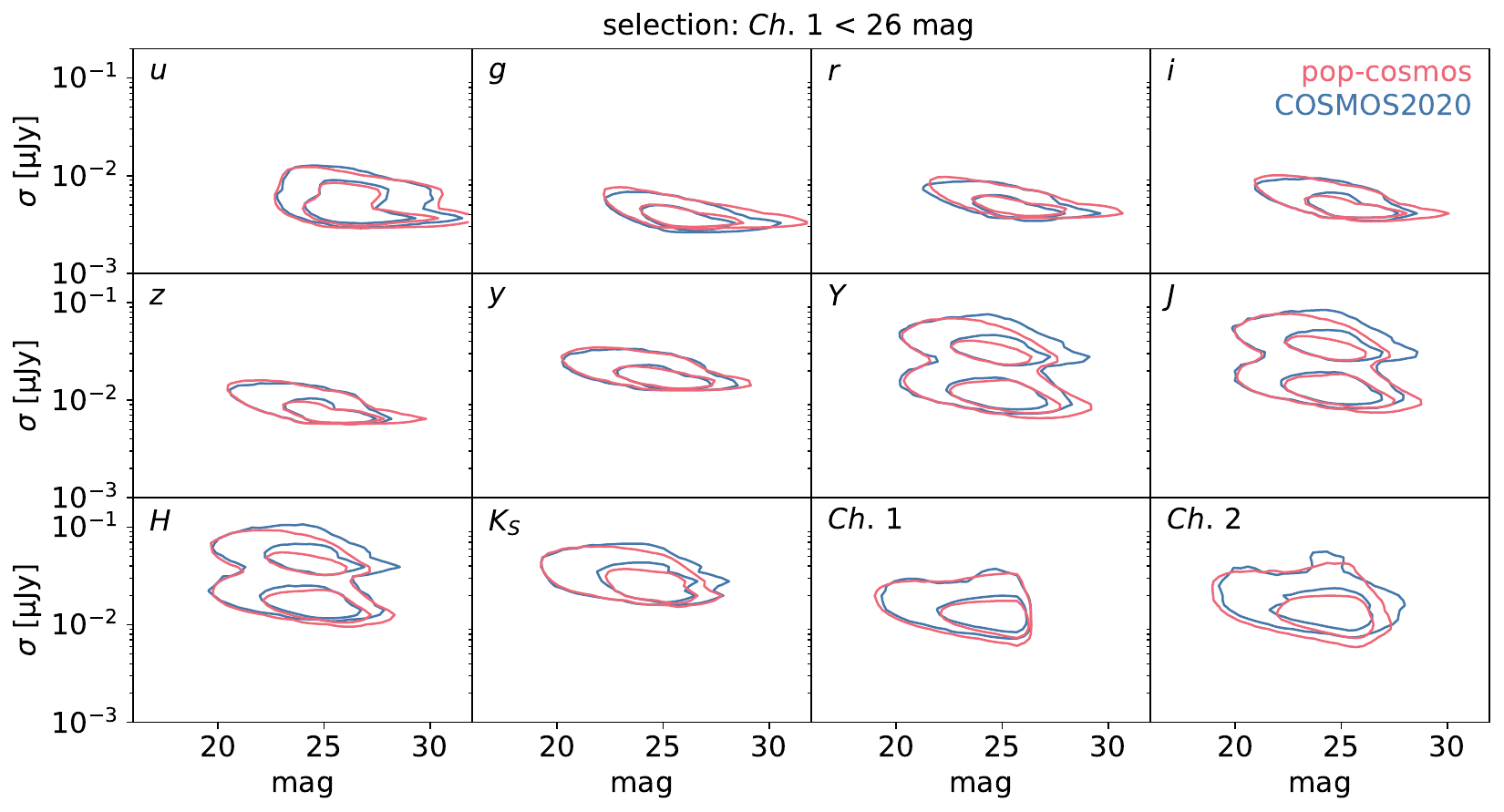}
    \caption{Flux uncertainty vs.\ (logarithmic) magnitude for the COSMOS broad bands, compared to the flux errors reported in the COSMOS2020 catalog \citep{weaver22} for $\textit{Ch.\,1}<26$. Note that the \texttt{pop-cosmos} uncertainty predictions are made conditional on the \texttt{pop-cosmos} model magnitudes. Contours enclose the 68 and 95\% highest probability density regions.}
    \label{fig:uncs_ch126}
\end{figure*}

\subsection{Colors and Magnitudes for $\textit{Ch.\,1}<26$}
\label{sec:results_photo}

In Figure \ref{fig:mags_ch126} and Figure \ref{fig:colors_ch126}, we show the broadband color and magnitude marginals for the $\textit{Ch.\,1}<26$ mock galaxy sample, compared to COSMOS2020 \citep{weaver22}. There are some small offsets in the magnitude distributions. The most prominent of these is for $\textit{Ch.\,1}$, where the turnover in number counts at $\textit{Ch.\,1}>25$ is not well modeled. As discussed in Section~\ref{sec:photometry} here (and \citealp{davidzon17, weaver22, weaver23_smf}), there is likely a degree of unmodeled incompleteness at these faint magnitudes.

The marginal distributions of colors in Figure \ref{fig:colors_ch126} show generally very good agreement. The most substantial offsets are seen in $g-r$ and $K_S-\textit{Ch.\,1}$. The $K_S-\textit{Ch.\,1}$ color is known to be particularly sensitive to the photometric extraction \citep{weaver22}, and exhibits some of the largest offsets between the \texttt{Farmer} \citet{weaver23} profile fitting photometry and the \texttt{Classic} \texttt{SExtractor}-based \citep{bertin96} aperture photometry pipeline (comparable to COSMOS2015; \citealp{laigle16}), the latter of which tends to be systematically redder. The $u$- and $g$-bands, and $g-r$ color at the faint end are also known to be sensitive to the photometric extraction in COSMOS2020 \citep{weaver22}. We have found that even with a broader \texttt{Prospector}-$\alpha$-like prior, it is difficult to produce $K_S-\textit{Ch.\,1}<-1$ using our current SPS configuration (described in Section~\ref{sec:sps}).

We show color vs.\ color plots in Figure \ref{fig:color_color_ch126}, for the set of color pairs used by \citetalias{alsing24}. These color choices are motivated by \cite{weaver22}, who identified them as the most significantly informative COSMOS colors for galaxy evolution. We perform a detailed 9-dimensional validation of the broadband colors for $\textit{Ch.\,1}<26$ in Appendix \ref{sec:photo_validation} (following the scheme from \citealp{thorp25}). This validation includes a comparison based on quantile--quantile (Q--Q) and probability--probability (P--P) tests \citep{wilk68}, as well as statistical summaries in the form of Wasserstein distances \citep{kantorovichrubinstein58, vasserstein69} and the Kolmogorov--Smirnov (K--S) statistic \citep{kolmogorov33, smirnov48}. These tests confirm that the quantitative agreement between the model predictions and data is comparable to agreement between the two COSMOS2020 photometric reductions, \texttt{Classic} and \texttt{Farmer}. We show plots of colors, magnitudes, and flux uncertainties for the $r<25$ selection in Appendix \ref{sec:results_photo_extra}.

\subsection{Uncertainty Model}
\label{sec:unc_validation}
We now inspect the trained uncertainty model, introduced in Section~\ref{sec:unc}, by using it to generate flux uncertainties conditional on the \texttt{pop-cosmos} model magnitudes. Figure \ref{fig:uncs_ch126} shows the joint distribution of flux uncertainties and magnitudes drawn from the trained \texttt{pop-cosmos} model, compared to the reported magnitudes and flux errors from the COSMOS2020 \texttt{Farmer} catalog \citep{weaver22}. The agreement here is good by construction, as the uncertainty model is trained directly on the \texttt{Farmer} catalog, with any visible offsets arising due to the offsets in the magnitude distributions for this catalog (see Figure \ref{fig:mags_ch126}). The bimodal distributions of flux errors resulting from the mixed depths in the $uYJHK_S$ bands are successfully captured by the trained uncertainty model.

\subsection{Individual Galaxy Inference: Redshifts}
\label{sec:results_mcmc_redshift}
We obtain full posterior distributions for the properties of each galaxy, including redshift. For each galaxy, we take its posterior median redshift as a point estimate of $z^\text{phot}$ when comparing to the $z^\text{spec}$ collated by \citet{khostovan25}. For the 39,588 objects with spectroscopic cross-matches in \citet{khostovan25} we follow \citetalias{thorp24} and \citetalias{leistedt23} by characterising the mismatch using $\Delta_z=(z^\text{phot} - z^\text{spec})/(1 + z^\text{spec})$. We define an outlier as a galaxy with $|\Delta_z|>0.15$, following the widely used convention in the literature. We find a total of 2,844 outliers (7.18\% of all galaxies). We find $\sigma_\text{MAD}=1.48\times\mathrm{median}(|\Delta_z|)=0.0132$, and $\mathrm{median}(\Delta_z)=-8\times10^{-4}$. Despite the new spectroscopic sample being a factor of $3.2\times$ larger than used in \citetalias{thorp24}, and the redshift range being considerably larger, the outlier fraction here is comparable; the median and $\sigma_\text{MAD}$ are both improved.

\begin{figure}
    \centering
    \includegraphics[width=\linewidth]{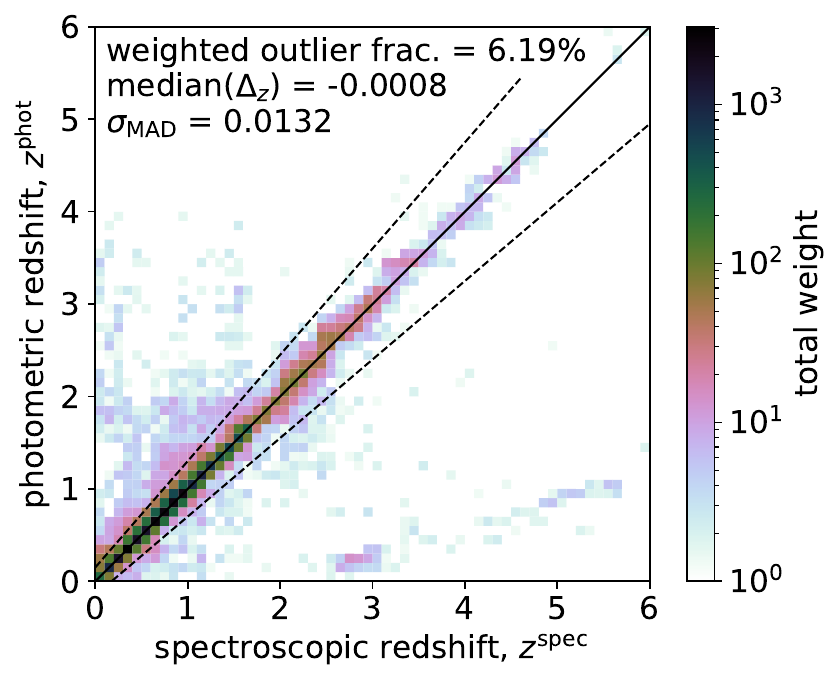}
    \caption{Comparison of $z^\text{phot}$ vs.\ $z^\text{spec}$ based on $z^\text{phot}$  (posterior median $z$) inferred under the new \texttt{pop-cosmos} prior. Each galaxy is weighted by its spectroscopic confidence level from \citet{khostovan25}. Solid and dashed lines show $z^\text{phot}=z^\text{spec}$ and $|\Delta_z|=0.15$, respectively.}
    \label{fig:specz}
\end{figure}

In Figure \ref{fig:specz}, we show a 2D histogram of $z^\text{phot}$ vs.\ $z^\text{spec}$. In the histogram, we weight each galaxy based on the standardized $z^\text{spec}$ confidence level assigned by \citet{khostovan25} (so a galaxy with $\text{CL}=50\%$ will have weight 0.5, and so forth). The 39,588 galaxies have a total weight of 33,616.55 in this scheme. We show the weighted outlier fraction (6.19\%) on Figure \ref{fig:specz}, defining this as the sum of weights of galaxies with $|\Delta_z|>0.15$. We summarize the results for the \texttt{pop-cosmos} and \texttt{Prospector} priors in Table \ref{tab:mcmc_z_comparison}. We also show the results for \texttt{LePhare} \citep{arnouts99, ilbert06} and \texttt{EAZY} \citep{brammer08} on the same spectroscopic sample, based on the photometric redshifts from \citet{weaver22}.

\begin{table}
    \centering
    \caption{Statistics on $\Delta_z=(z^\text{phot} - z^\text{spec})/(1 + z^\text{spec})$. Outlier rate is weighted by spectroscopic confidence level \citep{khostovan25}.}
    \label{tab:mcmc_z_comparison}
    \begin{tabular}{l c c c}
    \toprule\toprule
        Model & Median & $\sigma_\text{MAD}$ & Outlier Rate (\%) \\ \midrule
        \texttt{pop-cosmos} & $-8\times10^{-4}$ & 0.0132 & 6.19\% \\
        \texttt{Prospector} & $-3\times10^{-4}$ & 0.0148 & 6.57\% \\
        \texttt{LePhare} & $-2\times10^{-4}$ & 0.0164 & 8.44\% \\
        \texttt{EAZY} & $-4\times10^{-4}$ & 0.0132 & 6.84\% \\
        \bottomrule
    \end{tabular}
\end{table}

\subsection{Individual Galaxy Inference: Stellar Masses}
\label{sec:results_mcmc_mass}

\begin{figure}
    \centering
    \includegraphics[width=\linewidth]{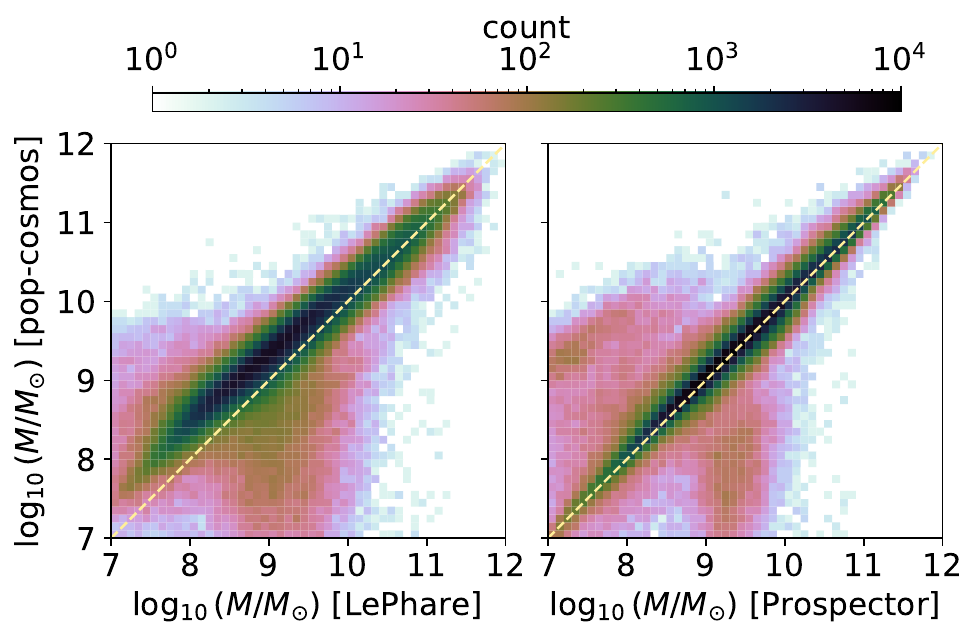}
    \caption{Comparison of stellar mass estimates for the 423,262 COSMOS2020 galaxies with $\textit{Ch.\,1}<26$. The vertical axis shows the posterior median $\log_{10}(M/M_\odot)$ under the new \texttt{pop-cosmos} prior. The horizontal axis shows \textbf{(left)} the \texttt{LePhare} stellar mass estimates from \citet{weaver22}, and \textbf{(right)} the posterior median under the \texttt{Prospector} prior.}
    \label{fig:mcmc_mass}
\end{figure}

In Figure \ref{fig:mcmc_mass}, we compare the posterior median stellar masses estimated under the \texttt{pop-cosmos} prior to those estimated under the \texttt{Prospector} prior, and to the \texttt{LePhare} \citep{arnouts99, ilbert06} stellar masses published by \citet{weaver22}. We make this comparison for the full $\textit{Ch.\,1}<26$ sample of 423,262 galaxies. We see that for $\log_{10}(M/M_\odot)\gtrsim10$, the results under the new \texttt{pop-cosmos} prior agree closely with the \texttt{Prospector} prior and with \texttt{LePhare}. For lower masses, there is more scatter, and the \texttt{pop-cosmos} prior gives rise to slightly higher stellar mass estimates on average than \texttt{LePhare}. In the \citet{weaver22} \texttt{LePhare} fits, the template library used (from \citealp{polletta07, ilbert09, onodera12}, using SPS models from \citealp{silva98, bruzual03}) is based on a finite set of parametric SFH models\footnote{The 19 elliptical and spiral galaxy templates based on \citet{polletta07} use a power-law  $\text{SFR}(z)$ using the \citet{silva98}  \texttt{GRASIL} model. The additional star-forming galaxy templates from \citet{ilbert09} are based on 12 \citet{bruzual03} models that incorporate a smooth SFH with bursts. The 2 \citet{onodera12} elliptical galaxy templates use exponentially declining SFH models from \citet{bruzual03}.}. Several works \citep[e.g.,][]{leja19, lower20} have shown that parameteric SFH models tend to produce systematically lower stellar mass estimates than non-parametric models, likely accounting for the offset between our inferences and \texttt{LePhare}. That the difference is largest at low masses aligns with the results seen in \citet{leja19}.

\begin{table}
    \centering
    \caption{Statistics on $\Delta\log_{10}(M/M_\odot)$ between the \texttt{pop-cosmos} and \texttt{Prospector} inferences. Galaxies are sorted into mass bins based on their \texttt{pop-cosmos} mass estimates.}
    \label{tab:mcmc_mass_comparison}
    \begin{tabular}{l r r r r r}
    \toprule\toprule
        Mass Bin & Count & Median & $\sigma_\text{MAD}$ & \multicolumn{2}{c}{Outlier Rate (\%)} \\ \cmidrule(l){5-6}
        & & & & 0.2\,dex & 0.5\,dex \\ \midrule
        $[7,12)$ & 421630 & 0.037 & 0.102 & 15.7 & 5.9\\
        $[7,8)$ & 16422 & -0.077 & 0.293 & 49.8 & 36.6\\
        $[8,9)$ & 118585 & 0.049 & 0.113 & 17.7 & 7.3\\
        $[9,10)$ & 211883 & 0.032 & 0.096 & 13.2 & 4.0\\
        $[10,11)$ & 68494 & 0.039 & 0.088 & 12.2 & 2.2\\
        $[11,12)$ & 6246 & 0.039 & 0.074 & 12.1 & 2.4\\
        \bottomrule
    \end{tabular}
\end{table}

Table \ref{tab:mcmc_mass_comparison} contains summaries of the difference in $\log_{10}(M/M_\odot)$ inferred under the \texttt{pop-cosmos} and \texttt{Prospector} priors. The quoted statistics are similar to those used in the spectroscopic redshift comparison: $\text{median}[\Delta\log_{10}(M/M_\odot)]$, $\sigma_\text{MAD}=1.48\times\text{median}[|\Delta\log_{10}(M/M_\odot)|]$, and the fraction of $0.2$~dex and $0.5$~dex outliers (the latter motivated by \citealp{mobasher15}). The logarithmic mass residuals are computed for \texttt{pop-cosmos} minus \texttt{Prospector}; i.e.\ a positive $\text{median}[\Delta\log_{10}(M/M_\odot)]$ would imply systematically higher masses under the \texttt{pop-cosmos} prior. Galaxies are sorted into the mass bins based on their inferred mass under the \texttt{pop-cosmos} prior. Across the full sample, the median offset in the inferred stellar mass is $<0.04$~dex, with $\sigma_\text{MAD}\sim1$~dex. The fraction of galaxies where the two priors find $|\Delta\log_{10}(M/M_\odot)|>0.5$~dex is very small at $5.9\%$. As seen in Figure \ref{fig:mcmc_mass}, the outliers are predominantly at the low mass end, with the $0.5$~dex outlier rate for $\log_{10}(M/M_\odot)\geq10$ being $\sim2\%$

\begin{figure}
    \centering
    \includegraphics[width=\linewidth]{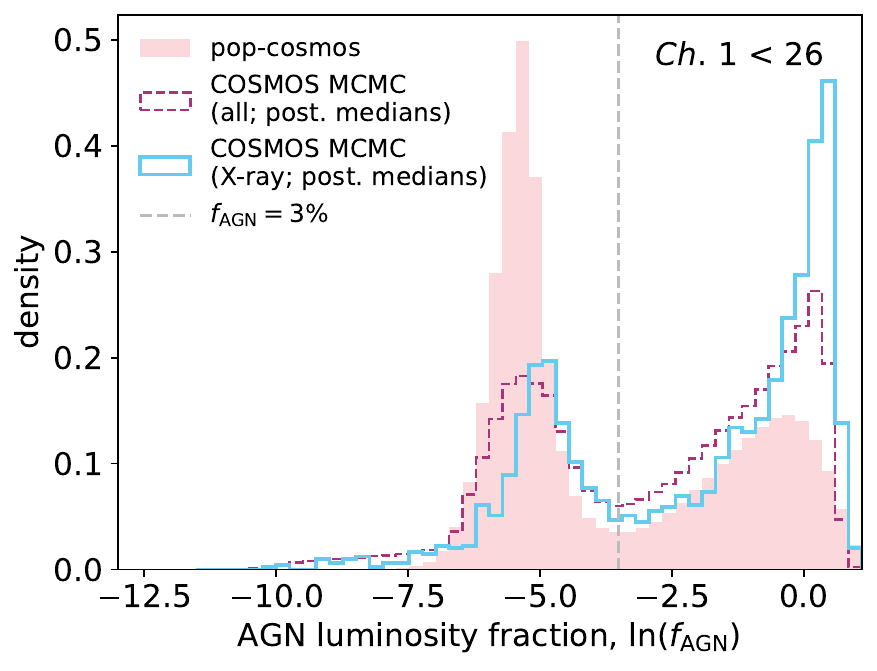}
    \caption{Distribution of AGN bolometric luminosity fraction $f_\text{AGN}$, for galaxies with $\textit{Ch.\,1}<26$. The filled red histogram shows the prior distribution computed from \texttt{pop-cosmos} model draws. The purple dashed histogram shows the posterior medians for 423,262 COSMOS2020 galaxies, with the cyan histogram showing posterior medians for the subset of 1,951 COSMOS2020 galaxies with \textit{Chandra} X-ray detections.}
    \label{fig:nfAGN}
\end{figure}

\subsection{Individual Galaxy Inference: AGN}
\label{sec:results_mcmc_fagn}
To test the robustness of our inferences for AGN host galaxies, we look briefly at our results for the 1,951 COSMOS2020 $\textit{Ch.\,1}<26$ galaxies with \textit{Chandra} X-ray detections (based on the \citealp{weaver22} cross-match against the \citealp{civano16} catalog). In Figure \ref{fig:nfAGN}, we compare the posterior median estimates of $\ln(f_\text{AGN})$ for these galaxies to the posterior medians for the full COSMOS2020 catalog, and to the \texttt{pop-cosmos} prior over $\ln(f_\text{AGN})$. Whilst the overlap between X-ray- and IR-bright AGN is not expected to be perfect \citep[see, e.g.,][]{lamassa19, carroll21}, both of these observational signatures are thought to correspond to radiatively efficient modes of accretion so a degree of correlation is expected. Examining the \texttt{pop-cosmos} model draws for all $\textit{Ch.\,1}<26$ galaxies, we see a bimodal distribution in $\ln(f_\text{AGN})$ similar to \citetalias{alsing24}, with the two modes corresponding roughly to a low-AGN ($f_\text{AGN}\lesssim3\%$) and high-AGN ($f_\text{AGN}\gtrsim3\%$) state. For the \texttt{pop-cosmos} model trained in this work, we find that $\sim40\%$ of model galaxies have $f_\text{AGN}>3\%$. Looking at the $f_\text{AGN}$ posteriors for the 1,951 \textit{Chandra} X-ray detected galaxies, we find that $\sim67\%$ of these galaxies have a posterior median AGN luminosity fraction greater than $3\%$, with the distribution of posterior medians in Figure \ref{fig:nfAGN} skewing strongly towards the high-AGN mode of the prior. Our posterior medians for the full COSMOS2020 $\textit{Ch.\,1}<26$ catalog show a weaker skew towards the high-AGN mode.

\begin{figure}
    \centering
    \includegraphics[width=\linewidth]{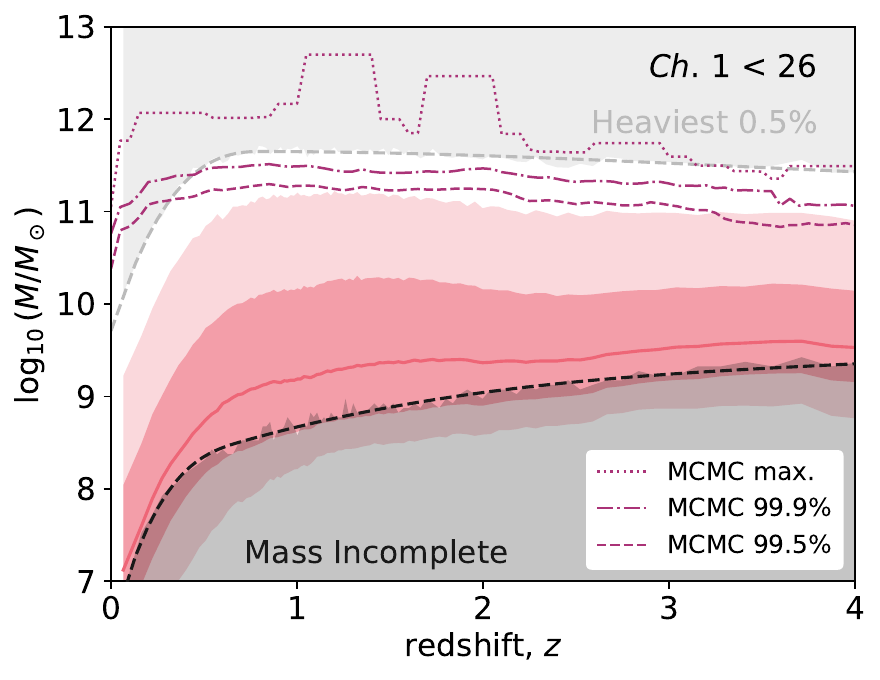}
    \caption{Stellar mass vs.\ redshift relation computed from the \texttt{pop-cosmos} model draws. Red shaded regions show the 68 and 95\% conditional credible intervals, with the solid red line showing the median. The gray shaded region and dashed line correspond to the 99.5th percentile of stellar mass at a given redshift. The black shaded region and dashed line show the estimated mode or `` turnover'' of the stellar mass distribution as a function of redshift. Purple lines show upper limits on the masses of COSMOS2020 galaxies, estimated from the galaxy-level SED fits presented in Section~\ref{sec:results_mcmc_mass}.}
    \label{fig:mass_redshift}
\end{figure}

\subsection{Mass Completeness}
\label{sec:mass_completeness}
It is important to quantify the regime of validity of the trained population model. We do this in two ways: (1) by estimating a mass completeness limit for the population described by the model; and (2) by considering the tails of the stellar mass distribution where a COSMOS-sized field may be subject to large sample variance \citep{jespersen24}. Figure \ref{fig:mass_redshift} shows the conditional distribution of stellar mass as a function of redshift for $0\leq z\leq 4$, computed using the draws from the \texttt{pop-cosmos} generative model. The 99.5th percentile (gray shaded region) is shaded in gray. We follow \citetalias{alsing24} in estimating the mass completeness limit of the \texttt{pop-cosmos} mock galaxy catalog by identifying the mode or ``turnover'' of the stellar mass distribution as a function of redshift. This is illustrated with a dark shaded region, with a spline fit to this completeness limit shown as a black dashed line. In subsequent analyses, we focus on the mass complete part of the mock catalogs, omitting the model galaxies that fall below the estimated completeness threshold. 

Our estimated mass completeness limit for \texttt{pop-cosmos} is consistent with the COSMOS2020 analysis by \citet{weaver23_smf}, despite the difference in approach. \citet{weaver23_smf} estimate their stellar mass completeness limits using an approach similar to \citet{pozzetti10}, with a limiting mass that is quadratic in $(1+z)$. Our estimate tends to be slightly more conservative; around $\sim0.1$--$0.3$~dex higher across the redshift range. For example, at redshifts $z=[1.0, 2.0, 4.0, 6.0]$ our estimated limiting mass is $\log_{10}(M/M_\odot)=[8.68, 9.05, 9.36, 9.70]$, compared to $\log_{10}(M/M_\odot)=[8.40, 8.78, 9.25, 9.56]$ using equation 3 from \citet{weaver23_smf}.

Overlaid on Figure \ref{fig:mass_redshift} with purple dashed and dash-dotted lines, we show respectively 99.5th and 99.9th percentiles of stellar masses estimated for individual COSMOS2020 galaxies from the SED fits in Section~\ref{sec:results_mcmc_mass}. This comparison shows that the most massive 0.5\% of \texttt{pop-cosmos} model draws are likely not well-represented in the training data. Therefore, we excise these from subsequent analyses. Appendix \ref{sec:mmax} provides a fitting function for masking these galaxies, shown as a gray dashed line in Figure \ref{fig:mass_redshift}. Appendix \ref{sec:mcomplete} provides a fitting function for the mass incompleteness, shown as a black dashed line in \ref{fig:mass_redshift}.

\begin{figure}
    \centering
    \includegraphics[width=\linewidth]{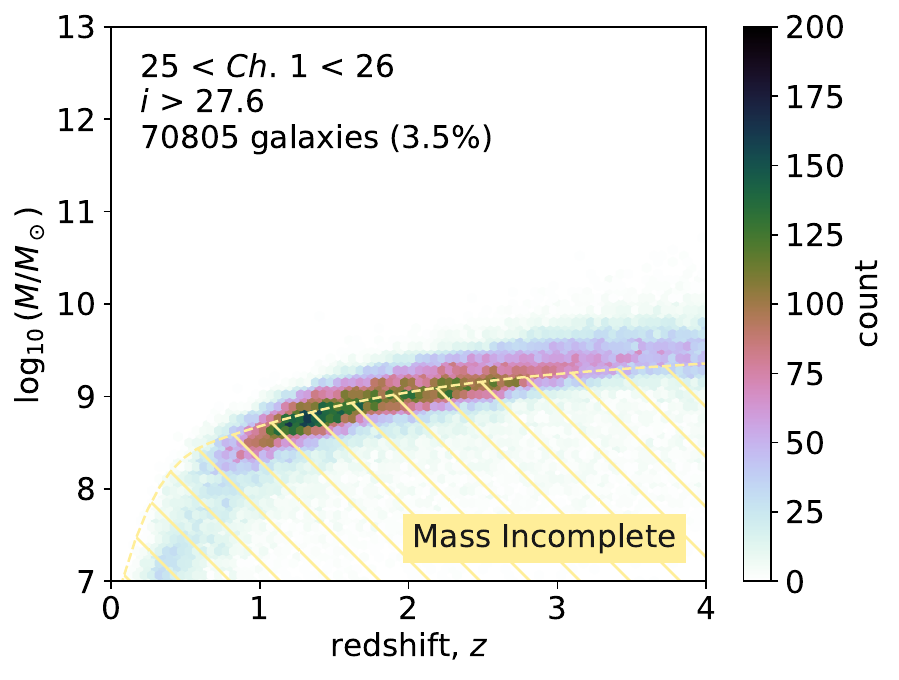}
    \caption{Stellar mass vs.\ redshift computed from the \texttt{pop-cosmos} model draws for galaxies with $25<\textit{Ch.\,1}<26$ and $i>27.6$ ($3.5\%$ of the $\textit{Ch.\,1}<26$ mock catalog). The mass completeness limit from Figure \ref{fig:mass_redshift} is shown as a yellow dashed line and hatched region.}
    \label{fig:mass_redshift_25-26}
\end{figure}

To test the impact of the COSMOS2020 incompleteness at $\textit{Ch.\,1}\gtrsim25$, we examine the stellar masses and redshifts of the model galaxies at the faint end of the $\textit{Ch.\,1}$ magnitude distribution (i.e., in the region where \texttt{pop-cosmos} and COSMOS2020 begin to diverge in Figure \ref{fig:mags_ch126}). For the \texttt{pop-cosmos} model, $36.8\%$ of galaxies with $\textit{Ch.\,1}<26$ have $\textit{Ch.\,1}>25$. For COSMOS2020, we find that $32.8\%$ of detected galaxies lie within the $25<\textit{Ch.\,1}<26$ bin. We further require that model galaxies are optically faint, with $i>27.6$ (based on the COSMOS2020 3$\sigma$ depth; \citealp{weaver22}). We find that $3.5\%$ of mock galaxies have $25<\textit{Ch.\,1}<26$ and $i>27.6$, compared to just $0.9\%$ of COSMOS2020 galaxies\footnote{The $i$-band is the deepest in the COSMOS detection image, so requiring an $i$ band magnitude fainter the $3\sigma$ depth is an imperfect proxy for a galaxy's detectability. As discussed in depth by \citet{weaver23_smf}, such a cut will not perfectly model the detection process, which can detect $i>27.6$ galaxies that are bright in the shallower but redder $K_S$-band, or which have adequate S/N in the $izYJHK_S$ co-add (see Section \ref{sec:photometry}).}. In Figure \ref{fig:mass_redshift_25-26}, we show the stellar masses and redshifts of \texttt{pop-cosmos} model galaxies with $25<\textit{Ch.\,1}<26$ and $i>27.6$. We see that the majority of these galaxies are at $z\gtrsim1$ and have low stellar masses, generally within $\sim0.5$~dex of the completeness threshold defined in Figure \ref{fig:mass_redshift}. We calculate that $\sim56\%$ of model galaxies with $25<\textit{Ch.\,1}<26$ and $i>27.6$, or $\sim69\%$ of galaxies with $25.5<\textit{Ch.\,1}<26$ and $i>27.6$ lie below the mass completeness limit. We therefore infer that the $\textit{Ch.\,1}<26$ galaxies missed by the COSMOS2020 detection pipeline are most likely to be low-mass galaxies in the mass-incomplete regime. Missing such galaxies from our training data is unlikely to significantly impact our conclusions, which are based on the mass-complete regime.

\section{Astrophysical Predictions}
\label{sec:results_astro}

The data-driven calibration of the  \texttt{pop-cosmos} model to the IRAC \textit{Ch.\,1}-selected COSMOS2020 catalog yields a population prior over the galaxy population probed by this selection. Samples drawn from this prior can be interpreted as synthetic galaxy populations, from which we can make predictions of population-level trends in galaxy-evolution. In effect, by marginalizing over different parameters in the learned SPS parameter distribution (or derived quantities thereof), we are able to make a range of astrophysical predictions. A selection of these are showcased in Section~\ref{sec:nz}--\ref{sec:agn_results}. A complete corner plot over the SPS parameter space is shown in in Appendix \ref{sec:corner}.

These predictions from the trained generative model serve twin purposes. In the cases where well-explored trends in galaxy evolution exist in the literature, comparing the predictions from \texttt{pop-cosmos} with these relationships serves to further validate the model and increase trust in its robustness as a generative model for other galaxy surveys well-represented by the $\textit{Ch.\,1}<26$ selection. However, as the model encapsulates the full joint distribution of the SPS parameters, which is highly challenging to access through other methods, its predictions can also provide insight into open questions in galaxy evolution. 

\subsection{Redshift Distribution}
\label{sec:nz}
\begin{figure}
    \centering
    \includegraphics[width=\linewidth]{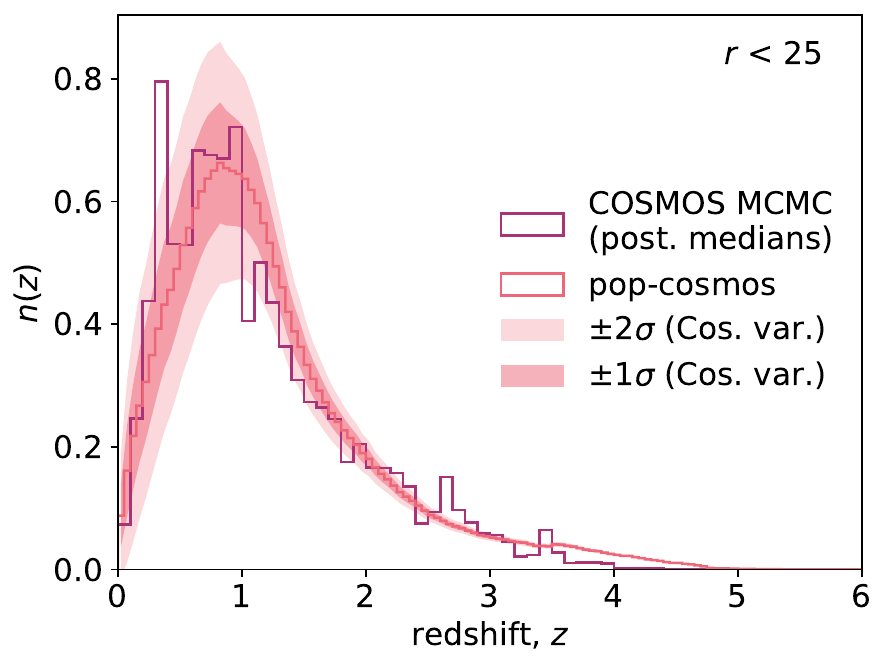}
    \includegraphics[width=\linewidth]{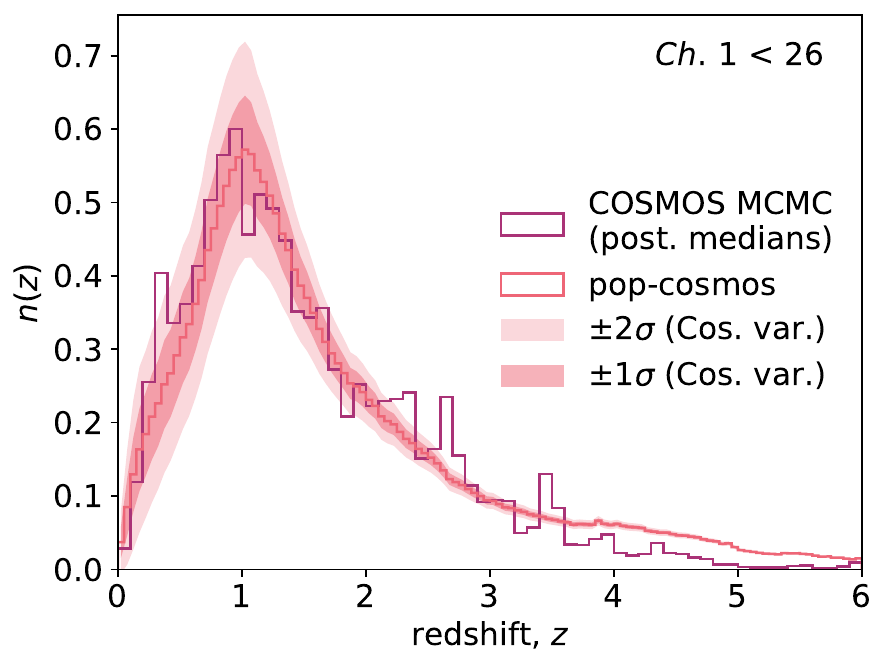}
    \caption{Normalized redshift distribution for \textbf{(top)} $r<25$ and \textbf{(bottom)} $\textit{Ch.\,1}<26$. Shaded regions show the fractional effect of cosmic variance that would be expected for a COSMOS-sized field. The purple histogram shows posterior median redshifts for \textbf{(top)} 146,252, and \textbf{(bottom)} 423,262 COSMOS galaxies that pass each selection cut (see Section~\ref{sec:results_mcmc_redshift}).}
    \label{fig:nz}
\end{figure}

Inferring the redshift distribution of a population of galaxies, subject to selection, is one of the most important statistical challenges of doing cosmology with photometric survey data \citep[for background, see][]{leistedt16, malz21, malz22, newman22}. In Figure \ref{fig:nz}, we show the predicted redshift distribution for the limited COSMOS-like mock catalog with $\textit{Ch.\,1}<26$. We compute the redshift distribution of our mock galaxies with a redshift bin width of $\Delta z=0.05$. For each bin, we compute the fractional cosmic variance based on \citet{moster11}. In this calculation, we assume a COSMOS-sized field ($84'\times84'$), and use the galaxy bias parameters for a stellar mass threshold of $\log_{10}(M/M\odot)>8.5$ in each bin (from table 4 of \citealp{moster11}).

Overlaid on Figure \ref{fig:nz}, we show a histogram of the posterior median redshifts for the actual COSMOS2020 galaxies, based on the inference presented in Section~\ref{sec:results_mcmc_redshift}. For $r<25$, we see a prominent spike for the $0.3\leq z<0.4$ redshift bin, corresponding to large-scale structure (LSS) in the COSMOS field at  $z\sim0.35$ \citep[see, e.g.,][]{scoville07_lss, sochting12, cherouvrier25}. This feature is less prominent in the $\textit{Ch.\,1}<26$ distribution. Several higher-$z$ spikes are visible prominently in the \textit{Ch.\,1}-selected redshift histogram, potentially correlating with $z\geq2$ LSS and proto-clusters \citep[for an extensive review see, e.g.,][]{toni25, mcconachie25}. Confirmed structures at $z\sim2.5$ \citep[e.g.,][]{diener15, chiang15, cucciati18, sikorski25}, $z\sim2.7$ \citep{ito23}, and $z\sim3.4$ \citep{mcconachie22, forrest23} may be responsible for the two most prominent features.

Additionally visible when comparing the galaxy-level posterior medians to the \texttt{pop-cosmos} predictions is a less steep drop-off in sources in the model $n(z)$ at $z\gtrsim4$ when compared to the galaxy-level inferences. This is seen prominently for the $\textit{Ch.\,1}<26$ selection in the lower panel of Figure \ref{fig:nz}, and less strongly in the $r<25$ selection for $4\lesssim z\lesssim 5$. We hypothesize that the galaxies ``missing'' from the COSMOS sample at $z\gtrsim4$ are those lost due to the incompleteness of the data at $25\lesssim\textit{Ch.\,1}\lesssim26$ (see Sections~\ref{sec:photometry} and \ref{sec:results_photo}, and \citealp{davidzon17, weaver22, weaver23_smf}). We find that $\sim57\%$ of \texttt{pop-cosmos} model galaxies with $z>5$ have $25<\textit{Ch.\,1}<26$, placing them in the range where the observed and model $\textit{Ch.\,1}$ magnitude distributions being to diverge (see Figure \ref{fig:mags_ch126}).

\subsection{Stellar Mass Function}
\label{sec:smf}
\begin{figure}
    \centering
    \includegraphics[width=\linewidth]{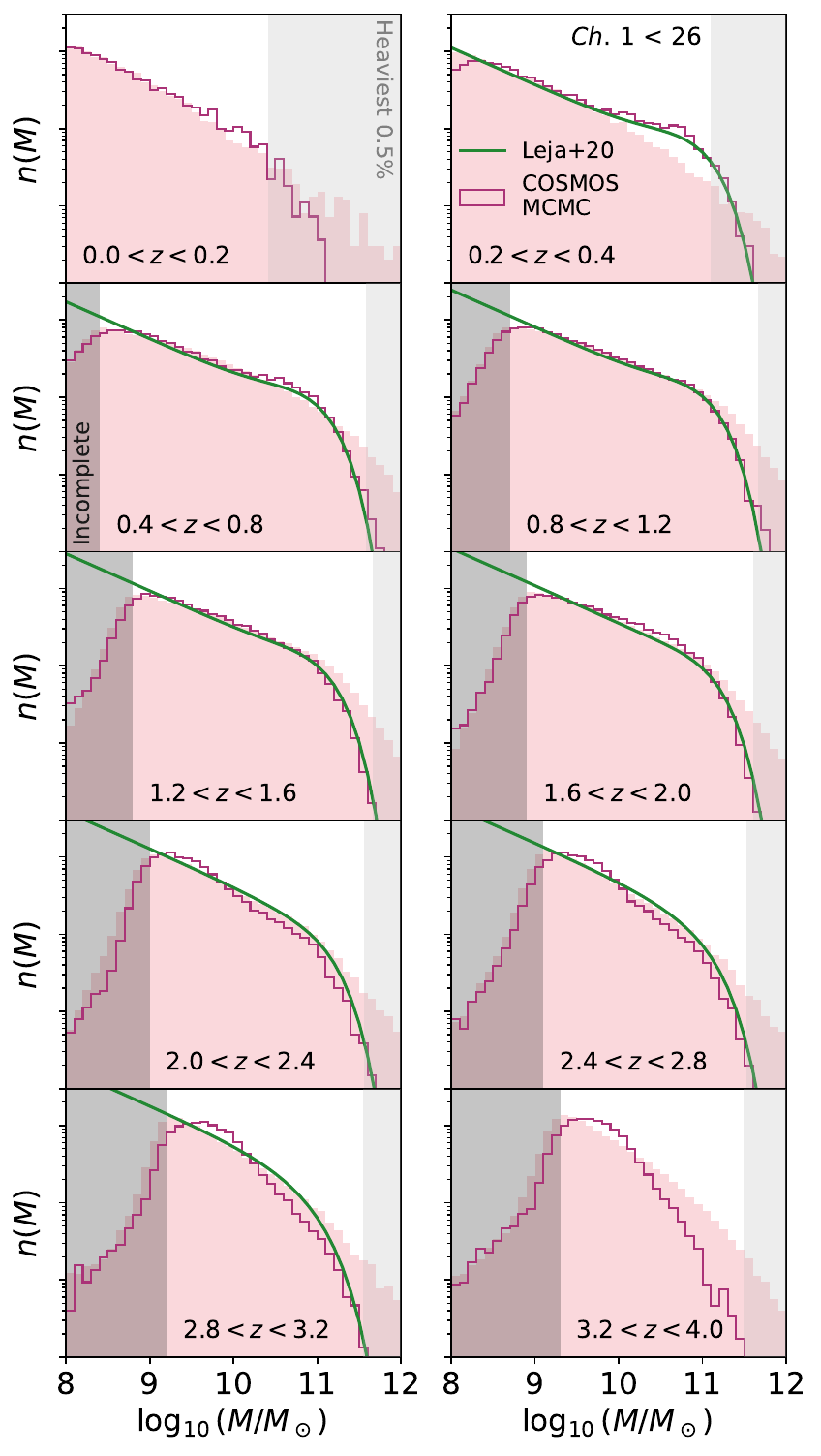}
    \caption{Stellar mass function (normalized, logarithmic axes) predicted by \texttt{pop-cosmos} (solid red histogram), compared to the stellar masses from Bayesian SED fitting (purple) to COSMOS2020 galaxies selected on $\textit{Ch.\,1}<26$ (see Section~\ref{sec:results_mcmc_mass}). For the \texttt{pop-cosmos} draws, we consider the most massive 0.5\% of galaxies at a given redshift (rightmost gray shaded region) to be an extrapolation due to the rarity of such galaxies in a COSMOS-sized field. The histograms are normalized to integrate to 1 between the vertical shaded regions. In green we plot the double Schechter mass function from \citet{leja20}, normalized within the same limits.}
    \label{fig:nM}
\end{figure}

The redshift evolving probability density function over stellar mass --- the stellar mass function --- has been a key observational target \citep[for recent examples see, e.g.,][]{leja20, driver22, weaver23_smf, shuntov24, zalesky25}, and is a fundamental test for cosmological models \citep[see, e.g.,][]{somerville15}. Figure \ref{fig:nM} shows the stellar mass distribution of the draws from the \texttt{pop-cosmos} generative model for $\textit{Ch.\,1}<26$. We shade out the most massive 0.5\% of galaxies in each redshift bin on the basis of the results in Figure \ref{fig:mass_redshift}. For $M\lesssim10^{11}M_\odot$, we find that the distribution of stellar masses agrees well with the distribution implied by SED fits to the COSMOS2020 galaxies from Section~\ref{sec:results_mcmc_mass}. At $M\gtrsim10^{11}M_\odot$, we see that \texttt{pop-cosmos} has a much heavier tail than found by the COSMOS SED fits from Section~\ref{sec:results_mcmc_mass}, although the fraction of the 2 million mock galaxies that lie in this region is very small, generally within the grayed-out 99.5th percentile. The completeness limit (based on the turnover of the \texttt{pop-cosmos} mass function) is shown in black at the left-hand edge of each panel. In the $0.2<z<0.4$ bin, the mass distribution implied by the SED fitting results shows a noticeable shoulder around $10^{10}\lesssim M/M_\odot\lesssim10^{11}$. This may be driven by the higher prevalence of large scale structures seen in the COSMOS field around $z\sim0.35$ \citep{scoville07_lss, sochting12, cherouvrier25}; see also Section~\ref{sec:nz} for further discussion.

Overlaid on Figure \ref{fig:nM} is the double Schechter mass function constrained by \citet{leja20} using data from COSMOS2015 \citep{laigle16} and 3D-HST \citep{skelton14, whitaker14, momcheva16}. The \citet{leja20} mass function is evaluated based on the posterior median parameter values given in their figure 3. We chose to show the \citet{leja20} result as our comparison here, as it is determined based on a deep sample selected from the COSMOS field, and as a self-consistent constraint on the star-forming main sequence is available for comparison (see our Section~\ref{sec:sfs}) from \citet{leja22}. Additionally, the \citet{leja20} mass function is inferred in a Bayesian framework, with full propagation of uncertainties, making it a particularly interesting comparison. 

For $z>0.4$, the low-mass slope of the \texttt{pop-cosmos} mass function shows excellent agreement with \citet{leja20} for masses above our estimated completeness limit. The high-mass slope of the \citet{leja20} mass function agrees well with the mass distribution implied by our galaxy-level inference (the purple unshaded histogram in Figure \ref{fig:nM}), declining more steeply past the ``knee'' of the mass function than the distribution implied by the \texttt{pop-cosmos} predictions. Interestingly, the results from \citet{leja20} are closer to the galaxy-level inference in the $0.2<z<0.4$ bin, with a more pronounced knee than is seen in the \texttt{pop-cosmos} predictions. At $z<0.5$, the \citet{leja20} stellar mass function is only constrained by COSMOS2015, so any large-scale structure present in the field at low-$z$ (e.g., the $z\sim0.35$ feature seen in Figure \ref{fig:nz} and by \citealp{scoville07_lss, sochting12, cherouvrier25}) could influence our galaxy-level inferences and the \citet{leja20} mass function. Inspecting the stellar mass function from our galaxy-level inference in narrower redshift bins, we find that the $0.3\leq z<0.4$ slice has a noticeable excess of $10\lesssim \log_{10}(M/M_\odot)\lesssim 11$ galaxies compared to the $0.2\leq z<0.3$ and $0.4\leq z<0.5$ bins, consistent with the presence of a cluster. Within this mass range, we find that the $0.3\leq z<0.4$ bin also has an excess of galaxies with $\log_{10}(\text{SFR}/M_\odot\,\text{yr}^{-1})$ compared to the neighbouring bins, further aligning with this picture.

We forgo a more detailed to other mass function estimates in the literature (e.g., the COSMOS2020 results from \citealp{weaver23_smf}, or the Cosmic Dawn Survey results from \citealp{zalesky25}), but will revisit the stellar mass function predicted by \texttt{pop-cosmos} in future work by S.\ Deger et al. (in prep.). Recent observational studies (\citealp{weaver23_smf, xiao24, weibel24, zalesky25}; but see also \citealp{shuntov24}) and simulations \citep[e.g.,][]{bennett24} have favoured the presence of non-negligible numbers of very massive galaxies with $\log_{10}(M/M_\odot)\gtrsim11.5$, and simulations \citep[e.g.,][]{lagos18, lagos19, dave19} often predict shallower high-mass slopes than observations (see, e.g., comparisons in \citealp{adams21, weaver23_smf}). However, observational constraints are sensitive to the strategy adopted for correcting Eddington bias \citep{eddington13, eddington40}, with this substantially influencing the inferences made about the high-mass end of the mass function \citep[see, e.g.,][]{ilbert13, caputi15, grazian15, davidzon17, adams21, weaver23_smf, zalesky25}.

\citet{leja20} argues that their hierarchical Bayesian approach with principled uncertainty propagation from the photometry to the mass function ought to naturally avoid Eddington bias. A similar argument can be made for the population-level approach we have taken here with \texttt{pop-cosmos}. Irrespective of Eddington bias, the high-mass end of the stellar mass function is also generally more susceptible to the effects of cosmic variance when constrained based on small-area surveys \citep[see, e.g., the discussion in][]{jespersen24}. The calibration of the \texttt{pop-cosmos} stellar mass function at the massive end can likely be improved in future work by incorporating data from a wider area (e.g., the full 670~deg$^2$ HSC survey area; \citealp{aihara22}; or the 1347~deg$^2$ KiDS catalog; \citealp{wright24}).

\begin{figure}
    \centering
    \includegraphics[width=\linewidth]{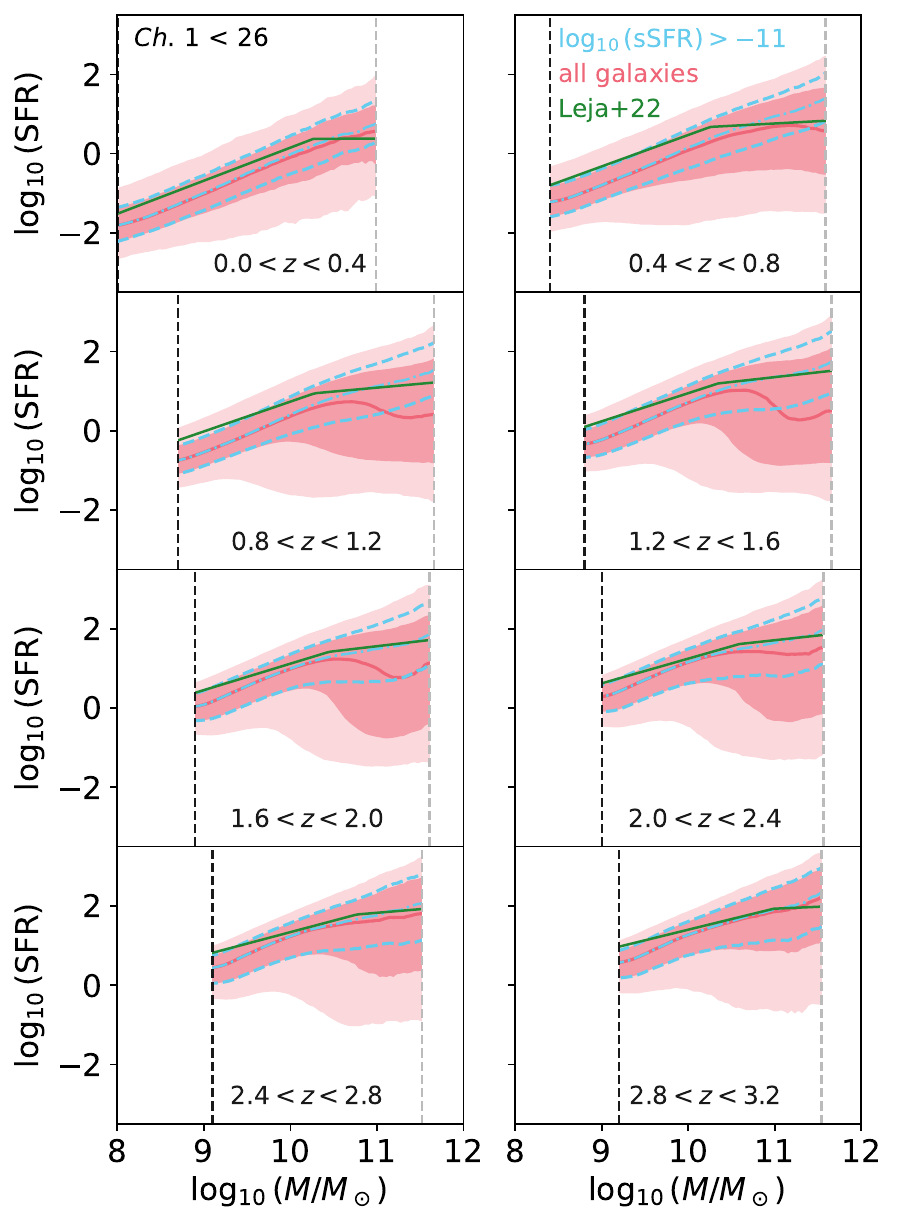}
    \caption{Star forming main sequence predicted by \texttt{pop-cosmos}. The solid red line shows the rolling median star formation rate ($M_\odot\,\text{yr}^{-1}$) conditional on stellar mass. Red shaded regions show the rolling 68 and 95\% credible intervals. Rolling quantities are computed in a stellar mass window of width $\delta\log_{10}(M/M_\odot)=0.4$~dex. The mass limits from Figure \ref{fig:nM} are applied and indicated with vertical dashed lines. The overlaid cyan lines show the median and 68\% interval for the subset of star-forming model galaxies with $\log_{10}(\text{sSFR}/\text{yr}^{-1})>-11$. The overlaid green line shows the double power-law from \citet{leja22}.}
    \label{fig:sfs}
\end{figure}

\subsection{Star-Forming Main Sequence}
\label{sec:sfs}
The tight correlation between stellar mass and SFR amongst star forming galaxies \citep[see][]{daddi07, whitaker12, whitaker14, speagle14, popesso23} has important implications for their evolutionary pathways \citep[e.g.,][]{peng10, leitner12, abramson15}, and has seen a growing qualitative consensus over the past 15 years of observational studies, albeit with some quantitative tensions \citep[see discussion in][]{wuyts11, speagle14, leja15, leja22, sandles22, popesso23, fu24}. Figure \ref{fig:sfs} shows the star-forming main sequence predicted by \texttt{pop-cosmos} for $\textit{Ch.\,1}<26$. In the lowest redshift bin, this is highly consistent with the results from \citetalias{alsing24} (whose results have shown good consistency at $z=0$ with simulations; \citealp{bennett24}). At higher redshifts, we see that the 68\% credible interval encompasses lower SFRs, particularly at the massive end, than were included previously. This broadening of the distribution is likely reflective of the greater completeness of the $\textit{Ch.\,1}<26$ selection with respect to quiescent galaxies \citep{weaver23_smf}. In Figure \ref{fig:sfs} we also show the median and 68\% interval for star-forming galaxies --- selected based on $\log_{10}(\text{sSFR}/\text{yr}^{-1})>-11$ \citep[e.g.,][]{ilbert10, ilbert13, dominguez11} --- which shows the expected monotonic rise. The star-forming/quiescent split of \texttt{pop-cosmos} model galaxies, and the demographics of these samples will be explored in more detail by S.\ Deger et al.\ (in prep.).

\begin{figure}
    \centering
    \includegraphics[width=\linewidth]{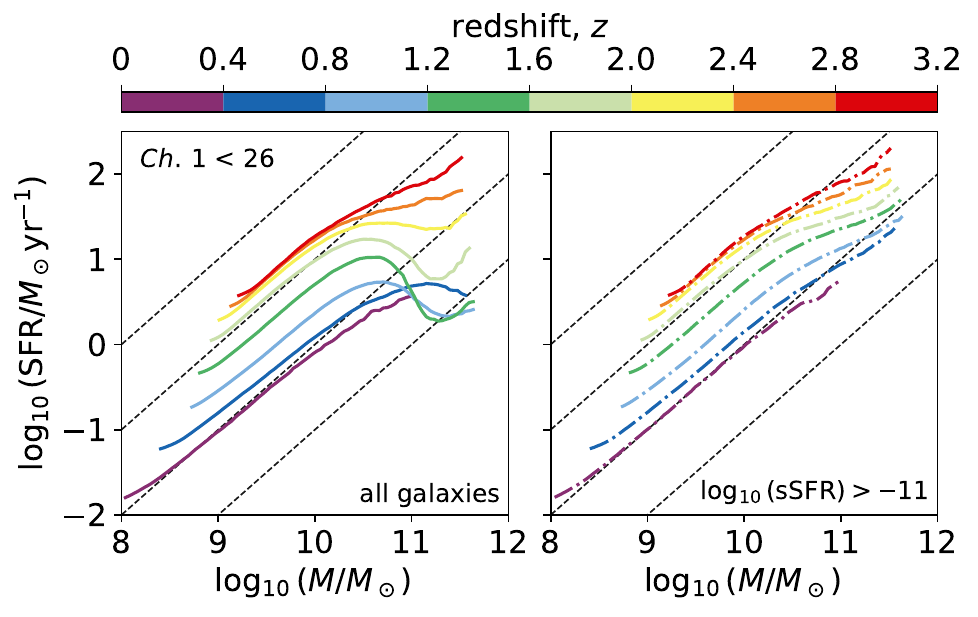}
    \caption{Median star forming sequence predictions from Figure \ref{fig:sfs} for \textbf{(left)} all galaxies, and \textbf{(right)} galaxies with $\log_{10}(\text{sSFR})>-11$. The black dashed lines show constant sSFR; top to bottom these correspond to $\log_{10}(\text{sSFR}/\text{yr}^{-1})=[-8, -9, -10, -11]$.}
    \label{fig:sfs_medians}
\end{figure}

Overlaid on Figure \ref{fig:sfs} is the double power-law SFS model constrained by \citet{leja22} using the same COSMOS2015 and 3D-HST sample as was used in the \citet{leja20} stellar mass function analysis, which agrees well with other literature results with regard to the shape and slope of the SFS (for an exhaustive review and comparison of these, see \citealp{speagle14, popesso23}). It is worth noting that \citet{leja22} found a consistently lower normalization at a given stellar mass than previous literature results; likely due to improved stellar mass and SFR estimates arising from their use of a non-parametric SFH \citep[see][]{carnall19, leja19, leja19_sfh, lower20, sandles22}. We show the \citet{leja22} result by evaluating the fitting formula in their table 1 at the central redshift of each bin in Figure \ref{fig:sfs}. We find overall very good agreement with \citet{leja22} across the full redshift range covered in Figure \ref{fig:sfs}. At $z>1$, we find a slightly steeper relation at the low mass end than \citet{leja22}; this aligns well with recent empirical modeling results based on abundance matching \citep{fu24}, and predictions from cosmological simulations \citep{chaikin25}. For $z<2$, we find that the low-mass slope remains fairly consistent, and that the median relation at low masses has a gradient fairly close to unity in this redshift range. At $1.6<z<2$, the median $\log_{10}(\text{SFR})$ vs.\ $\log_{10}(M/M_\odot)$ relation for $\log_{10}(M/M_\odot)\lesssim10$ corresponds to $\text{sSFR}\sim10^{-9}\,\text{yr}^{-1}$. By the lowest redshift bin at $0.0<z<0.4$, the sSFR implied by the median relation has dropped by almost an order of magnitude to $\text{sSFR}\sim10^{-10}\,\text{yr}^{-1}$, with this being the median across the full mass range $\log_{10}(M/M_\odot)<11$. For $z>2$, the low-mass slope predicted by \texttt{pop-cosmos} falls somewhere between the shallower slope inferred by \citet{leja22}, and the slightly steeper slope that would be implied by a constant $\text{sSFR}=10^{-9}\,\text{yr}^{-1}$. The median $\log_{10}(\text{SFR})$ vs.\ $\log_{10}(M/M_\odot)$ relations from Figure \ref{fig:sfs} are shown in Figure \ref{fig:sfs_medians}, with lines of constant sSFR highlighted.

\subsection{Mass--Metallicity Relation}
\label{sec:mass_metallicity}
\begin{figure}
    \centering
    \includegraphics[width=\linewidth]{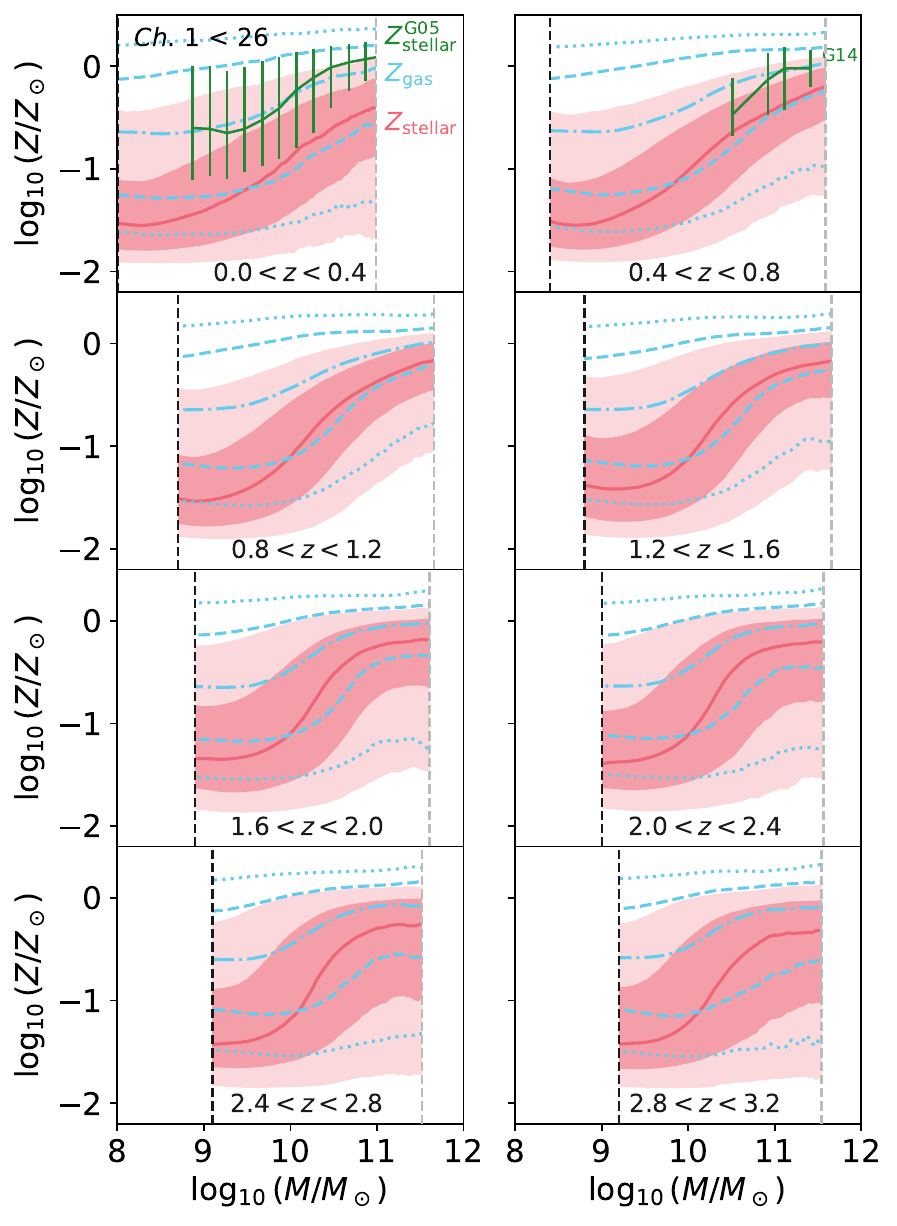}
    \caption{Stellar mass--metallicity relation (red) predicted by \texttt{pop-cosmos} in the redshift bins from Figure \ref{fig:sfs}, for galaxies with $\textit{Ch.\,1}<26$. The gas-phase metallicity vs.\ stellar mass relation is overlaid in cyan. Overlaid in green on the first panel, we show the \citet[\citetalias{gallazzi05}]{gallazzi05} stellar metallicity vs.\ mass relation (median and 68\% interval) based on SDSS spectra. On the second panel, the green shows the $z\sim0.7$ results from \citet[\citetalias{gallazzi14}]{gallazzi14}.}
    \label{fig:mass_metallicity}
\end{figure}

Both the gas-phase \citep{tremonti04} and stellar metallicity \citep{gallazzi05} of galaxies have been observed (predominantly using spectroscopic indicators of these quantities) to correlate with stellar mass, with important implications for the chemical evolution of galaxies \citep[for a review, see][]{maiolino19}. In Figure \ref{fig:mass_metallicity}, we show the stellar metallicity vs.\ stellar mass relation in increasing redshift bins. In each bin, we show the median metallicity as a function of stellar mass, as well as the 68 and 95\% credible intervals. We overlay in cyan the gas-phase metallicity vs.\ stellar mass relation (median, 68\%, and 95\% credible intervals). As in \citetalias{alsing24}, the gas-phase metallicity is consistently higher than the stellar metallicity at all redshifts. In all redshift bins, the gas-phase metallicity decreases less steeply towards low masses than the stellar metallicity. At the high mass end, we typically see an offset of $\sim+0.2$~dex between gas-phase and stellar metallicity, with a larger offset at the low-mass end.

\begin{figure}
    \centering
    \includegraphics[width=\linewidth]{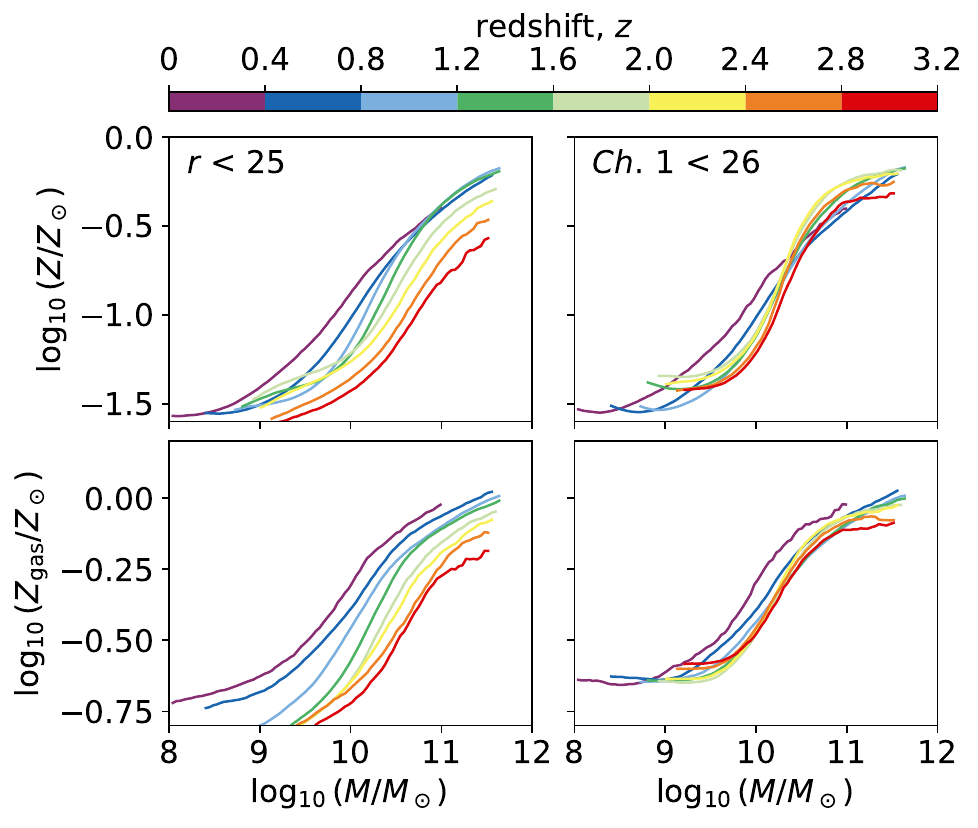}
    \caption{Median mass--metallicity relations predicted by \texttt{pop-cosmos} in the redshift bins from Figure \ref{fig:mass_metallicity}. We show the predictions for \textbf{(left)} the $r<25$ selection, \textbf{(right)} the $\textit{Ch.\,1}<26$ selection; and \textbf{(top)} the stellar metallicity, \textbf{(bottom)} the gas-phase metallicity. }
    \label{fig:mass_metallicity_medians}
\end{figure}

Figure \ref{fig:mass_metallicity_medians} shows the median stellar and gas-phase metallicity--mass relations from Figure \ref{fig:mass_metallicity}. We show these for both the $r<25$ and $\textit{Ch.\,1}<26$ mock galaxy catalogs generated with \texttt{pop-cosmos}. We find a much stronger redshift evolution for the optically-selected sample compared to the MIR-selected one; potentially due to the MIR selection including a wider diversity of galaxy types at any given redshift. In both cases however, the evolution of the median relation is much smaller than the inferred scatter in any given redshift bin. For both selections, we observe a steepening of the relation with redshift. The results for the $r<25$ sample are in good general agreement with the trends seen by \citetalias{alsing24}, although we see less of an obvious convergence in metallicities at high stellar mass. 

Comparing our $0.0<z<0.4$ results to the well known stellar mass vs.\ metallicity relation (shown in green and indicated \citetalias{gallazzi05} on Figure \ref{fig:mass_metallicity}) from \citet{gallazzi05}, our median relation is $\sim0.5$~dex lower at the high-mass end, and $\sim1.0$~dex lower at $\log_{10}(M/M_\odot)\approx9$ (the lowest stellar mass reported by \citealp{gallazzi05}). The \citet{gallazzi05} analysis is based on SDSS galaxies \citep{york00, abazajian04} with $r<17.77$; brighter than almost anything in the COSMOS field. In the $0.4<z<0.8$ redshift bin, the \texttt{pop-cosmos} median relation shows a smaller offset when compared to the $z\sim0.7$ results from \citet{gallazzi14}, which were based on a similar methodology to \citet{gallazzi05}. Due to the selection involved in the \citet{gallazzi05, gallazzi14} analyses, and the very different methodology used to constrain stellar metallicity, interpretation of the offset is challenging. \citet{gallazzi05, gallazzi14} base their constraints on comparison between a set of absorption line diagnostics \citep[e.g.,][]{worthey94} to model spectra from the \citet{bruzual03} stellar population models. Results at low-$z$ from full spectral fitting \citep[e.g.,][]{panter08, zahid17}, or from measurements of individual supergiant stars \citep[e.g.,][]{kudritzki16}, have differed in the asymptote at high masses, and the predicted behavior at low masses\footnote{Additionally, studies of local dwarf galaxies have suggested extrapolation to extremely subsolar metallicities at very low mass \citep{kirby13}.}. 

\begin{figure}
    \centering
    \includegraphics[width=\linewidth]{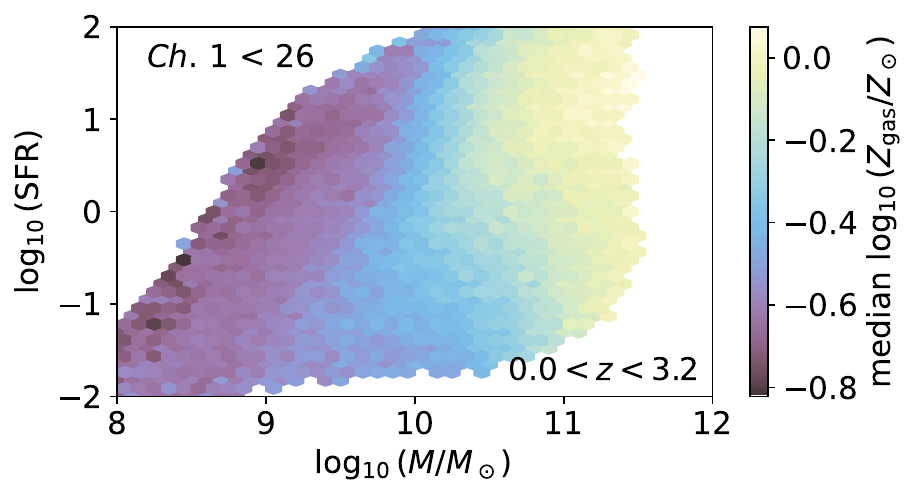}
    \caption{Fundamental metallicity relation for $\textit{Ch.\,1}$ galaxies at $0.0<z<3.2$, visualized in the style of \citet{curti20}. Mock galaxies drawn from the generative model are binned by stellar mass and SFR, with the cells shaded by the median gas-phase metallicity. Cells are shaded only if they contain $>100$ model galaxies.}
    \label{fig:FMR}
\end{figure}

High-$z$ studies typically rely on different spectroscopic absorption \citep[e.g.,][]{rix04, halliday08, sommariva12, calabro21}, or full spectral fitting \citep[e.g.,][]{cullen19, kashino22, chartab24}. The high-$z$ spectroscopic constraints typically find much lower stellar metallicities than the low-$z$ relations inferred by e.g., \citet{gallazzi05}, \citet{kudritzki16}, or \citet{zahid17}. Results from \citet[$2.5\leq z\leq5.0$]{cullen19}, \citet[$1.6\leq z\leq3.0$]{kashino22} and \citet[$2\lesssim z \lesssim 3$]{chartab24} suggest that the median stellar metallicity at a given mass is shifted downwards by $\sim0.5$--$1.0$~dex relative to the relation at low-$z$, with a fairly similar slope between the low- and high-mass ends. This is in good qualitative agreement with the trend we predict for the $r<25$ selection (which should be more reflective of the depth of the high-$z$ spectroscopic surveys used; \citealp{lilly07, pentericci18, mclure18, newman20, kashino21}). As shown by \citet{cullen19}, the normalization of the stellar metallicity vs.\ mass relation predicted by simulations \citep[e.g.,][]{ma16, dave19} is often lower than the spectroscopic constraints suggest, although predictions for the shape and redshift evolution are similar.

As it is traced by emission --- rather than absorption --- features, the stellar mass dependence of gas-phase metallicity has been studied more extensively with spectroscopic observations. Of particular interest has been the fundamental metallicity relation (FMR) between stellar mass, SFR, and gas-phase metallicity \citep{mannucci10, lara10}. We visualize the FMR implied by draws from the \texttt{pop-cosmos} model in Figure \ref{fig:FMR}, by plotting SFR against stellar mass, with each cell shaded based on the median $\log_{10}(Z_\text{gas}/Z_\odot)$ in that part of the $\log_{10}(\text{SFR})$ vs.\ $\log_{10}(M/M_\odot)$ plane (inspired by figure 5 of \citealp{curti20}). In Figure \ref{fig:FMR}, we include all \texttt{pop-cosmos} model draws with $z<3.2$ and $\textit{Ch.\,1}<26$, motivated by observational \citep[e.g.,][]{mannucci10, zahid14, hunt16, cresci19, curti24} and theoretical \citep[e.g.,][]{lilly13} results suggesting approximate invariance of the FMR out to at least $z\sim3$. We find that the \texttt{pop-cosmos} result in Figure \ref{fig:FMR} is fairly insensitive to the upper redshift limit chosen, or the choice of an $r<25$ selection, although in future work a more detailed investigation of this would be worthwhile. We see good qualitative agreement between our Figure \ref{fig:FMR} and figure 5 of \citet{curti20}. The strongest trend is one of metal enrichment with increasing stellar mass at a fixed SFR. We also see the expected gradient from the upper left towards the lower right of the Figure, implying that the most star-forming galaxies at a given stellar mass have the lowest gas-phase metallicity (with a weaker SFR dependence at the high mass end, reflecting the saturation). As noted by \citet{curti20}, quantitative comparisons of the FMR even amongst spectroscopic measurements can be challenging due to differing selection criteria and metallicity calibrations, so we forgo a more detailed comparison here.

\begin{figure}
    \centering
    \includegraphics[width=\linewidth]{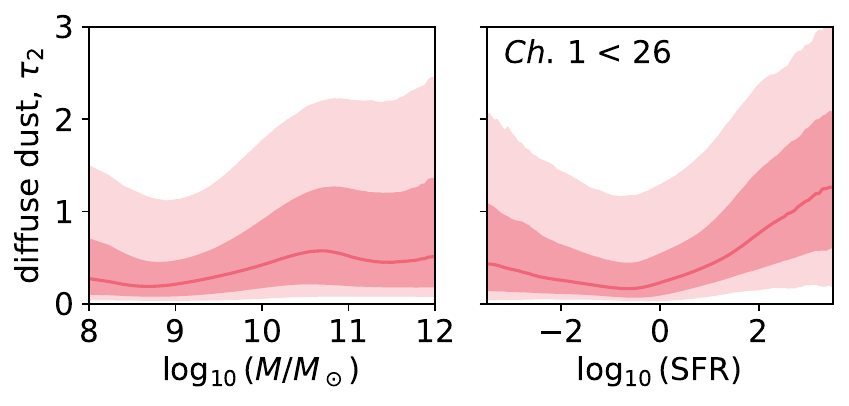}%
    \caption{Diffuse dust attenuation $\tau_2$ vs.\ \textbf{(left)} stellar mass, and \textbf{(right)} star formation rate, as predicted by \texttt{pop-cosmos} for galaxies with $\textit{Ch.\,1}<26$.}
    \label{fig:dust2}
\end{figure}

\subsection{Dust Attenuation}
Modeling the dust attenuation of starlight (for a review see \citealp{calzetti01, salim20}) is crucial for the interpretation of other observables \citep[see, e.g.,][]{salim19, lower22, nagaraj22_tng}. Correlations between attenuation properties and other quantities such as stellar mass and SFR have been investigated in detail both observationally \citep[e.g.,][]{garn10, zahid13, salim16, salim18, nagaraj22} and theoretically \citep[e.g.,][]{witt00, chevallard13, sommovigo25}. Figure \ref{fig:dust2} shows the distribution of diffuse dust attenuation, as a function of stellar mass and star formation rate. The results here are consistent with \citetalias{alsing24}, albeit with a slightly heavier tail towards high dust attenuation at a given stellar mass or star formation rate. Again, this reflects the change of the \texttt{pop-cosmos} training set to include extremely dusty galaxies that are too optically faint to pass an $r<25$ selection. The upturn in the level of dust attenuation for $\text{SFR}\gtrsim1~M_\odot\,\text{yr}^{-1}$ reproduces the trends seen by, e.g., \citet{zahid13} and \citet{nagaraj22}. The \texttt{pop-cosmos} predictions of dust attenuation are evaluated further by \citet{petri25}, who find that the median attenuation predicted by \texttt{pop-cosmos} is consistently lower than \citet{nagaraj22} for all SFRs, but that the upturn at high SFR occurs in the same place and has a comparable slope. \citet{petri25} show that this has significant implications for the selection of Lyman-break galaxies, highlighting the importance of constraining this relationship at high-$z$.

\begin{figure}
    \centering
    \includegraphics[width=\linewidth]{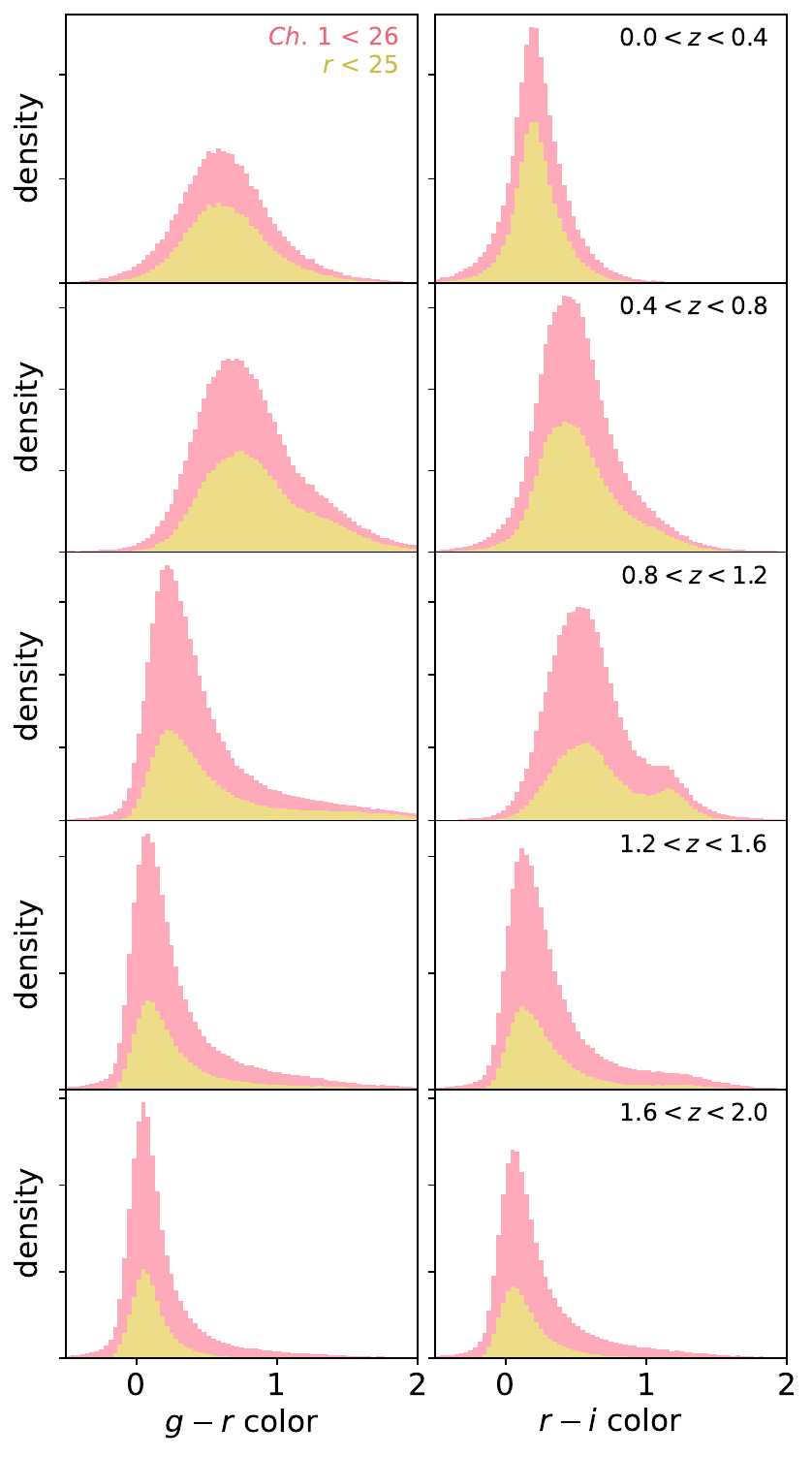}
    \caption{Redshift evolution of the observer-frame \textbf{(left)} $g-r$ and \textbf{(right)} $r-i$ color distributions predicted by \texttt{pop-cosmos}. Histograms are plotted for the $\textit{Ch.\,1}<26$ mock catalog, with the $r<25$ portion of this catalog shaded in yellow.}
    \label{fig:color_redshift_binned}
\end{figure}

\subsection{Observer-Frame Color--Redshift Relations}
\label{sec:color_redshift}
As well as looking at the survey-wide distribution of observer-frame colors, as we did in Section \ref{sec:results_photo}, we can also examine the redshift evolution of these distributions and their associations with physical parameters. Figure \ref{fig:color_redshift_binned} shows the observer-frame $g-r$ and $r-i$ color distributions predicted by \texttt{pop-cosmos} in a series of redshift bins for $z<2.0$.  Within this redshift range, the typical $g-r$ colors are reddest around $z\sim0.6$. At $z\sim1.0$ the $r-i$ color shows a noticeable bimodality --- more pronounced for the $r<25$ selection\footnote{In Figure \ref{fig:color_redshift_binned}, we are plotting the $\textit{Ch.\,1}$-selected mock catalog, with the $r<25$ portion of this highlighted in yellow.} --- that is seen prominently in spectroscopic surveys that cover this range \citep[e.g.,][]{newman13}. The expected $g-r$ bimodality is less strongly visible.

\begin{figure}
    \centering
    \includegraphics[width=\linewidth]{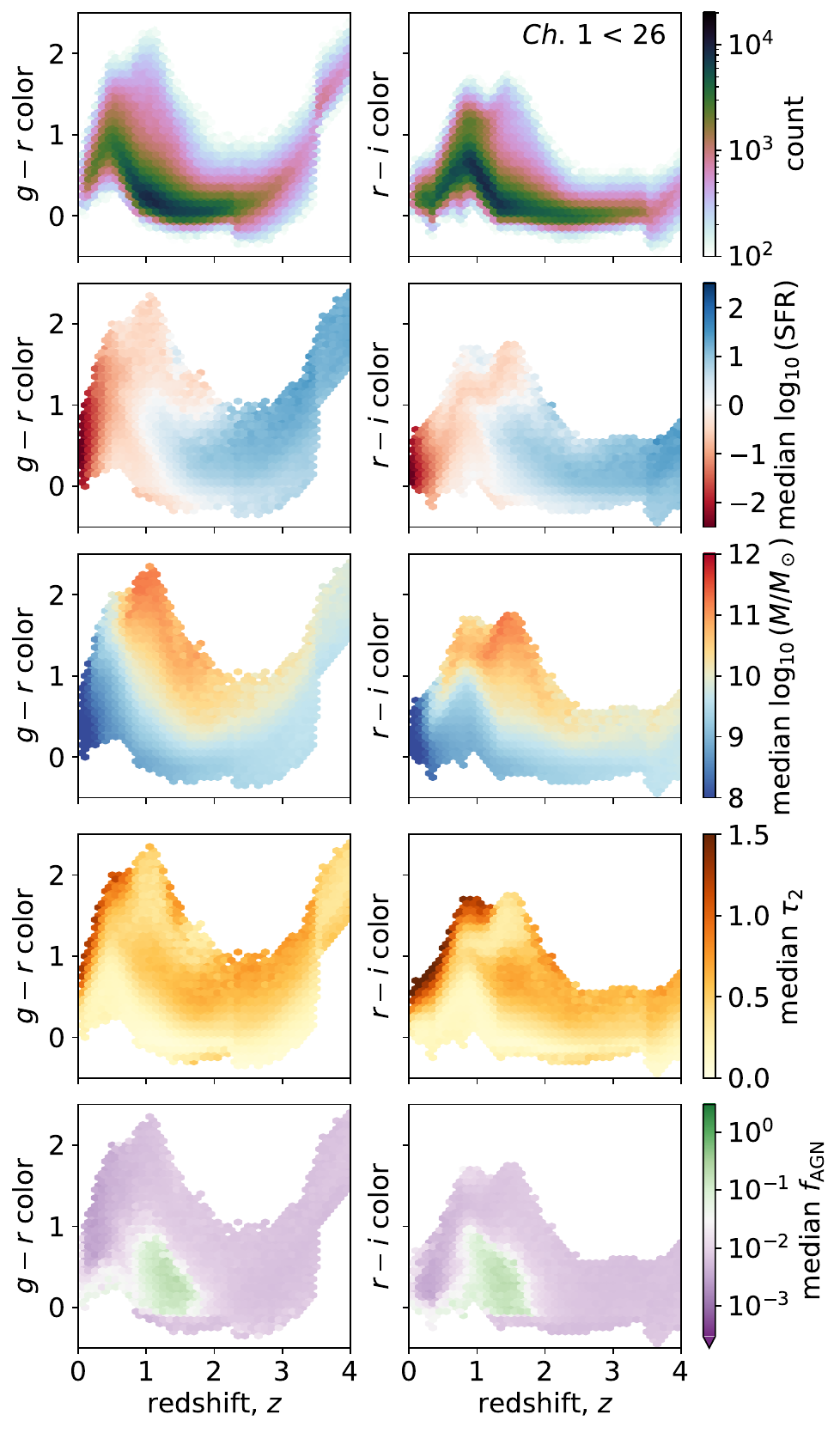}
    \caption{Observer-frame color vs.\ redshift relations, for \textbf{(left)} $g-r$ and \textbf{(right)} $r-i$, predicted by \texttt{pop-cosmos} for $\textit{Ch.\,1}<26$. A cell in each panel is shaded only if it contains $>100$ model galaxies ($0.005\%$ of the full mock catalog), and galaxies are only included above the mass completeness limit. Panels are shaded by: \textbf{(row 1)} model galaxy count; \textbf{(row 2)} median $\log_{10}(\text{SFR}/M_\odot\text{yr}^{-1})$; \textbf{(row 3)} median $\log_{10}(M/M_\odot)$; \textbf{(row 4)} diffuse dust optical depth at 5500\AA, $\tau_2$; and \textbf{(row 5)} median AGN bolometric luminosity fraction, $f_\text{AGN}$.}
    \label{fig:color_redshift}
\end{figure}

In Figure \ref{fig:color_redshift}, we show the $g-r$ and $r-i$ color vs.\ redshift relations predicted by \texttt{pop-cosmos} for $0<z<4$. The density (top row) at $z<1.5$ shows the structure expected from deep spectroscopic surveys \citep[e.g.,][]{newman13}. At $z\approx3.5$ in the $g-r$ color, we see a striking ``pinch'' point, coinciding with Ly-$\alpha$ 1215\AA\ reaching the red edge of the $g$-band. In rows 2--4 of Figure \ref{fig:color_redshift}, we shade each cell based on the median physical properties of galaxies in that cell. The correlation between $g-r$ and $r-i$ and SFR is strongest at $z\sim1$. For $1<z<3$, the model predicts a strong correlation between optical color and mass, seen in the third row of Figure \ref{fig:color_redshift}. The fourth row illustrates that the optical depth of diffuse dust strongly influences $r-i$ color at $z<1.5$, and $g-r$ color at $z<1$. From the fifth row, we see that there are small regions of color--redshift space with high median AGN fractions ($\gtrsim10\%$) and blue colors at $z<2$. The AGN model in \texttt{pop-cosmos} \citep{nenkova08i, nenkova08ii, leja18} concerns re-radiation from a dust torus, so the association with blue color is not due to optical emission from an accretion disk. As was suggested by \citet{jespersen25} recently, IR luminosity can be very tightly coupled to optical color, which could be influenced by feedback or AGN--host co-evolution. We now explore this result in greater detail.

\subsection{AGN Luminosity}
\label{sec:agn_results}
\begin{figure}
    \centering
    \includegraphics[width=\linewidth]{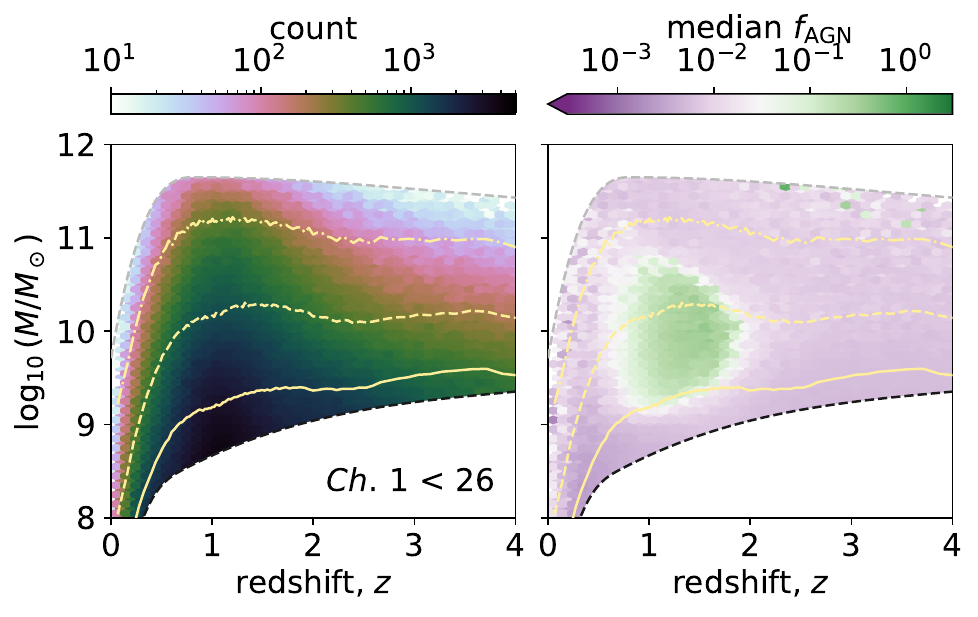}
    \caption{Stellar mass vs.\ redshift for $\textit{Ch.\,1}<26$. Cells are shaded by \textbf{(left)} model galaxy count, and \textbf{(right)} median AGN bolometric luminosity fraction, $f_\text{AGN}$. Cells are shaded only if they contain $>10$ model galaxies from the total sample of 2 million. The black and gray dashed lines show respectively the mass completeness limit and upper mass limit (see Figures \ref{fig:mass_redshift}, \ref{fig:mass_redshift_fit}, and Appendix \ref{sec:mcuts}). Solid, dashed, and dash-dotted yellow lines show respectively the 50th, 84th, and 97.5th percentiles of $\log_{10}(M/M_\odot)$ given $z$ (same as Figure \ref{fig:mass_redshift}).}
    \label{fig:mass_redshift_agn}
\end{figure}

One potentially significant source of feedback in galaxies is AGN emission.  We investigate this possibility in Figure~\ref{fig:mass_redshift_agn}, which shows the stellar mass vs.\ redshift relation from Figure~\ref{fig:mass_redshift}, but with the right-hand panel shaded based on the median AGN bolometric luminosity fraction, $f_\text{AGN}$, in each bin. In \citetalias{alsing24} and here (see the full SPS parameter distribution in Appendix \ref{sec:corner}), we generally see a bimodal population distribution in $f_\text{AGN}$, with the two modes broadly corresponding to negligible and non-negligible AGN contributions. We set the color scale in the right panel of Figure \ref{fig:mass_redshift_agn} to have its pivot point at $3\times10^{-2}$, approximately between the two modes of the population distribution\footnote{This corresponds to an AGN torus with a bolometric luminosity that contributes $3\%$ as much as the stellar component.} (c.f., Figures \ref{fig:nfAGN} and \ref{fig:corner}). This shows that there is a region with $9\lesssim\log_{10}(M/M_\odot)\lesssim11$ and $1\lesssim z\lesssim2$ where the median $f_\text{AGN}$ falls within the non-negligible range.

\begin{figure}
    \centering
    \includegraphics[width=\linewidth]{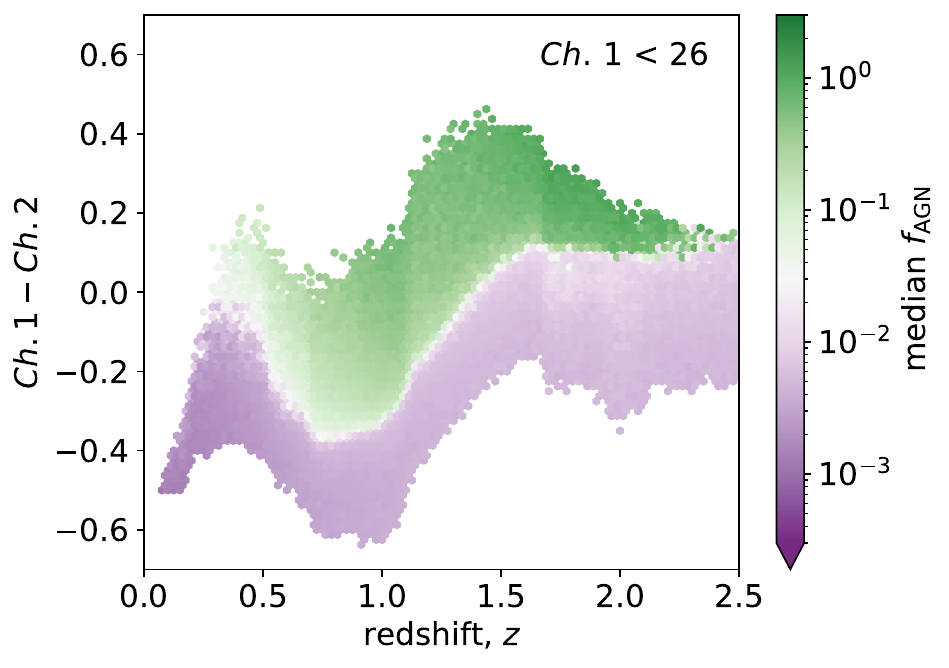}
    \caption{Same as row 5 of Figure \ref{fig:color_redshift}, but showing observer-frame IRAC $\textit{Ch.\,1}-\textit{Ch.\,2}$ color vs.\ redshift.}
    \label{fig:color_redshift_agn}
\end{figure}

\begin{figure}
    \centering
    \includegraphics[width=\linewidth]{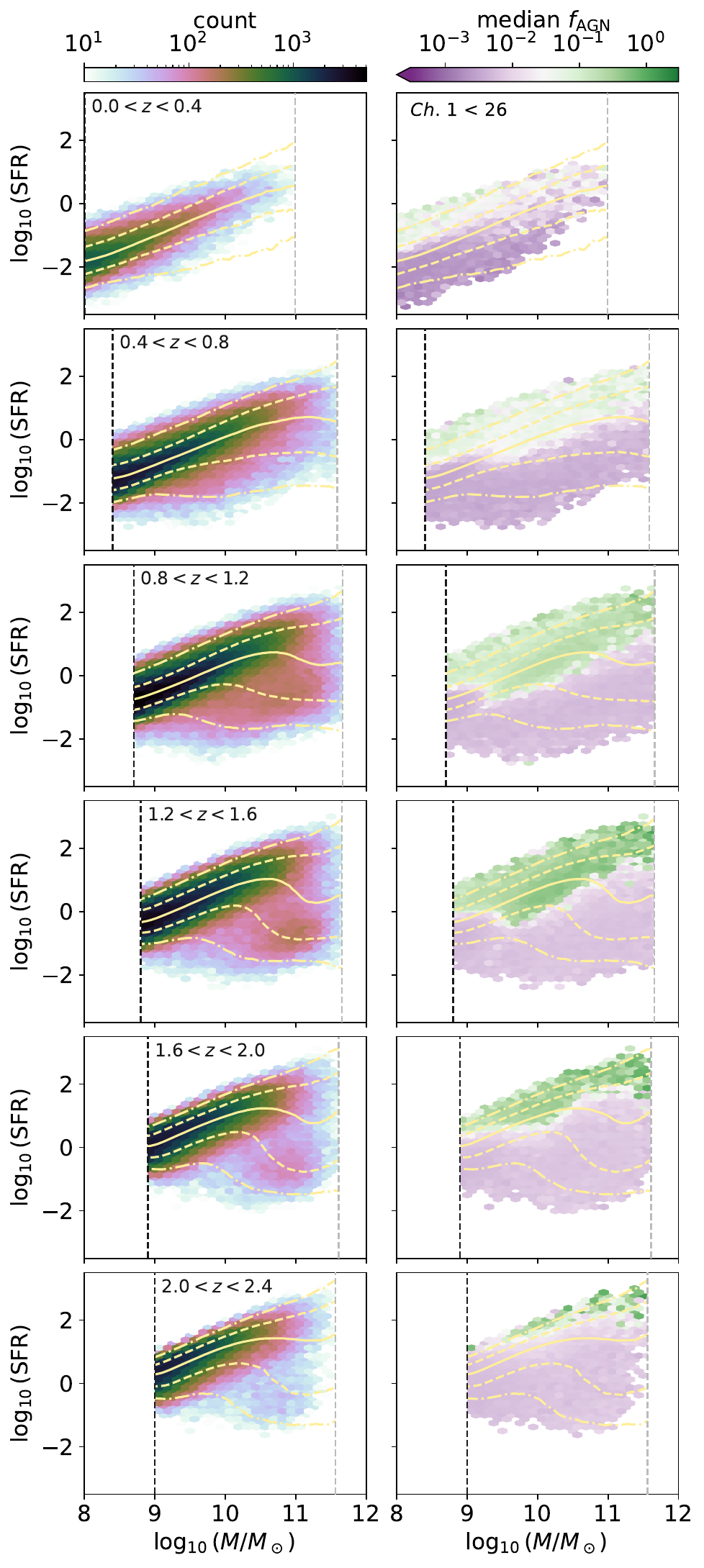}
    \caption{Star formation rate vs.\ mass binned by redshift and shaded by \textbf{(left)} model galaxy count, and \textbf{(right)} median AGN bolometric luminosity fraction, $f_\text{AGN}$. Cells are shaded only if they contain $>10$ model galaxies. The black and gray dashed lines show respectively the mass completeness limit and upper mass limit. Solid yellow lines shows the median from Figure \ref{fig:sfs}, and dashed (dash-dotted) yellow lines show 68 (95)\% intervals.}
    \label{fig:sfs_agn}
\end{figure}

In Figure \ref{fig:sfs_agn}, we show SFR vs.\ stellar mass, binned by redshift and shaded based on $f_\text{AGN}$. As with Figure \ref{fig:mass_redshift_agn}, we see the most pronounced AGN contributions around $1\lesssim z\lesssim 2$, where the upper edge of the star forming main sequence correlates with a high median $f_\text{AGN}$; i.e.\ the model predicts that galaxies with the highest sSFR have the strongest AGN-related emission in the IR. A non-stellar IR excess towards the upper edge of the star forming sequence was previously identified by \citet{lange16} for $0.5\leq z\leq1.5$, with a very similar tendency to the behavior we see in Figure \ref{fig:sfs_agn}. Observational works have showed that heavily-obscured IR-luminous AGN are ubiquitous within the galaxy population \citep[e.g.,][]{juneau13, kirkpatrick12, kirkpatrick15}, and make up a substantial fraction of all AGN \citep[e.g.,][]{lacy13, assef13, assef15, assef18, hickox18, carroll21}, making them a plausible source of such a non-stellar excess \citep{leja18}. The IR signature of AGN tends to be reasonably distinguishable from that of star formation, due to the former's tendency to produce a power-law SED (identifiable at $z\lesssim2$ by very red $\textit{Ch.\,1}-\textit{Ch.\,2}$ colors; see, e.g., \citealp{stern05, stern12, assef10, donley12, juneau13}). In Figure \ref{fig:color_redshift_agn}, we show the \texttt{pop-cosmos} observer-frame $\textit{Ch.\,1}-\textit{Ch.\,2}$ vs.\ $z$ relation, shaded by median $f_\text{AGN}$, highlighting that this color is a strong discriminant of AGN activity for $z\lesssim2$ \citep[c.f.,][figure 2]{stern05}. For $z\gtrsim2$, the AGN IR power law is shifted too far redward in the observer frame for this color to be a strong diagnostic.

The redshift window in Figure \ref{fig:mass_redshift_agn} where we identify the strongest MIR signature correlates roughly with the redshift range where the cosmic SFR density and the black hole accretion rate density peak and begin to turn over \citep[e.g.,][]{hopkins07, aird10, delvecchio14, madau14, hickox18}. In a unified picture of AGN--host co-evolution \citep[see, e.g.,][]{heckman14, padovani17, harrison17, alexander25}, the presence of a gas reservoir would be favorable for both star formation and rapid black hole growth via radiatively efficient accretion modes, potentially correlating the two. Moreover, it is theoretically expected that the interplay between stellar (i.e. supernova) feedback and AGN feedback conspires to set a characteristic mass scale where non-linear black hole growth via rapid, efficient accretion will activate \citep[see, e.g.][]{dubois15, bower17, mcalpine17, mcalpine18}; this stellar mass scale of $\sim10^{10.3}M_\odot$ aligns with our Figure \ref{fig:mass_redshift_agn}. For galaxies much below this characteristic mass, stellar feedback is expected to dominate gas dynamics and prevent radiatively efficient accretion. For galaxies much above this mass, AGN feedback is sufficient to remove gas from the halo, regulating the accretion onto the black hole and shutting down star formation \citep[see][]{bower17, lucie25}.

Obscured or IR-selected AGN have typically been found to correlate more strongly with SFR than other AGN types \citep[e.g.,][]{hickox09, juneau13, azadi17}, with this population lying typically on or above the star-forming main sequence \citep[e.g.,][]{juneau13, ellison16, chang17, hatcher21, cunha25}; this aligns well with the trend we see in Figure \ref{fig:sfs_agn}. For X-ray AGN (which only have partial overlap with IR-detected sources; see, e.g., \citealp{donley07, donley10, donley12, eckart10, lamassa19, carroll21}), a wider diversity of hosts is observed \citep[e.g.,][]{hickox09, aird12, mullaney15, azadi15, azadi17, suh19}, but some works have found a connection with enhanced SFR \citep[e.g.,][]{florez20, florez21, mountrichas24}, and many X-ray AGN lie on the main sequence \citep[e.g.,][]{kartaltepe12, mullaney12_goods, suh19}. Whilst some theoretical \citep[e.g.,][]{hopkins06} and observational \citep[e.g.,][]{kartaltepe12, ellison16, weston17, goulding18, ellison19} studies have made a connection between galaxy interactions or mergers as driving AGN and star formation (for a review see \citealp{hickox18}), others have suggested that the correlation is driven by secular processes linked to a common gas reservoir \citep{mullaney12, mullaney12_goods, heckman14, chang17, dai18}.

With future work, we will be able to revisit the correlations seen in Figures \ref{fig:mass_redshift_agn} and \ref{fig:sfs_agn} in further detail. As highlighted in Appendix \ref{sec:agn}, it will be possible to take advantages of recent improvements to \texttt{FSPS} \citep{wang24_agn} that refine the joint AGN--galaxy SED model by incorporating UV--optical and broad-line emission from the accretion disk \citep[via, e.g., the model of][]{temple21}\footnote{Recent studies of blue dust obscured galaxies at intermediate redshifts (``BlueDOGs''; e.g., \citealp{assef16, assef20, assef22, noboriguchi19, noboriguchi22, kim25}), and ``little red dots'' at high redshifts (see e.g., \citealp{barro24, baggen24, setton24, matthee24, wang24_agn, ma25, kocevski25}) have highlighted the challenges and importance of detailed modeling to separate AGN and SFR signatures (see extensive discussion in \citealp{kim25}).}. Moreover, by folding in data from further in the IR (e.g., the \textit{JWST F770W} photometry in COSMOS2025; \citealp{shuntov25}), it may be possible to probe the rest-frame IR and to trace the relationship between SFR and AGN activity at higher-$z$.

\section{Discussion and Conclusions}
\label{sec:discussion}
In this work we have demonstrated that an SPS-based forward model, trained on NUV--MIR photometry from COSMOS2020 with \textit{Spitzer} IRAC $\textit{Ch.\,1}<26$, is able to faithfully describe the redshift-evolving galaxy population over $\sim12.8$~Gyr of cosmic time. To enable population fitting using such deep data, we have developed a new, more flexible, noise modeling approach that uses a conditional diffusion model to predict the photometric uncertainties of an object given its brightness. This approach can accurately represent the complicated, multimodal uncertainty distributions that arise from variable depths in the COSMOS field. With the trained population model, we are able to rapidly synthesize large mock galaxy catalogs ($\sim1,000,000\,\text{galaxies}/\text{GPU-hr}$) with realistic multi-band photometry and physical properties. With an efficient SPS emulator, and hardware accelerated MCMC, we are able to perform rapid parameter inference (with a throughput of $<10\,\text{GPU-sec}/\text{galaxy}$ for full MCMC chains) using the trained model as an informative prior, achieving superior photometric redshift accuracy for $0<z<6$. 

We validate the learned generative model by comparing it to well-studied scaling relations within galaxy evolution, and explore correlations that are normally extremely difficult to access from photometry alone. Where literature results are available for comparison, we find generally very good agreement. Our stellar mass function agrees well with the expectation from other recent results \citep[e.g.][]{leja20, weaver23_smf, zalesky25}, albeit with a tail (comprising $\sim0.5\%$ of the model galaxy population) at very high stellar masses $\gtrsim10^{11}M_\odot$. There may be reasons to investigate the extrapolation of the model in this regime further. For instance, recent surveys have made tentative detections of such massive galaxies even at $z\gtrsim5$. \citet{xiao24} report three candidate galaxies with masses in excess of $10^{11}M_\odot$ at $5<z<6$ in the $124\,\text{arcmin}^2$ FRESCO survey field \citep{oesch23}, one of which has been confirmed and studied extensively \citep[GN10; see e.g.,][]{riechers20, sun24, herard25, lagache25}. This number is consistent with the expectation from re-scaling a COSMOS-sized mock galaxy catalog from \texttt{pop-cosmos} to a FRESCO-sized field. Our learned star-forming main sequence agrees well with the deep photometric study by \citet{leja22}, with a similar normalization (i.e., relatively low median SFRs compared to earlier work based on parameteric SFH models). Other galaxy evolution trends we explore in depth include metallicity scaling relations, dust attenuation properties, and the correlation between AGN activity and SFR.

Taken together, the high fidelity representation of astrophysics and the computational efficiency of the model gives us the ability to model the quantities needed for Stage IV cosmology, with the required level of realism, and at the required throughput. Because the \texttt{pop-cosmos} population model is learned as a distribution over SPS parameters, it is possible to deploy it as a generative model for \emph{any} other survey which has a depth compatible with the COSMOS2020 training sample. In upcoming work, we will demonstrate this with two applications of the model to the KiDS survey \citep{kuijken19}: B.\ Leistedt et al. (in prep.) will show the use of the trained model's predictions to determine tomographic redshift distributions, whilst A.\ Halder et al. (in prep.) will show the use of the model as a prior for inferring KiDS galaxy properties and redshifts at scale.

Complementary approaches are being developed in parallel to meet related goals. \citet{hahn23, hahn24} and \citet{li24} have presented a forward modeling approach based on emulated SPS optimized for handling bright, low-redshift galaxies with spectro-photometry from the Dark Energy Spectroscopic Instrument (DESI; \citealp{hahn23_bgs, adame24}). Large-scale spectroscopic data, such as that provided at low-$z$ by DESI, or that expected in future at high-$z$ from programs such as MOONRISE \citep{maiolino20} on the Very Large Telescope's Multi-Object Optical and NIR Spectrograph (MOONS; \citealp{cirasuolo11, cirasuolo20}), will be important for future extensions to the \texttt{pop-cosmos} model. Such data will offer the possibility of further refining our modeling of emission lines \citep[see, e.g.,][]{khederlarian24} and gas physics.

In future work, we plan to incorporate the galaxy--halo connection into \texttt{pop-cosmos} in order to further improve the utility of the model within cosmological analyses. Conditional abundance matching \citep[for a review see][]{wechsler18} offers one of the most attractive paths to achieving this. Recent work has shown that sub-halo abundance matching can be expressed elegantly as the solution to an optimal transport problem \citep{fischbacher25}. Empirical models such as \texttt{UniverseMachine} \citep{behroozi19} have demonstrated the predictive power that can be gained from modeling the galaxy--halo connection. The \texttt{Diffsky} forward model\footnote{\url{https://github.com/ArgonneCPAC/diffsky}} builds on this foundational work, linking halo mass assembly \citep{hearin21, hearin22} with galaxies' SFHs \citep{alarcon23} and SPS within a differentiable framework \citep{hearin23}.

Modeling galaxy sizes and morphologies simultaneously with their other physical parameters is a further extension to \texttt{pop-cosmos} that we plan to develop. We will take a featurized approach to this, rather than carrying out full image simulation. The \texttt{GalSBI} forward modeling framework \citep{herbel17, moser24, fishbacher24, fischbacher24_py} is a complementary effort that connects a fast simulator of full images \citep{berge13, fishbacher24_ufig} to a parametric model for the galaxy population, and a template-based representation of galaxy SEDs. Recently, \citet{tortorelli25} demonstrated an SPS-based version of this model using an empirically calibrated population model.

In addition to the directions highlighted above, in future iterations of the \texttt{pop-cosmos} model, we will revisit our SPS modeling choices to take advantage of the newest advances made in this space (some of the most promising avenues are highlighted in depth in Appendix \ref{sec:discussion_sps}). Improvements such as incorporating $\alpha$-enhancement \citep[e.g.,][]{byrne22, byrne25, park24}, emulated photoionization calculations \citep[e.g.,][]{li24_cue, morisset25}, binary stellar evolution pathways (for a review see \citealp{eldridge19}), and re-calibrated IGM attenuation \citep[e.g.,][]{inoue14}, will be important for meeting the challenges of large spectroscopic datasets \citep[e.g.,][]{maiolino20, adame24}, or the redshift frontier being probed by the deepest \textit{JWST} surveys (e.g., \citealp{shuntov25}; for a review see, e.g., \citealp{robertson22, gardner23, merlin24}).

\section*{Data Availability}
\noindent The COSMOS catalog \citep{weaver22} is publicly available at the\dataset[ESO Archive]{https://doi.org/10.18727/archive/52} \citep{dunlop16} and the\dataset[COSMOS2020 webpage]{https://cosmos2020.calet.org}. We have made COSMOS-like mock galaxy catalogs with two million sources drawn from our trained model available on\dataset[Zenodo]{https://doi.org/10.5281/zenodo.15622324}. All plots in this paper are produced using v1.0.1 \citep{mocks_v1} of these catalogs with\dataset[doi:10.5281/zenodo.15757717]{https://doi.org/10.5281/zenodo.15757717}. We have also released on\dataset[Zenodo]{https://doi.org/10.5281/zenodo.13627488} MCMC chains for 429,669 COSMOS2020 galaxies based on the results in Section~\ref{sec:results_mcmc_redshift}--\ref{sec:results_mcmc_mass}. The results in this paper are based on v2.1.1 \citep{mcmc_v2} of this catalog with \dataset[doi:10.5281/zenodo.15623082]{https://doi.org/10.5281/zenodo.15623082}. These results supersede the results of \citet{thorp24}, contained in v1.3.0 \citep{mcmc_v1} with\dataset[doi:10.5281/zenodo.14832248]{https://doi.org/10.5281/zenodo.14832248}. Software for working with the \texttt{pop-cosmos} model is available via the \texttt{Cosmo-Pop} organization on GitHub\footnote{\url{https://github.com/Cosmo-Pop}}.

\begin{contribution}
We outline the different contributions below using keywords based on the Contribution Roles Taxonomy (CRediT; \citealp{brand15}). %
\textbf{ST:} 
conceptualization; 
methodology; 
software; 
validation; 
investigation; 
data curation; 
writing -- original draft.
\textbf{HVP:} Conceptualization; 
methodology; 
investigation; 
visualization; 
validation; 
supervision; 
project administration;
funding acquisition;
writing -- review and editing. 
\textbf{GJ:} 
methodology; 
software; 
validation.
\textbf{SD:} 
methodology; 
validation; 
investigation; 
writing -- review and editing.
\textbf{JA:}
conceptualization;
methodology;
software;
investigation.
\textbf{BL:} 
methodology; 
validation; 
investigation; 
writing --  review and editing. 
\textbf{DJM:} 
methodology; 
investigation; 
visualization;  
writing -- review and editing.
\textbf{AH:} 
validation.
\textbf{JL:} 
methodology; 
writing -- review and editing.
\end{contribution}

\begin{acknowledgments}
We thank Joel Johansson, Ben Johnson, Katherine Kauma, Jeff Newman, Francesco Petri, Sandro Tacchella, Madalina Tudorache, Benedict van den Bussche, and Vivienne Wild for useful discussions. We are also grateful to Ben Johnson and the developers of \texttt{fsps}, \texttt{Prospector}, \texttt{python-fsps}, and \texttt{sedpy} for making these tools available for the community, and for responding to our queries about them. We also thank John Weaver for helpful correspondence about the COSMOS2020 catalog, and all members of the COSMOS team for publishing this outstanding dataset. We thank Mikica Kocic for computing support. We thank the referee and ApJ data editor for constructive and helpful comments that have improved the manuscript.

ST, HVP, SD and JA have been supported by funding from the European Research Council (ERC) under the European Union's Horizon 2020 research and innovation programmes (grant agreement no.\ 101018897 CosmicExplorer). This work has been enabled by support from the research project grant ‘Understanding the Dynamic Universe’ funded by the Knut and Alice Wallenberg Foundation under Dnr KAW 2018.0067. HVP was additionally supported by the G\"{o}ran Gustafsson Foundation for Research in Natural Sciences and Medicine. BL was supported by the Royal Society through a University Research Fellowship. 
\end{acknowledgments}

%


\facilities{This study utilizes observations collected at the European Southern Observatory under ESO programme ID 179.A-2005 and 198.A-2003 and on data products produced by CALET and the Cambridge Astronomy Survey Unit on behalf of the UltraVISTA consortium. This research utilized the Sunrise HPC facility supported by the Technical Division at the Department of Physics, Stockholm University. This work was performed using resources provided by the Cambridge Service for Data Driven Discovery (CSD3) operated by the University of Cambridge Research Computing Service (\url{www.csd3.cam.ac.uk}), provided by Dell EMC and Intel using Tier-2 funding from the Engineering and Physical Sciences Research Council (capital grant EP/T022159/1), and DiRAC funding from the Science and Technology Facilities Council (\url{www.dirac.ac.uk}).}

\software{\texttt{NumPy} \citep{harris20}; \texttt{SciPy} \citep{virtanen20}; \texttt{Matplotlib} \citep{hunter07}; \texttt{astropy} \citep{astropy13, astropy18, astropy22}; \texttt{corner} \citep{dfm16}; \texttt{PyTorch} \citep{paszke19}; \texttt{Speculator} \citep{alsing20}; \texttt{torchdiffeq} \citep{chen18}; \texttt{Prospector} \citep{johnson21}; \texttt{FSPS v3.2} \citep{conroy09, conroy10a, conroy10b}; \texttt{python-fsps v0.4.1} \citep{pythonfsps}; \texttt{sedpy} \citep{sedpy}; \texttt{tqdm} \citep{tqdm}; \texttt{wandb} \citep{wandb}; \texttt{RayTune} \citep{ray,tune}; \texttt{quantile\_utilities} \citep{quantile_utilities, thorp25}; \texttt{affine}\footnote{\url{https://github.com/justinalsing/affine}}; \texttt{flowfusion v0.1}\footnote{\url{https://github.com/Cosmo-Pop/flowfusion}}; \texttt{pop-cosmos v0.1}\footnote{\url{https://github.com/Cosmo-Pop/pop-cosmos}} (\citealp{alsing24, thorp24}; this work); \texttt{hist\_contour}\footnote{\url{https://github.com/stevet40/hist_contour}}; \texttt{ffjord-lite}\footnote{\url{https://github.com/jackgoffinet/ffjord-lite}}. We have used color schemes by \citet{tol21} and \citet{green11}.}
%

%
\appendix
\begin{table}
    \centering
    \caption{Band-dependent calibration parameters.}
    \label{tab:band_calibration}
    \begin{tabular}{l c c r}
        \toprule\toprule
        Band & $f_b$ (nMgy) & Eff.\ Depth (mag) & $\alpha_\text{ZP}$ \\ \midrule
         $u$ & 0.001080 & 29.916840 & 1.009450\\
        $g$ & 0.003085 & 28.776844 & 1.085712\\
        $r$ & 0.005081 & 28.235106 & 1.075836\\
        $i$ & 0.000879 & 30.139624 & 1.019943\\
        $z$ & 0.002913 & 28.839199 & 1.023337\\
        $y$ & 0.004059 & 28.478890 & 1.048863\\
        $Y$ & 0.002624 & 28.952726 & 0.999337\\
        $J$ & 0.004989 & 28.254908 & 1.004181\\
        $H$ & 0.003334 & 28.692439 & 0.970161\\
        $K_S$ & 0.001146 & 29.851584 & 1.042268\\
        \textit{Ch.\,1} & 0.003229 & 28.727212 & 0.960127\\
        \textit{Ch.\,2} & 0.003452 & 28.654855 & 0.921076\\
        IB427 & 0.005263 & 28.196843 & 0.976578\\
        IB464 & 0.003999 & 28.495201 & 0.972594\\
        IA484 & 0.006620 & 27.947804 & 1.020441\\
        IB505 & 0.004718 & 28.315492 & 1.008716\\
        IA527 & 0.006163 & 28.025453 & 0.994083\\
        IB574 & 0.001148 & 29.850477 & 0.929141\\
        IA624 & 0.004297 & 28.417198 & 1.012976\\
        IA679 & 0.006221 & 28.015279 & 1.131825\\
        IB709 & 0.001822 & 29.348750 & 0.976940\\
        IA738 & 0.004118 & 28.463242 & 0.969744\\
        IA767 & 0.008231 & 27.711334 & 0.965354\\
        IB827 & 0.008909 & 27.625397 & 0.920645\\
        NB711 & 0.010834 & 27.413028 & 0.987654\\
        NB816 & 0.007929 & 27.752018 & 0.929771\\
        \bottomrule
    \end{tabular}
\end{table}

\begin{table}
    \centering
    \caption{Emission line-dependent calibration parameters.}
    \label{tab:line_calibration}
    \begin{tabular}{l c c c}
        \toprule\toprule
        Line & $\lambda_\text{EM}$ (\AA) & $\beta_\text{EM}$ & $\gamma_\text{EM}$ \\ \midrule
         C \textsc{ii}] 2326 & 2326.11 & $1.336\times 10^{-3}$ & $1.425\times 10^{-13}$\\\relax 
        [O \textsc{iii}] 2321 & 2321.66 & $2.927\times 10^{-3}$ & $1.424\times 10^{-13}$\\\relax 
        [O \textsc{i}] 6302 & 6302.05 & $8.403\times 10^{-3}$ & $1.464\times 10^{-13}$\\\relax 
        [S \textsc{ii}] 4070 & 4069.75 & $-8.081\times 10^{-3}$ & $1.433\times 10^{-13}$\\\relax 
        H \textsc{i} (Ly-$\alpha$) & 1215.67 & $6.178\times 10^{-3}$ & $1.455\times 10^{-13}$\\\relax 
        [Al \textsc{ii}] 2670 & 2669.95 & $2.853\times 10^{-3}$ & $1.425\times 10^{-13}$\\\relax 
        [Ar \textsc{iii}] 7753 & 7753.19 & $-1.335\times 10^{-2}$ & $1.427\times 10^{-13}$\\\relax 
        H \textsc{i} (Pa-7) & 9017.80 & $-1.219\times 10^{-2}$ & $1.425\times 10^{-13}$\\\relax 
        [Al \textsc{ii}] 2660 & 2661.15 & $2.905\times 10^{-3}$ & $1.425\times 10^{-13}$\\\relax 
        [S \textsc{iii}] 6314 & 6313.81 & $4.801\times 10^{-3}$ & $1.426\times 10^{-13}$\\\relax 
        H \textsc{i} (Pa-6) & 9232.20 & $-1.025\times 10^{-2}$ & $1.426\times 10^{-13}$\\\relax 
        [S \textsc{iii}] 3723 & 3722.75 & $-4.322\times 10^{-3}$ & $1.425\times 10^{-13}$\\\relax 
        Mg \textsc{ii} 2800 & 2803.53 & $-6.834\times 10^{-3}$ & $1.434\times 10^{-13}$\\\relax 
        H \textsc{i} (Pa-5) & 9548.80 & $-8.112\times 10^{-3}$ & $1.427\times 10^{-13}$\\\relax 
        He \textsc{i} 7065 & 7067.14 & $2.562\times 10^{-3}$ & $1.427\times 10^{-13}$\\\relax 
        [N \textsc{ii}] 6549 & 6549.86 & $1.167\times 10^{-2}$ & $2.163\times 10^{-13}$\\\relax 
        [S \textsc{ii}] 6732 & 6732.67 & $9.945\times 10^{-3}$ & $3.600\times 10^{-13}$\\\relax 
        C \textsc{iii}] & 1908.73 & $1.850\times 10^{-3}$ & $1.424\times 10^{-13}$\\\relax 
        He \textsc{i} 6680 & 6679.99 & $9.079\times 10^{-3}$ & $1.437\times 10^{-13}$\\\relax 
        Mg \textsc{ii} 2800 & 2796.35 & $-5.579\times 10^{-3}$ & $1.460\times 10^{-13}$\\\relax 
        [S \textsc{ii}] 6717 & 6718.29 & $8.977\times 10^{-3}$ & $1.028\times 10^{-13}$\\\relax 
        [Ar \textsc{iii}] 7138 & 7137.77 & $2.573\times 10^{-3}$ & $1.490\times 10^{-13}$\\\relax 
        [C \textsc{iii}] & 1906.68 & $1.905\times 10^{-3}$ & $1.425\times 10^{-13}$\\\relax 
        He \textsc{i} 4472 & 4472.73 & $4.407\times 10^{-3}$ & $1.437\times 10^{-13}$\\\relax 
        [O \textsc{iii}] 4364 & 4364.44 & $2.081\times 10^{-3}$ & $1.425\times 10^{-13}$\\\relax 
        [N \textsc{ii}] 6585 & 6585.27 & $-5.913\times 10^{-1}$ & $1.000\times 10^{-13}$\\\relax 
        [S \textsc{iii}] 9071 & 9071.10 & $-1.011$ & $1.702\times 10^{-13}$\\\relax 
        H-8 3798 & 3798.99 & $-5.927\times 10^{-4}$ & $1.465\times 10^{-13}$\\\relax 
        He \textsc{i} 3889 & 3889.75 & $-6.603\times 10^{-3}$ & $1.500\times 10^{-13}$\\\relax 
        H-7 3835 & 3836.49 & $-1.916\times 10^{-3}$ & $1.485\times 10^{-13}$\\\relax 
        [Ne \textsc{iii}] 3968 & 3968.59 & $-1.005\times 10^{-2}$ & $1.448\times 10^{-13}$\\\relax 
        He \textsc{i} 5877 & 5877.25 & $3.240\times 10^{-1}$ & $1.555\times 10^{-13}$\\\relax 
        H-6 3889 & 3890.17 & $-7.012\times 10^{-3}$ & $1.536\times 10^{-13}$\\\relax 
        [S \textsc{iii}] 9533 & 9533.20 & $-1.008$ & $2.017\times 10^{-13}$\\\relax 
        H-5 3970 & 3971.20 & $-1.504\times 10^{-1}$ & $1.689\times 10^{-13}$\\\relax 
        [O \textsc{ii}] 3726 & 3727.10 & $4.595\times 10^{-3}$ & $1.000\times 10^{-13}$\\\relax 
        H-$\delta$ 4102 & 4102.89 & $-5.235\times 10^{-1}$ & $1.912\times 10^{-13}$\\\relax 
        [O \textsc{ii}] 3729 & 3729.86 & $2.642\times 10^{-1}$ & $1.000\times 10^{-13}$\\\relax 
        [Ne \textsc{iii}] 3870 & 3869.86 & $-8.795\times 10^{-2}$ & $1.889\times 10^{-13}$\\\relax 
        H-$\gamma$ 4340 & 4341.69 & $-3.297\times 10^{-1}$ & $1.013\times 10^{-13}$\\\relax 
        [O \textsc{iii}] 4960 & 4960.30 & $2.037\times 10^{-4}$ & $1.000\times 10^{-13}$\\\relax 
        H-$\beta$ 4861 & 4862.71 & $-5.577\times 10^{-1}$ & $1.000\times 10^{-13}$\\\relax 
        H-$\alpha$ 6563 & 6564.60 & $-3.314\times 10^{-1}$ & $1.000\times 10^{-13}$\\\relax 
        [O \textsc{iii}] 5007 & 5008.24 & $1.160\times 10^{-1}$ & $2.642\times 10^{-2}$ \\ \bottomrule
    \end{tabular}
\end{table}

\section{Calibration Parameters}
\label{sec:calibration}
Our model has a large number of calibration parameters which need to be included to ensure the galaxy parameters of interested are not biased. Table \ref{tab:band_calibration} lists the set of band-dependent parameters $(f_b, \alpha_\text{ZP})$, and Table \ref{tab:line_calibration} lists the emission line-dependent parameters $(\beta_\text{EM}, \gamma_\text{EM})$. The tabulated values of $\alpha_\text{ZP}$, $\beta_\text{EM}$, and $\gamma_\text{EM}$ are optimized during the population model training process described in Section~\ref{sec:training}. The values of $f_b$ are fixed and determined using the method outlined in Section~\ref{sec:data}. In Table \ref{tab:band_calibration}, the $f_b$ values are quoted in nano-maggies (i.e.\ $10^9\times$ flux relative to a standard source; e.g., \citealp{stoughton02, abazaijan03}), assuming an AB standard source with a flux of 3631~Jy. The effective depths these correspond to are logarithmic AB magnitudes, i.e.\ $\text{eff.\ depth} = 22.5 - 2.5\log_{10}(f_b / \text{nMgy})$. The emission line strength corrections ($\beta_\text{EM}$) and standard deviations ($\gamma_\text{EM}$) in Table \ref{tab:line_calibration} are dimensionless fractional quantities.

\begin{figure}
    \centering
    \includegraphics[width=\linewidth]{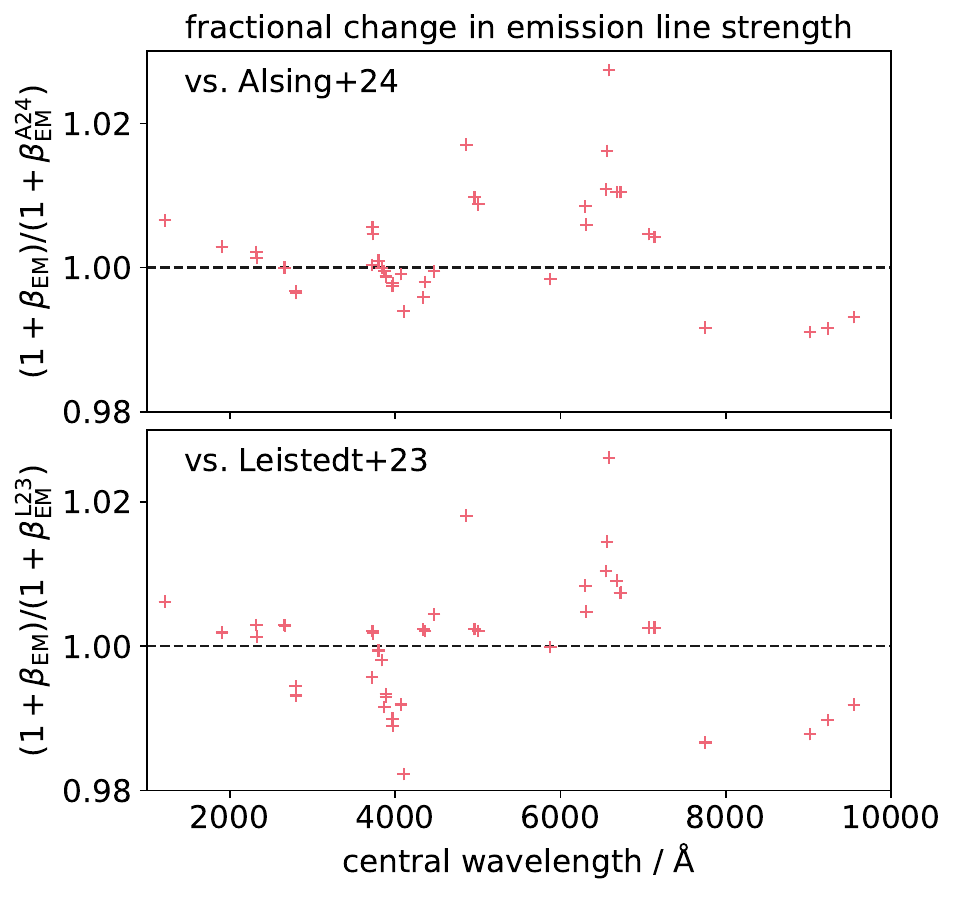}
    \caption{Effective change in predicted emission line strength for \texttt{pop-cosmos} relative to the line strengths inferred in \citetalias{leistedt23} and \citetalias{alsing24}.}
    \label{fig:line_strengths}

    \centering
    \includegraphics[width=\linewidth]{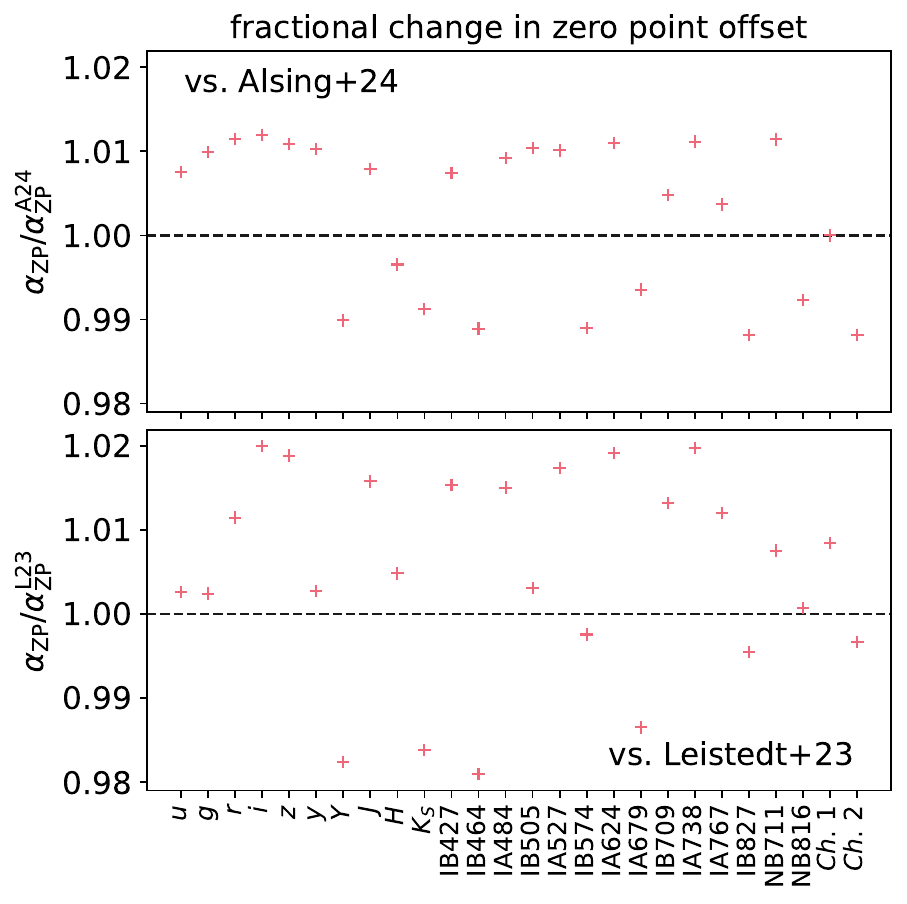}
    \caption{Effective change in photometric fluxes in the the 26 COSMOS bands, relative to the corrections inferred in \citetalias{leistedt23} and \citetalias{alsing24}.}
    \label{fig:zp_corrections}
\end{figure}

In Figure \ref{fig:line_strengths}, we show the effective change in emission line strength implied by the values of $\beta_\text{EM}$ in Table \ref{tab:line_calibration}, relative to the earlier calibrations by \citetalias{leistedt23} and \citetalias{alsing24}. The mean flux of any line will be given by $(1+\beta_\text{EM})\times F_\text{EM}(\bm{\vartheta})$, where $F_\text{EM}(\bm{\vartheta})$ is the uncorrected line strength from the \citet{byler17} \texttt{Cloudy} model grid in FSPS. The \texttt{pop-cosmos} line strengths from this work relative to \citetalias{leistedt23} will thus be given by $(1+\beta_\text{EM})/(1+\beta_\text{EM}^\text{L23})$, and similarly for this work relative to \citetalias{alsing24}. For the 44 lines in Table \ref{tab:line_calibration}, the effective change relative to \citetalias{leistedt23} is within $\pm3\%$. The pair of doubly-forbidden sulphur lines, [S \textsc{iii}] 9071 and [S \textsc{iii}] 9533, are omitted from Figure \ref{fig:line_strengths}, as these are essentially negated by their inferred values of $\beta_\text{EM}=-1$ (for \citetalias{leistedt23}, \citetalias{alsing24}, and this work).

Figure \ref{fig:zp_corrections} shows the effective change in photometric flux implied by our zero-point corrections, $\bm{\alpha}_\text{ZP}$, relative to the inferences from \citetalias{leistedt23} and \citetalias{alsing24}. For all bands, our corrections leave the fluxes within $\pm2\%$ of the \citetalias{leistedt23} and \citetalias{alsing24} results. The absolute values of the zero-point corrections we infer in Table \ref{tab:band_calibration} are typically significantly smaller than the offsets estimated by \citet{skelton14}, where our sets of passbands overlap, reflecting the progress that has been made in both data and modeling in the past decade.

\section{Ambiguous Cross-matches in the COSMOS Spectroscopic Archive}
\label{sec:ambiguous_spec}
From the unique objects in the \citet{khostovan25} spectroscopic archive, we find 39,702 that have a $\text{CL}\geq50\%$ redshift, and a COSMOS2020 \texttt{Farmer} ID that is within our $\textit{Ch.\,1}<26$ catalog. We find that these 39,588 sources correspond to 39,579 unique COSMOS2020 \texttt{Farmer} IDs. There are thus 9 of our selected \texttt{Farmer} IDs (30647, 58379, 117181, 159826, 189699, 257074, 656499, 942331, 948097) that have two spectra associated with them in \citet{khostovan25}. One of these (257074) has two spectra with distinct COSMOS2020 \texttt{Classic} IDs (576772, 577956) associated with them, but for the other eight there is one spectrum with a \texttt{Classic} source match and one with no \texttt{Classic} counterpart. The two spectra / \texttt{Classic} sources associated with \texttt{Farmer} ID 257074 have distinct spectroscopic redshifts (both from PRIMUS; \citealp{coil11}): $z=0.62133$ and $z=0.1717$, with $\text{CL}=80\%$ and $\text{CL}=50\%$ respectively. For \texttt{Farmer} ID 30647, the two spectroscopic redshifts are substantially different: $z=1.51598$ (from DESI EDR; \citealp{adame24}) and $z=5.621$ (from 10k--DEIMOS; \citealp{hasinger18}), both with $\text{CL}=80\%$. Only the former of these has an associated \texttt{Classic} ID in \citet{khostovan25}.

\section{Derived Parameters}
\label{sec:derived}
In this Appendix, we describe the computation of (specific) star formation rate and mass-weighted age given the base SPS parameters in the upper part of Table \ref{tab:sps_notation}. For a model galaxy at redshift $z$, we compute the age of the Universe at that redshift, $t_\text{Univ}(z)$, under a flat $\Lambda\text{CDM}$ cosmology with \citet{planck18} parameters. The SFH for a galaxy is piecewise constant and follows the sequence of 7 lookback time bins from \citet{leja19_sfh, leja19}. The first two bin edges are fixed at $t_{\text{edge},1}=0~$Myr, and $t_{\text{edge},2}=30$~Myr, with the last placed at $t_\text{edge}^8=t_\text{Univ}(z)$. The five inner edges are spaced logarithmically \citep{ocvirk06} between $100$~Myr and $0.85\times t_\text{Univ}(z)$, so the $i$th edge is given by 
\begin{equation}
    t_{\text{edge},i} = 100\,\text{Myr} + \frac{i-3}{4}\times[0.85\times t_\text{Univ}(z) - 100\,\text{Myr}]
\end{equation}
for $i=3,\dots,7$.

The SFR in bin $j$, relative to the first (most recent) bin can be computed from the SFR ratios as
\begin{equation}
    \log_{10}\left(\frac{\text{SFR}_j}{\text{SFR}_1}\right) = -\sum_{k=1}^{j-1}\Delta\log_{10}(\text{SFR})_k
\end{equation}
for $j=2,\dots7$. We can then define a normalized SFR for bin $j$,
\begin{equation}
    \begin{split}
        \frac{\text{SFR}_j}{M_{\text{form}}} = \frac{\text{SFR}_j/\text{SFR}_1}{\Delta t_1 + \sum_{k=2}^7 (\Delta t_k\times \text{SFR}_k/\text{SFR}_1)} \\
        = \frac{\text{SFR}_j}{\sum_{k=1}^7 (\Delta t_k\times \text{SFR}_k)},
    \end{split}
    \label{eq:sfr_norm}
\end{equation}
where $\Delta t_j = t_{\text{edge},j+1} - t_{\text{edge},j}$ is the width of the $j$th bin, and $M_\text{form}$ is the total mass the galaxy will form by redshift $z$. The fraction of stellar mass formed in bin $j$ is then simply given by $M_{\text{form},j}/M_\text{form}=\Delta t_j\times \text{SFR}_j/M_{\text{form}}$.

From this, the SFR per unit mass formed, averaged over the most recent 100~Myr of a galaxy's life can be computed by summing the fractional mass formed over the first two bins of the SFH and dividing by 100~Myr:
\begin{equation}
    \frac{\text{SFR}}{M_\text{form}} = \left(\frac{M_{\text{form},1}}{M_\text{form}} + \frac{M_{\text{form},2}}{M_\text{form}}\right)/100\,\text{Myr}.
    \label{eq:ssfr_formed}
\end{equation}
We compute the SFR by multiplying Equation \ref{eq:ssfr_formed} by $M_\text{form}$. The sSFR values we quote throughout the paper follow the conventional definition of SFR per unit mass remaining. This is found by dividing Equation \ref{eq:ssfr_formed} by the fraction of stellar mass remaining, $M/M_\text{form}$, where this quantity is emulated as described in Section~\ref{sec:sps}.

The normalized SFR defined in Equation \ref{eq:sfr_norm} sets the fractional contribution to a galaxy's total stellar content of the stars formed at any given instant. In effect, it gives the weighting of all possible stellar ages within the galaxy's SFH. As the SFR is piecewise constant, taking the weighted mean of stellar age across the full SFH can be done by summing the weighted mean in each bin:
\begin{equation}
    \begin{split}
        t_\text{age} &= \sum_{j=1}^7 \left[\int_{t_{\text{edge},j}}^{t_{\text{edge},j+1}} \frac{\text{SFR}_j}{M_{\text{form}}}\times t\,\mathrm{d}t\right]\\
        &=\sum_{j=1}^7 \left[\frac{1}{2}(t_{\text{edge},j+1}^2 - t_{\text{edge},j}^2) \times \frac{\text{SFR}_j}{M_{\text{form}}}\right].
    \end{split}
\end{equation}

\begin{figure}
    \centering
    \includegraphics[width=\linewidth]{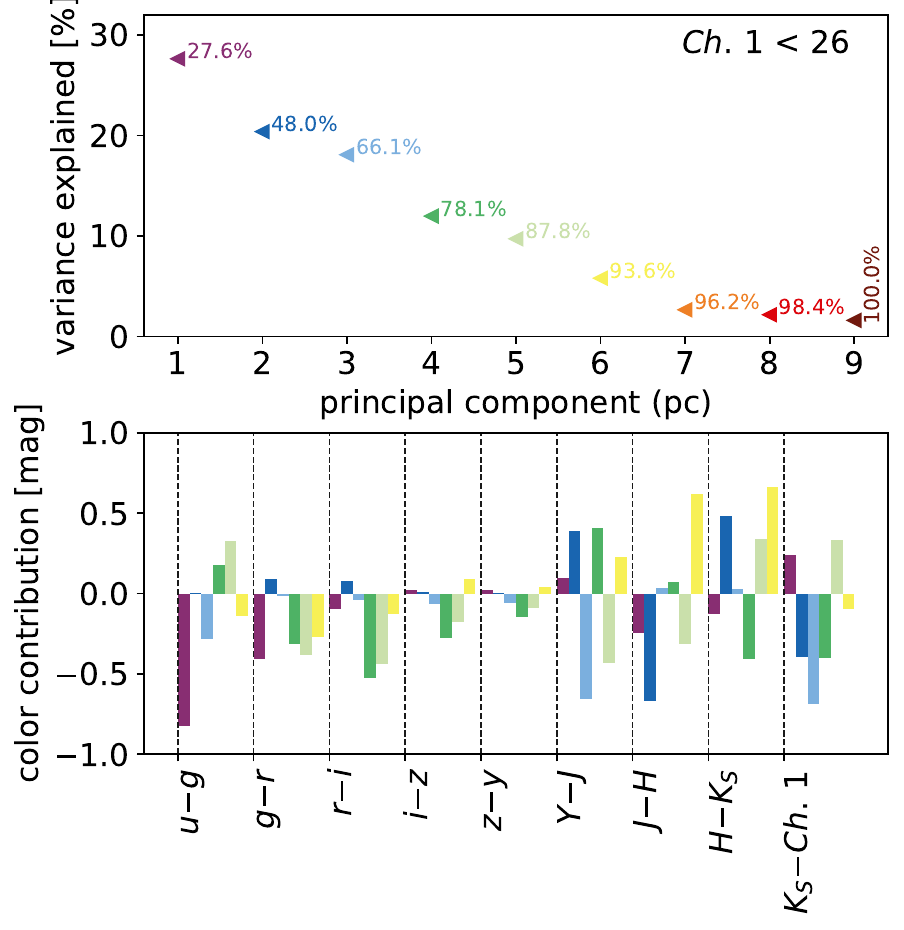}
    \caption{Principal components (PCs) of the broadband colors of the COSMOS2020 catalog with $\textit{Ch.\ 1}<26$, showing \textbf{(upper panel)} the eigenvalues, and \textbf{(lower panel)} the principal eigenvectors. The vertical axis in the upper panel shows the fraction of variance explained by each PC, with the annotation showing the cumulative variance explained by each PC and those leftward of it in the plot. The lower panel shows the contribution of each color to the first 6 PCs, which cumulatively explain $>90\%$ of the variance in COSMOS2020.}
    \label{fig:PCA}
\end{figure}

\begin{figure}
    \centering
    \includegraphics[width=0.95\linewidth]{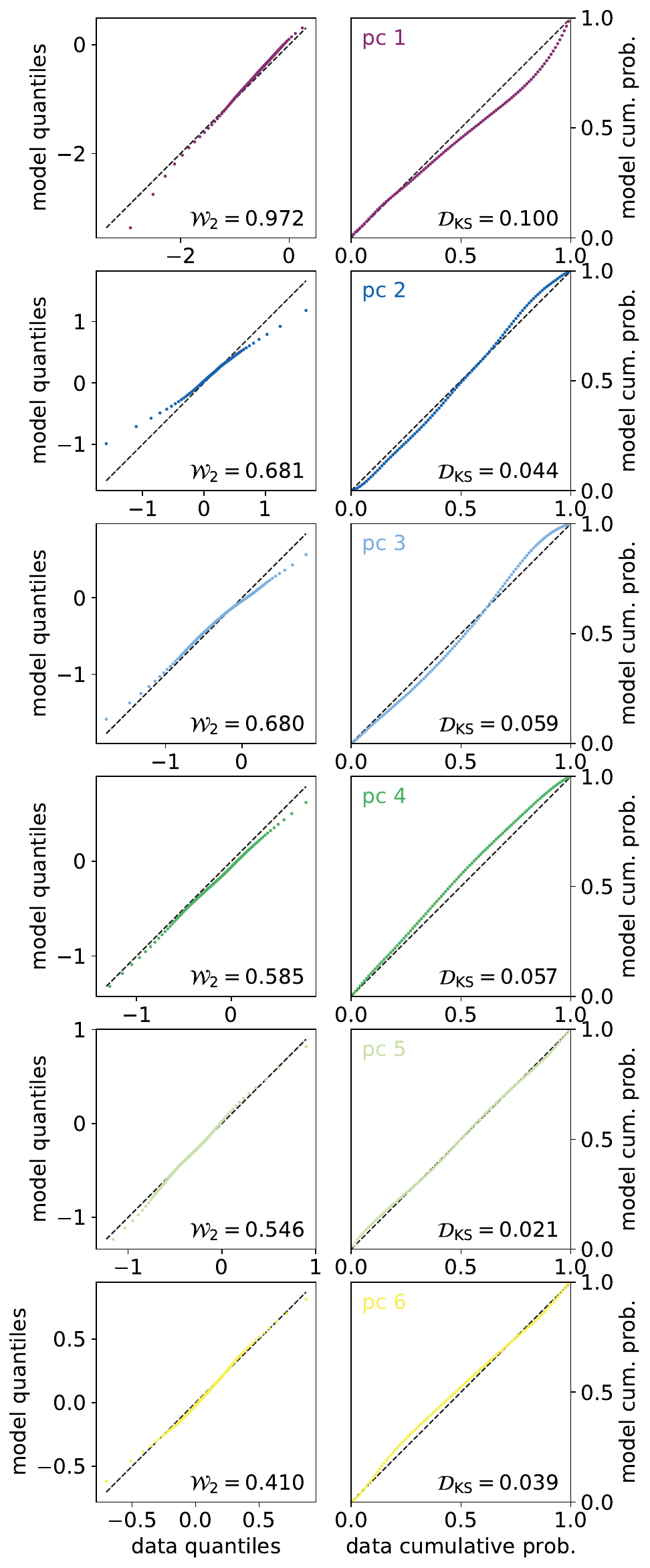}
    \caption{Quantile-based comparison --- \textbf{(left)} Q--Q plots, and \textbf{(right)} P--P plots --- of the model colors and COSMOS colors projected along the first six principal axes (93\% of total variance) identified in Figure \ref{fig:PCA}. On the lower right corners of the Q--Q plots, we show the marginal 2-Wasserstein distance $\mathcal{W}_2$; on the lower right corner of the P--P plots we show the K--S statistic $\mathcal{D}_\text{KS}$.}
    \label{fig:QQ}
\end{figure}

\section{Further Photometric Validation}
\label{sec:photo_validation}
In Figure \ref{fig:PCA}, we follow \citet{thorp25} in performing a principal component analysis (PCA) of the 9 broadband colors defined by adjacent passbands. Shown are the principal components (PCs) of the COSMOS2020 catalog itself. The upper panel shows a scree plot \citep{cattell66} ranking the PCs by the fraction of total variance explained, and with labels showing the cumulative variance explained. We see that 6 PCs explains $93.6\%$ of the total variance in the COSMOS2020 broadband colors, so we will focus on these going forward. In the lower panel of Figure \ref{fig:PCA}, we visualize the principal eigenvectors corresponding to each PC, with a bar plot showing the contribution of each color to the first 6 eigenvectors. We see that the first PC (explaining $\sim28\%$ of total variance) is dominated by $u-g$ and $g-r$, with the second and third (both explaining $\sim20\%$ of variance each) being IR-dominated.

In Figure \ref{fig:QQ}, we project the COSMOS2020 and \texttt{pop-cosmos} colors along the principal axes identified in Figure \ref{fig:PCA}. We then perform a comparison of the projected data and model colors using quantile--quantile (Q--Q) and probability--probability (P--P) plots \citep{wilk68}. As a numerical summary of discrepancy in quantile space, we quote the 2-Wasserstein distance $\mathcal{W}_2$, equal to the Euclidian distance between the data and model quantile functions \citep[see e.g.][]{munk98, levina01, ramdas17, peyre19}. In cumulative probability space, we use the K--S statistic $\mathcal{D}_\text{KS}$ as a numerical discrepancy measure, with this being equal to the maximal deviation in the P--P plot (for further discussion, see \citealp{thorp25}, and references therein). The agreement between model and data is generally good along all six directions, but with some mismatches. For the first PC, we see in the P--P plot a mismatch in the right-hand part of the distribution, suggesting that the model has more probability density in this direction than the data, implying bluer $u-g$ and $g-r$ colors than are seen in COSMOS2020 (see also Figure \ref{fig:colors_ch126}). The level of disagreement between model and data in the direction of these two colors was also largest in \citet{thorp25}; here the maximum deviation in the P--P plot is slightly larger ($\mathcal{D}_\text{KS}=0.100$; compared to $\mathcal{D}_\text{KS}=0.055$ there). The second and third PCs show similar mismatches to each other, suggestive of the model being under-dispersed relative to the data in these directions. For PC2, which is dominated by $Y-J$ and $K_S-\textit{Ch.\,1}$, this under-dispersion appears fairly symmetric, whereas for PC3 it is stronger in the right tail (which corresponds to red $Y-J$ and $H-K_S$, and blue $J-H$ and $K_S-\textit{Ch.\,1}$). For PC2 the K--S statistic, $\mathcal{D}_\text{KS}=0.044$, is relatively low compared to the 2-Wasserstein distance, $\mathcal{W}_2=0.681$, suggesting that the distribution is most discrepant in the tails (which the Q--Q plot/Wasserstein distance are most sensitive to). In the color marginals (Figure \ref{fig:colors_ch126}, the $K_S-\textit{Ch.\,1}$ color is visibly very discrepant in the blue tail, as was the case with the \citetalias{alsing24} model. As discussed in Section \ref{sec:results_photo}, this color shows the largest systematic difference between the different photometric extraction methods deployed in \citet{weaver22}.

The fourth, fifth, and sixth PCs each explain $\sim10\%$ of the total variance in COSMOS2020. For PC4, we see a consistent small bias in the negative direction, stronger towards the right tail of the distribution (c.f.\ the skew-normal in fig.\ 3 of \citealp{thorp25}). The closest match between model and data is seen for PC5, which corresponds to an even mixture of all colors that would imply a blue tilt in the SED between $g$- and $H$-band. A good match is also seen in PC6, which corresponds to a positive slope in the SED between the $i$- and $K_S$-bands.

\section{Redshift Dependent Mass Limits}
\label{sec:mcuts}

\begin{figure}
    \centering
    \includegraphics[width=\linewidth]{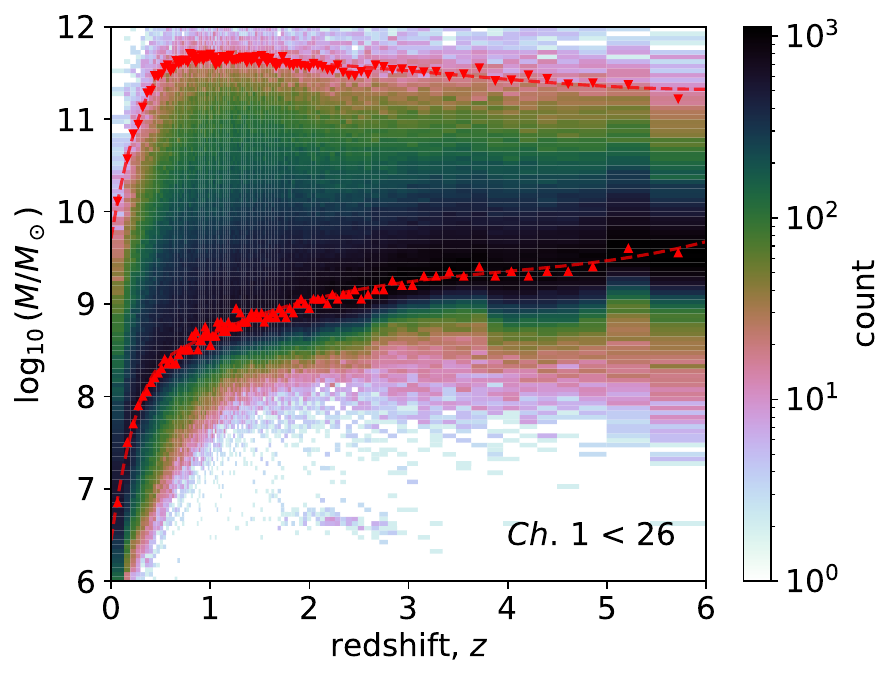}
    \caption{Histogram of stellar mass vs.\ redshift for \texttt{pop-cosmos} with $\textit{Ch.\,1}<26$. Redshift binning is based on percentiles of $z$, so each vertical slice of the histogram contains 20,000 mock galaxies. Red markers show the redshift binned turnover of the mass function (upward pointing triangles), and the 99.5th percentile in stellar mass (downward pointing triangles). Red dashed lines show the fitting functions of the form $\mathcal{M}(z)$ based on the results in Section~\ref{sec:mmax}--\ref{sec:mcomplete}.}
    \label{fig:mass_redshift_fit}
\end{figure}

In this Appendix, we define fitting formulae for the upper (Section~\ref{sec:mmax}) and lower (Section~\ref{sec:mcomplete}) limits of stellar mass where we expect \texttt{pop-cosmos} to be most reliable. These limits can be used to select \texttt{pop-cosmos} model galaxies that ought to be well represented in the COSMOS data itself (based on the $\textit{Ch.\,1}<26$ limit). The upper limit in Section~\ref{sec:mmax} is designed to mask very massive galaxies with $M\gtrsim10^{11}M_\odot$. Due to the size of the COSMOS field, these galaxies are not well represented in the \texttt{pop-cosmos} training data. The lower limit in Section~\ref{sec:mcomplete} is based on the estimated completeness of our model in stellar mass.

For both mass limits, we define a fitting function, $\mathcal{M}(z)=\log_{10}[M_\text{lim}(z)/M_\odot]$, using a three-knot cubic Hermite spline. This is defined by the redshifts $(z_0,z_1,z_2)$, log stellar masses $(\mathcal{M}_0, \mathcal{M}_1, \mathcal{M}_2)$, and gradients $(s_0, s_1, s_2)$ of $\mathcal{M}(z)$ at three positions. The general form of the fitting function for $z_k\leq z< z_{k+1}$ is
\begin{equation}
    \begin{split}
    \mathcal{M}(z) &= \mathcal{M}_k a(t_k) + \mathcal{M}_{k+1} b(t_k) \\&+ (z_{k+1}-z_k)s_kc(t_k) \\&+ (z_{k+1}-z_k)s_{k+1}d(t_k)
    \end{split}
\end{equation}
where $t_k = (z-z_k)/(z_{k+1}-z_k)$, and
\begin{align}
    a(t) &= 2t^3 -3t^2 +1 \\
    b(t) &= -2t^3 + 3t^2 \\
    c(t) &= t^3 - 2t^2 + t\\
    d(t) &= t^3 - t^2,
\end{align}
are the Hermite basis functions. Galaxies that lie beyond the limits defined by our fitted $\mathcal{M}(z)$ are included in the public catalogs, but are flagged. The fits described in Section~\ref{sec:mmax}--\ref{sec:mcomplete} are shown graphically in Figure \ref{fig:mass_redshift_fit}.

\begin{figure*}
    \centering
    \includegraphics[width=\linewidth]{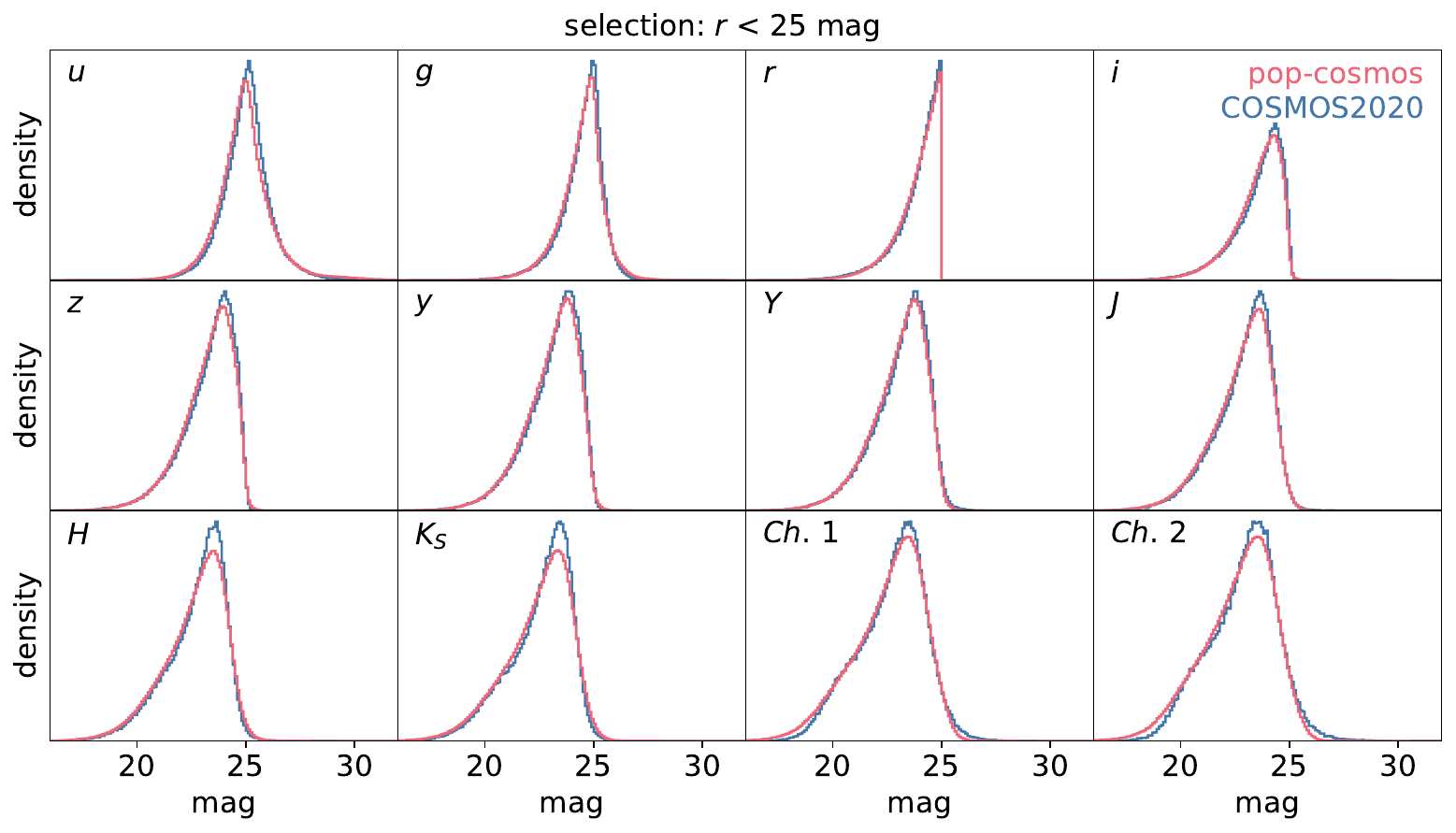}
    \caption{Broadband magnitudes (logarithmic) predicted by the new \texttt{pop-cosmos} generative model, compared to COSMOS2020 \citep{weaver22} for $r<25$.}
    \label{fig:mags_r25}
\end{figure*}

\begin{figure*}
    \centering
    \includegraphics[width=\linewidth]{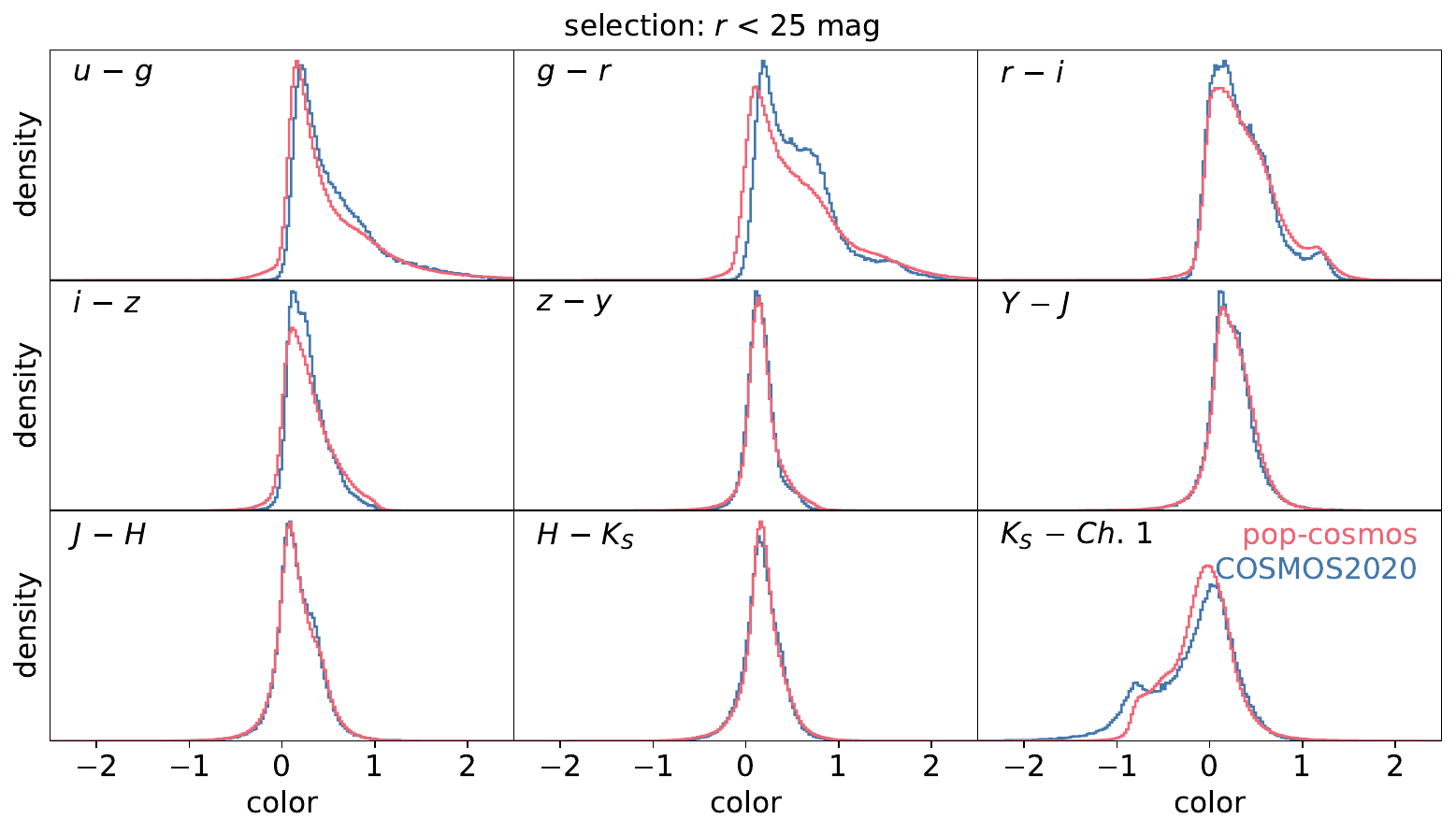}
    \caption{Broadband colors (logarithmic, adjacent bands) predicted by the new \texttt{pop-cosmos} generative model, compared to COSMOS2020 \citep{weaver22} for $r<25$.}
    \label{fig:colors_r25}

    \centering
    \includegraphics[width=\linewidth]{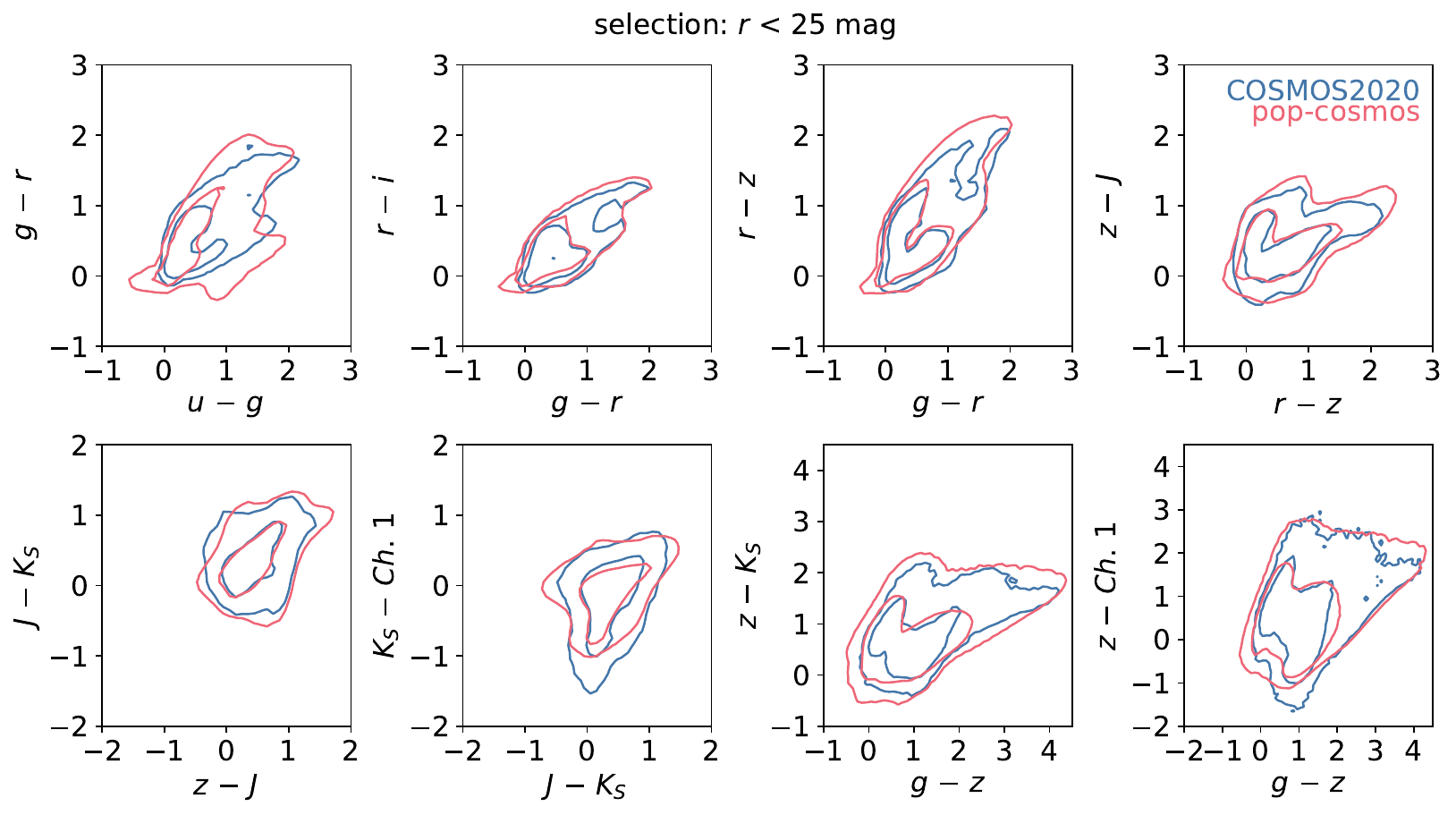}
    \caption{Broadband colors (logarithmic, \citetalias{alsing24} band combinations) predicted by the new \texttt{pop-cosmos} generative model, compared to COSMOS2020 \citep{weaver22} for $r<25$. Contours enclose the 68 and 95\% highest density regions.}
    \label{fig:color_color_r25}
\end{figure*}

\subsection{Mask for Over-Massive Galaxies}
\label{sec:mmax}
We find that in the extreme tails of the trained \texttt{pop-cosmos} model, there are mock galaxies more massive than almost any that are found in the COSMOS field (e.g., via the \texttt{LePhare} fits from \citealp{weaver22, weaver23_smf}; the individual-galaxy posterior inferences from \citetalias{thorp24}; or the updated individual-galaxy inferences in Section~\ref{sec:results_mcmc_mass}). We find that a mask for the \texttt{pop-cosmos} model draws based on the redshift-dependent 99.5th percentile of stellar mass is a reasonable cut for removing these. We divide our mock catalog of two million galaxies into 100 redshift bins each containing 1\% of the catalog. For each bin, we compute the 99.5th percentile of $\log_{10}(M/M_\odot)$.

We fit a two-segment cubic Hermite spline $\mathcal{M}(z)$ to these binned 99.5th percentiles via least squares. We fix $z_0\equiv0$, $z_2\equiv6$, $s_1\equiv s_2\equiv0$, and optimize the remaining parameters. The resulting fit has $z_1=0.7948$, $\mathcal{M}_0=9.7108$, $\mathcal{M}_1=11.6535$, $\mathcal{M_2}=11.3251$, and $s_0=6.4555$. This is shown as a gray dashed line on Figure \ref{fig:mass_redshift}, and the upper red dashed line on Figure \ref{fig:mass_redshift_fit}.

\subsection{Mask for Mass Incompleteness}
\label{sec:mcomplete}
We estimate a mass completeness limit for our mock galaxy catalogs by computing the rolling turnover of the stellar mass function in the same 100 redshift bins as in Section~\ref{sec:mmax}. We estimate the turnover based on the mode of the mass distribution computed with a bin size of $\Delta\log_{10}(M/M_\odot)=0.05$. We perform a least squares fit of $\mathcal{M}(z)$ to this estimated turnover, with $z_0\equiv0$ and $z_2\equiv6$ fixed. The resulting fit has $z_1=0.7404$, $\mathcal{M}_0=6.5184$, $\mathcal{M}_1=8.5429$, $\mathcal{M_2}=9.6966$, $s_0=6.9873$, $s_1=0.5549$, and $s_2=0.2956$. The completeness function is overplotted on Figures \ref{fig:mass_redshift} and \ref{fig:mass_redshift_fit} with these parameters. We find that the resulting curve for $0.2\lesssim z\leq 6.0$ is within $\pm0.3$~dex of the total mass completeness curve, $-3.23\times10^7 (1+z) + 7.83\times10^7(1+z)^2$, provided by \citet{weaver23_smf} for COSMOS2020. For $z\lesssim0.2$, we find our stellar mass function turns over at lower masses than the \citet{weaver23_smf} curve by up to $0.8$~dex.

\section{Further Color and Magnitude Distributions}
\label{sec:results_photo_extra}

\subsection{Broadband Colors and Magnitudes for $r<25$}
Figures \ref{fig:mags_r25}, and \ref{fig:colors_r25} show respectively the broadband magnitudes and colors predicted by the trained model for $r<25$. For this $r$-band limited mock catalog, the model magnitudes are in excellent agreement with the COSMOS2020 \texttt{Farmer} magnitudes. The $r<25$ agrees better with COSMOS2020 for the NIR and MIR colors than for the optical colors. Figure \ref{fig:color_color_r25} shows 2D color vs.\ color plots for the color pairs used by \citetalias{alsing24}.

\begin{figure*}
    \centering
    \includegraphics[width=\linewidth]{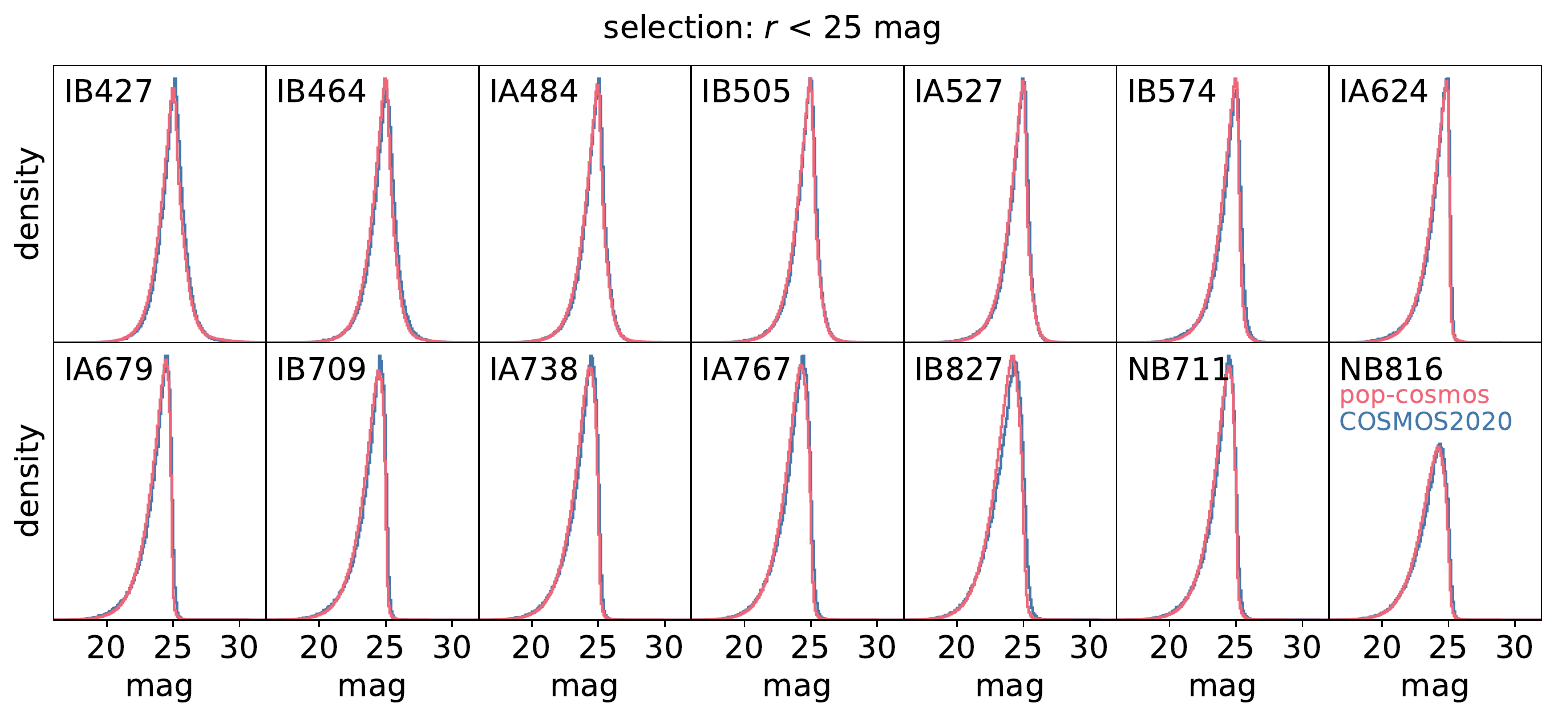}
    \includegraphics[width=\linewidth]{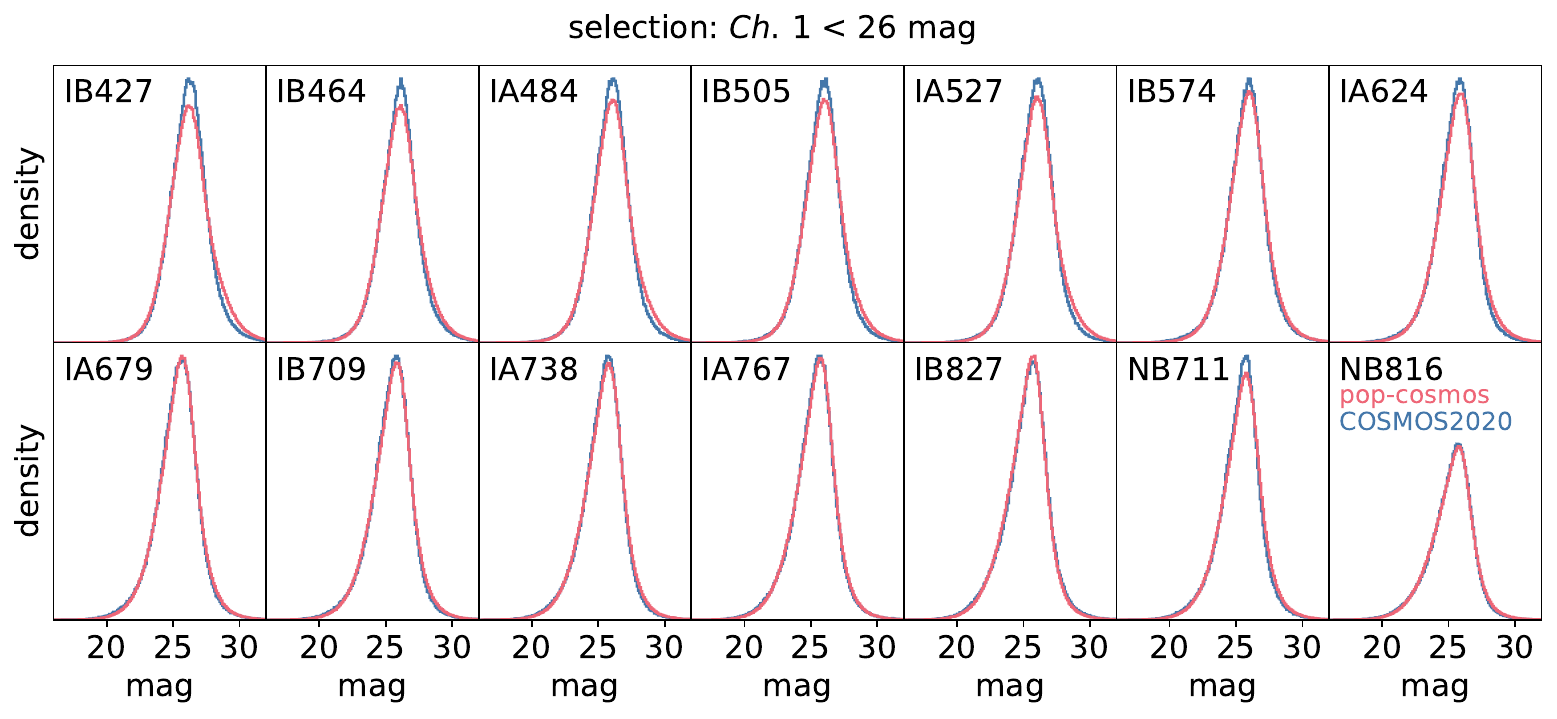}
    \caption{Narrow- and intermediate-band magnitudes from \texttt{pop-cosmos} vs.\ COSMOS2020 \citep{weaver22}.}
    \label{fig:mags_NB}
\end{figure*}

\begin{figure*}
    \centering
    \includegraphics[width=\linewidth]{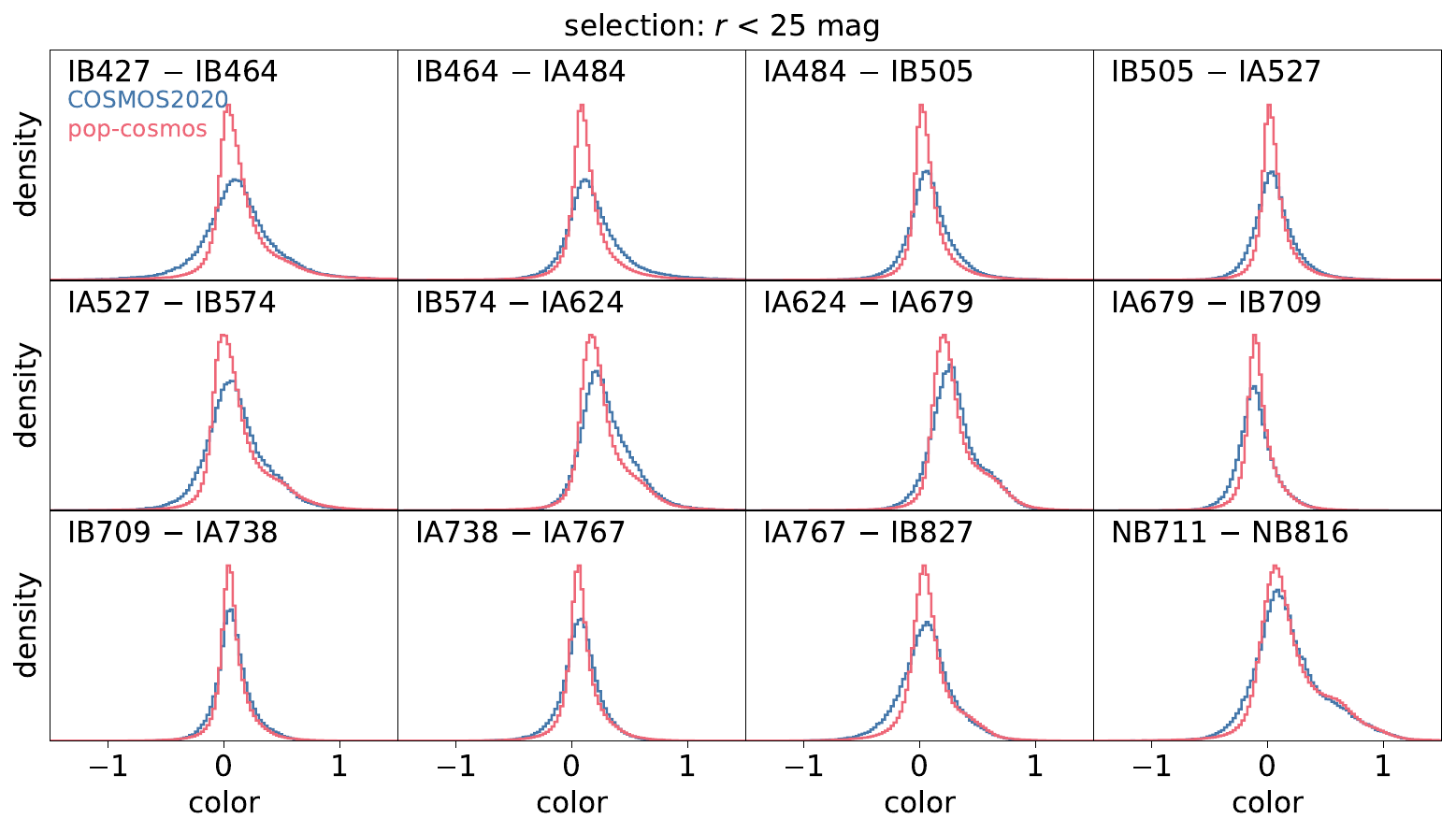}
    \includegraphics[width=\linewidth]{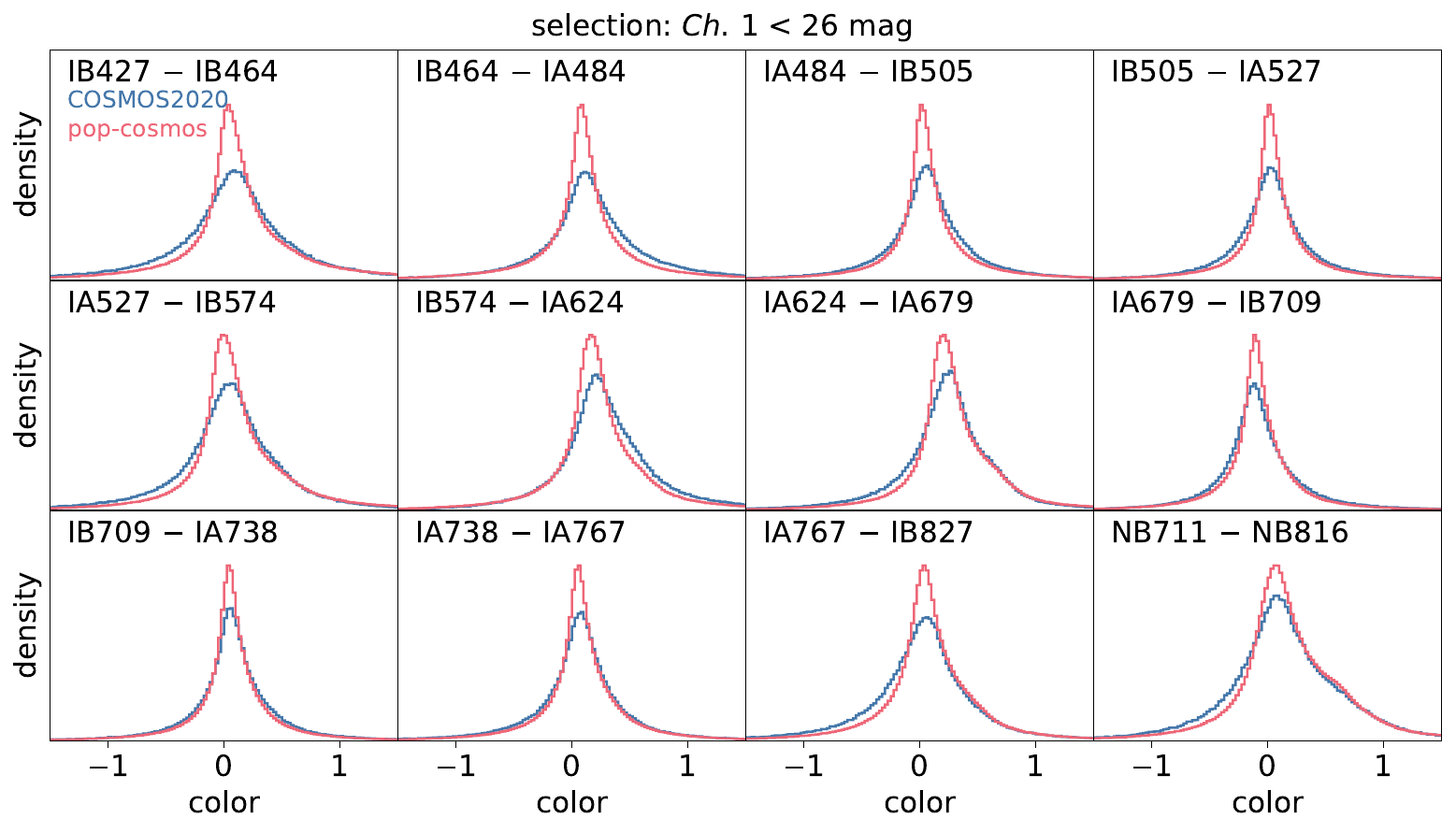}
    \caption{Narrow- and intermediate-band colors from \texttt{pop-cosmos} compared to COSMOS2020 \citep{weaver22}.}
    \label{fig:colors_NB}
\end{figure*}

\subsection{Narrowband Colors and Magnitudes}

Figure \ref{fig:mags_NB} shows the marginal distributions of the 14 narrow- and intermediate-band magnitudes predicted by \texttt{pop-cosmos} for the $r$-band- and $\textit{Ch.\,1}$-selected mock catalogs, compared to the COSMOS2020 catalog \citep{weaver22}. Figure \ref{fig:colors_NB} shows a set of 11 colors constructed from adjacent pairs of intermediate bands, and the pair of narrow bands. The magnitudes show excellent agreement for both selections. The \texttt{pop-cosmos} colors tend to be slightly under-dispersed relative to COSMOS2020, but the central tendency and widths of the distributions are in close agreement.

\subsection{Uncertainty Model Predictions for $r<25$}

\begin{figure*}[t!]
    \centering
    \includegraphics[width=\linewidth]{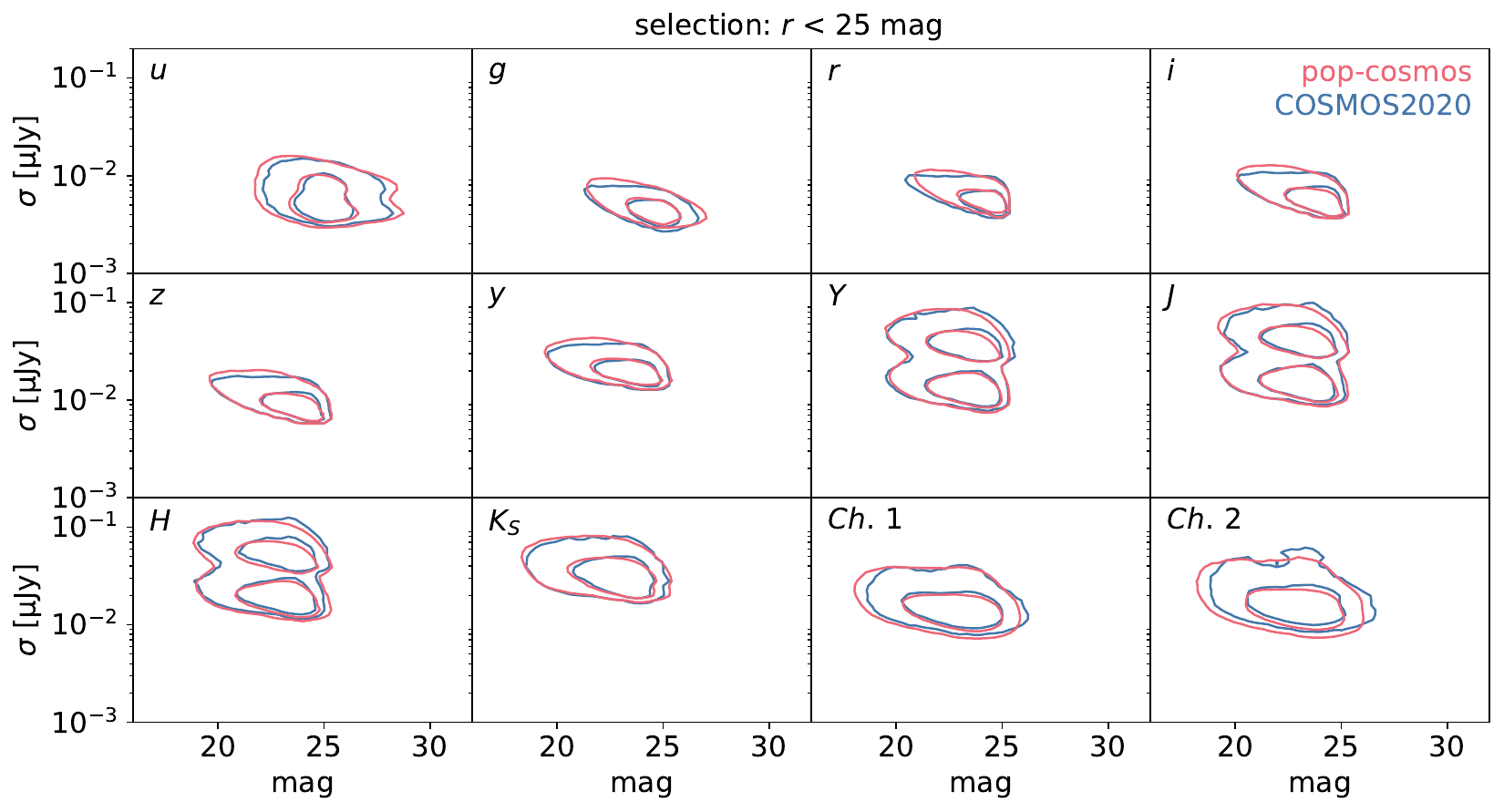}
    \caption{Flux uncertainty (micro Janskys, logarithmic axes) vs.\ magnitude (logarithmic) for the COSMOS broad bands, compared to the flux errors reported in COSMOS2020 for $r<25$. Contours enclose the 68 and 95\% highest density regions.}
    \label{fig:uncs_r25}
\end{figure*}

In Figure \ref{fig:uncs_r25}, we show the predictions made by the uncertainty model for the $r<25$ mock galaxy catalog, compared to the COSMOS2020 \texttt{Farmer} catalog \citep{weaver22}. The agreement here is excellent for all of the broadbands. Despite the $r<25$ selection being relatively shallow compared to the $\textit{Ch.\,1}<26$ cut, we still see substantial bimodality in the $YJHK_S$ bands from UltraVISTA, arising from the mix of deep and ultra-deep stripes \citep{mccracken12}.

\section{Full SPS Parameter Distribution}
\label{sec:corner}

\begin{figure*}
    \centering
    \includegraphics[width=\linewidth]{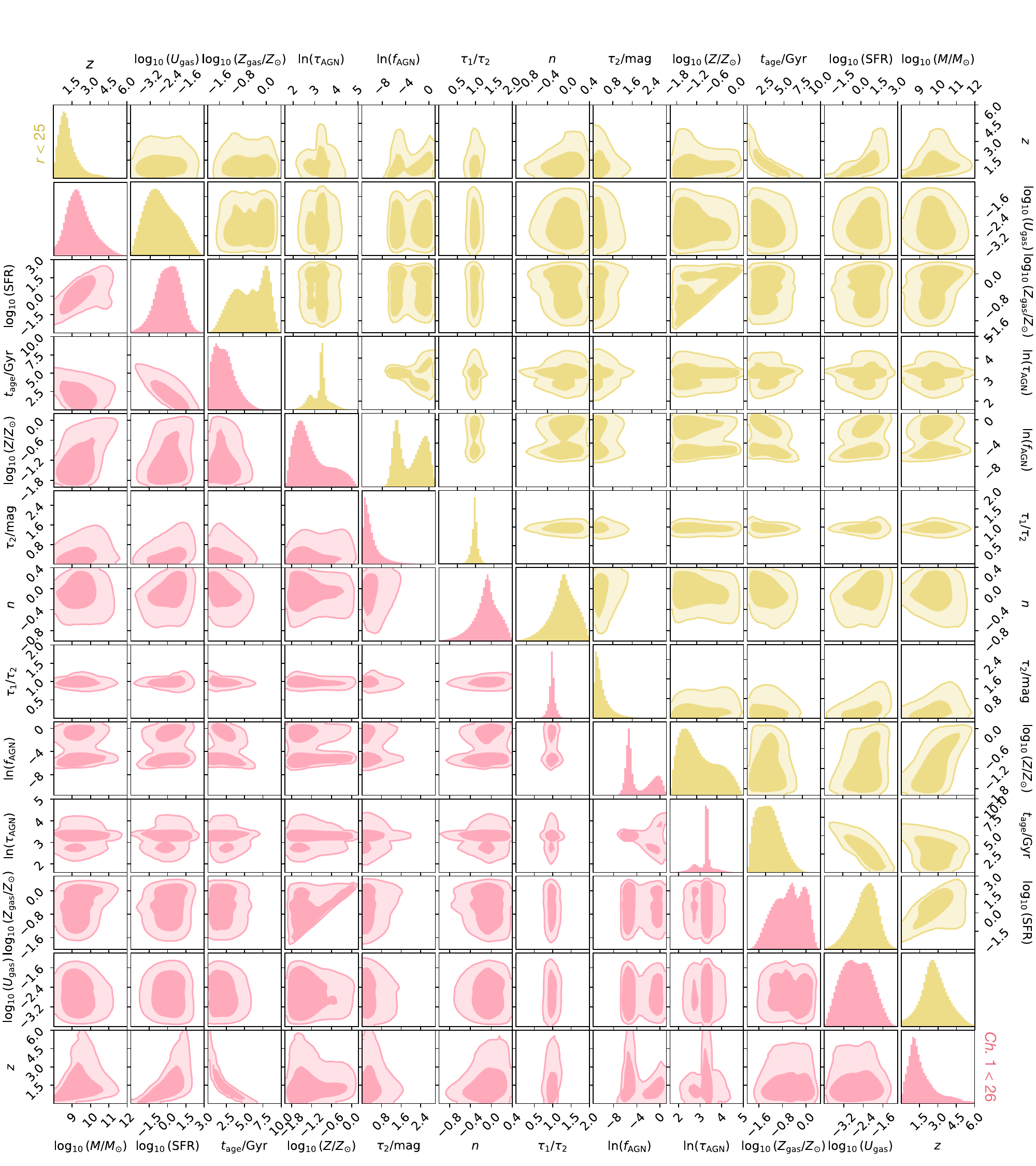}
    \caption{Marginals of the \texttt{pop-cosmos} population distribution over SPS parameters for \textbf{(top triangle, yellow)} $r<25$, and \textbf{(bottom triangle, red)} $\textit{Ch.\,1}<26$. Contours enclose the 68 and 95\% highest probability density regions.}
    \label{fig:corner}
\end{figure*}

In Figure \ref{fig:corner}, we show 1D and 2D marginals of the SPS parameter distribution implied by the \texttt{pop-cosmos} model. In place of the SFR ratios between the bins of our non-parametric SFH, we show SFR averaged over the last 100~Myr, and mass weighted age. 

\section{Possible Augmentations to the SPS Model}
\label{sec:discussion_sps}

Our existing SPS configuration has proven to be very successful, both within \texttt{pop-cosmos}, and as the basis of the widely used \texttt{Prospector} model family \citep{leja17, leja18, leja19, leja19_sfh, johnson21, wang23}. Nevertheless, these models are under active development, and we will continue to revisit the modeling choices, particularly in future extensions of the model to high-$z$ data or spectroscopy. In the following subsections, we highlight some aspects of the SPS model that will be particularly interesting to revisit.

\subsection{Stellar Templates}
\label{sec:stars}

Thus far, we have made use of the MILES empirical stellar template library \citep{sanchez06, falcon11}, combined with the MIST evolutionary tracks \citep{dotter16, choi16}, and a \citet{chabrier03} IMF as the basis of our SPS model. Whilst the MIST models --- based on the MESA stellar evolution code \citep{paxton11, paxton13, paxton15} --- include stellar rotation, they do not incorporate the effect of binary stars which can have significantly different evolutionary pathways \citep[see, e.g., the reviews by][]{eldridge19, eldridge22} and which are expected to be ubiquitous \citep{sana12}. The incorporation of binary stars can boost the strength of the ionizing continuum, leading to bluer galaxy SEDs and stronger nebular emission \citep[see, e.g.,][]{eldridge09, eldridge12, xiao18, gotberg17, gotberg19}; this is necessary to explain spectroscopic observations at high redshifts \citep[e.g.,][]{stanway14, steidel14, steidel16, larson23}. Future iterations of \texttt{pop-cosmos} could explore the impact of this using the Binary Population \& Spectral Synthesis models \citep{eldridge09, eldridge12, eldridge17, stanway18, byrne22}, which are included as template simple stellar populations (SSPs) in FSPS with \texttt{Cloudy}-based nebular emission \citep{garofali24}.

Another potential improvement would be the incorporation of $\alpha$-enhancement (super-solar $[\alpha/\mathrm{Fe}]$, where $\alpha=\{\mathrm{C}, \mathrm{O}, \mathrm{Ne}, \mathrm{Mg}, \mathrm{S}, \mathrm{Ar}\}$) into the underlying stellar libraries. This is observationally supported at $z\gtrsim2$ \citep[e.g.,][]{steidel16, strom17, strom22, cullen19, stanton24}, and can have effects on line strengths and ratios, as well as broadband colours \citep[see, e.g.,][]{park24}. Self-consistent SSPs with $\alpha$-enhancement are now available for the MIST isochrones \citep{park24} and single stars from BPASS \citep{byrne22, byrne25}.

\subsection{Nebular Emission}
As already noted in Section~\ref{sec:stars}, binary stellar evolution and the presence of $\alpha$-enhancement can modify the nebular emission properties of galaxies. Contributions from AGN (see Section~\ref{sec:agn}) can also introduce additional emission lines with variable widths and strengths that are not captured well by a single universal template \citep[see, e.g.,][]{marshall22}. Beyond improving the underlying stellar and AGN templates, we can incorporate more flexible emission line corrections into our forward-modeling to handle these effects. Currently, we use the approach introduced by \citetalias{leistedt23}, whereby the emission line strengths are based on FSPS \citep{byler17} and treated as delta functions pushed through approximations of the bandpasses. A more sophisticated model would allow for variation of the line widths as well as their amplitudes. Rather than inferring empirical corrections to a base set of precomputed nebular emission models, direct on-the-fly emulation of the photoionization calculations is a promising alternative route to more flexible emission line treatment \citep[e.g.,][]{li24_gs, li24_cue, naidu25, morisset25}.  A more detailed line model would be best calibrated directly with spectroscopic data, as has been explored recently by \citet{khederlarian24} using DESI Bright Galaxy Survey data \citep{hahn23_bgs}.

\subsection{AGN Modeling}
\label{sec:agn}

Our existing AGN treatment follows \citet{leja18} in using the \citet{nenkova08i, nenkova08ii} templates for emission from a dusty torus. As discussed in \citet{leja18}, the re-radiated emission from the AGN torus serves primarily to enhance the total MIR emission from a galaxy. In the absence of an AGN, MIR luminosity typically ascribed to the re-radiation of starlight absorbed at UV and optical wavelengths, with a tight correlation between UV attenuation and MIR emission following from a balancing of the energy budget \citep[e.g.,][]{rowan92, buat96, efstathiou00, witt00, buat02, buat05, burgarella05, dacunha08}. A dust-shrouded AGN can be invoked to explain MIR excess, with inhomogeneity or clumpiness in the torus (as in the \citealp{nenkova08i, nenkova08ii} models) being required to explain the diversity of emission scenarios \citep[see also][]{rowan95, nenkova02, mor09, honig10a, honig10b, honig17, siebenmorgen15, stavelski16, efstathiou13, efstathiou22}.

Significant emission in the MIR from the dust around AGN is highly prevalent ($\sim20$--40\% of galaxies; see, e.g., \citealp{kirkpatrick12, kirkpatrick15, juneau13}), and it has been shown to be a crucial component in accurate modeling of star formation histories \citep{cisela15, salim16, leja18}. However, depending on the orientation of the AGN and the amount of dust surrounding it \citep[see, e.g.,][]{antonucci93, urry95}, we may see significant flux from the accretion disk directly, which radiates primarily at UV and optical wavelengths \citep[see, e.g.,][]{harrison16, padovani17} with a UV--optical continuum that follows a power law \citep[e.g.,][]{koski78}. The impact of this on a galaxy SED can be non-negligible \citep[e.g.,][]{cardoso17}, and a variety of templates exist incorporating radiation from the disk, emission by hot dust, and broad and narrow emission lines \citep[e.g.,][]{maddox06, richards06, kubota18, temple21}. 

Although these templates have been incorporated alongside SPS-based galaxy models in SED fitting frameworks \citep[e.g.,][]{calistro16, cardoso17, boquien19, marshall22, martinez24, wang24_agn}, we have not yet included them within our forward model. Incorporating a more flexible and complete AGN model will be particularly important in future work to model spectroscopic data, where there will be increased sensitivity to the contributions of AGN-driven emission lines (which are highly diverse and dependent on black hole physics; e.g., \citealp{richards11, richards21, timlin20, rankine20, marshall22, temple21a, temple21, temple23}), as well as the suppression of stellar absorption lines due to the AGN UV continuum \citep[e.g.,][]{koski78, cid98, vega09, cardoso17}.  

\citet{tortorelli24} recently explored the impact on the broadband colors of galaxies drawn from the Prospector-$\beta$ prior \citep{wang23}, when adopting an AGN spectral template based on \citet{temple21} in place of the \citet{nenkova08i, nenkova08ii} torus model that is standard in FSPS. They found that the choice of AGN model had a moderate impact on observer-frame galaxy colors, and some difference of tomographic redshift bin assignments based on a self-organising map \citep{masters15, mccullough24}. They also showed that the optical colors ($g-r$, $r-i$, $i-z$) of SDSS quasars \citep{lyke20} are more closely matched by galaxies including the \citet{temple21} AGN models than those including the \citet{nenkova08i, nenkova08ii}. This is somewhat unsurprising given that (a) the \citet{temple21} models are calibrated against the \citet{lyke20} catalog, and (b) the \citet{nenkova08i,nenkova08ii} models only aim to model the MIR emission due to dust. Nevertheless, incorporating an AGN disk model alongside a torus model (as was done recently by \citealp{wang24_agn} in \texttt{Prospector}-$\beta$) would be a worthwhile extension of the \texttt{pop-cosmos} framework. 

\subsection{IGM Absorption}
We have so far used the \citet{madau95} model for the attenuation of SEDs by the IGM \citep{gunn65}. As the population of Ly-$\alpha$ absorbers in the Universe has become better understood, refined models for their impact have been developed \citep[see, e.g.,][]{bershady99, meiksin06, tepper08, inoue08, becker13, inoue14, kakiichi18}. The \citet{inoue14} update to the \citet{madau95} model predicts an IGM transmission function that agrees better with modern observations; particularly at $z\gtrsim5$ where the Ly-$\alpha$ optical depth increases very rapidly close to the epoch of reionization \citep[e.g.,][]{becker01, cen02, fan02, fan06, white03, paschos05}. Although galaxies with $z\gtrsim4$ make up a fairly small fraction of our current training set, a post-\citet{madau95} IGM absorption model \citep[e.g.,][]{inoue14} may be needed in future work to extrapolate \texttt{pop-cosmos} to higher redshifts. Moreover, there is considerable variance observed in the amount of IGM absorption along different sightlines towards high-$z$ galaxies (see, e.g., \citealp{thomas17, thomas20, thomas21}), which can in principle be accounted for in FSPS/\texttt{Prospector} \citep[see][]{johnson21}. Other factors such as the damping wing that attenuates light redward of the Ly-$\alpha$ line center \citep[see, e.g.,][]{miralda98, madau00, totani06, schroeder13, bach15, mortlock16, durovcikova24} may merit inclusion when extending \texttt{pop-cosmos} to $z\gtrsim6$.

%
%
\bibliography{main}
\bibliographystyle{aasjournal}
%


\end{document}